\documentclass[prd,superscriptaddress,amsfonts,amssymb,amsmath,showpacs,twocolumn]{revtex4-2}
\usepackage{bm}
\usepackage{amsfonts}
\usepackage{latexsym}
\usepackage{graphicx}
\usepackage{amsmath}
\usepackage{palatino}
\usepackage{mathpazo}
\usepackage{textcomp}
\usepackage{float}
\usepackage{booktabs}
\usepackage{dcolumn}
\usepackage{booktabs}
\usepackage{multirow}
\usepackage{hyperref}
\hypersetup{colorlinks,citecolor=blue}
\usepackage{amsmath}
\usepackage{xcolor}
\usepackage{orcidlink}
\usepackage{epsfig}
\usepackage{caption}
\usepackage{subcaption}
\usepackage{commath}
\hypersetup{colorlinks,citecolor=blue}
\hypersetup{colorlinks=true,linkcolor=magenta,filecolor=magenta,    urlcolor=blue}

\usepackage{amssymb}
\usepackage{relsize,exscale}

\def\be{\begin{equation}}
\def\ee{\end{equation}}
\def\bea{\begin{eqnarray}}
\def\eea{\end{eqnarray}}

\begin{document}

\title{\bf Observable Signatures of RN Black Holes with Dark Matter Halos via Strong Gravitational Lensing and Constraints from EHT Observations}

\author{Niyaz Uddin Molla}
\email{niyazuddin182@gmail.com}
\affiliation{Department of Mathematics,Indian Institute of Engineering Science and Technology, Shibpur, Howrah 711 103, India}
\author{Himanshu Chaudhary}
\email{himanshu.chaudhary@ubbcluj.ro}
\affiliation{Department of Physics, Babeș-Bolyai University, Kogălniceanu Street, Cluj-Napoca, 400084, Romania,}
\author{Salvatore~Capozziello}
\email{capozziello@na.infn.it (Corresponding author)}
\affiliation{Dipartimento di Fisica “E. Pancini", Universit‘a di Napoli “Federico II", Complesso Universitario di Monte Sant’ Angelo, Edificio G, Via Cinthia, I-80126, Napoli, Italy}
\affiliation{Scuola Superiore Meridionale, Largo S. Marcellino 10, I-80138, Napoli, Italy}
\affiliation{SIstituto Nazionale di Fisica Nucleare (INFN), Sez. di Napoli, Complesso Universitario di Monte Sant’ Angelo, Edificio G, Via Cinthia, I-80126, Napoli, Italy}
\author{Farruh Atamurotov}
\email{atamurotov@yahoo.com}
\affiliation{University of Tashkent for Applied Sciences, Str. Gavhar 1, Tashkent 100149, Uzbekistan}
\author{G. Mustafa}
\email{ gmustafa3828@gmail.com (Corresponding author)} 
\affiliation{Department of Physics,
Zhejiang Normal University, Jinhua 321004, Peoples Republic of China}
\affiliation{Research Center of Astrophysics and Cosmology, Khazar University, Baku, AZ1096, 41 Mehseti Street, Azerbaijan}
\author{Ujjal Debnath}
\email{ujjaldebnath@gmail.com} 
\affiliation{Department of Mathematics,Indian Institute of Engineering Science and Technology, Shibpur, Howrah 711 103, India}


\begin{abstract}
We explore the influence of dark matter halos on gravitational lensing produced by electrically charged, spherically symmetric black holes in the strong-field regime. This study delves into the strong gravitational lensing effects within the context of two significant dark matter models: the Universal Rotation Curve Model and the cold dark matter model. Initially, we derive the coefficients for the strong deflection limit and numerically analyze the substantial variations of the deflection angle. Additionally, we present graphical representations of these results. We find that the strong deflection angle, denoted as $\alpha_D$, increases with the rising charge parameter magnitude $Q$ in presence of a dark matter halo. Furthermore, we examine the various astrophysical consequences of the compact objects $M87^*$ and $SgrA^*$, as charged black holes, and compare results with those for other astrophysical black holes such as standard Reissner-Nordstr\"{o}m (RN) and  Schwarzschild black holes, via strong gravitational lensing observations. From our study, the observations point out that it may be possible to quantitatively differentiate and characterize charged black holes with dark matter halos from astrophysical black holes, such as the standard Reissner-Nordstr\"{o}m and Schwarzschild black holes. Finally, We constrain the charge  parameter $Q$ with the observational data by the Event Horizon Telescope Collaboration for the supermassive black holes $M87^*$ and $SgrA^*$. We  constrain the charge parameter $Q$ of the charged black holes with universal rotation curve dark matter halo as $0\leq |Q|\leq 0.366M$ for $M87^*$,  $0\leq |Q|\leq 0.586M$  for $SgrA^*$. The charged black hole with cold dark matter halo can be constrained with  $0\leq |Q|\leq 0.364M$ for $M87^*$ , and $0\leq |Q|\leq 0.584M$  for $SgrA^*$. It suggests that such charged black holes with dark matter halo satisfy the Event Horizon Telescope constraints. These results suggest, in principle, that it  could  be possible to identify  charged black holes with dark matter  halo in the future fine observational campaigns.
\end{abstract}

\maketitle

\section{Introduction}
The study of black hole (BH) spacetimes represents a crucial area of  investigation within the realm of general relativity and other gravitational theories \cite{BH1,BH2,BH3,hawking1976black}. BHs offer a unique window into our understanding of gravitation, thermodynamics, and quantum effects in curved spacetime. Recent years have witnessed significant progress in both theoretical and observational explorations of BHs. Notably, the LIGO and Virgo collaborations detected gravitational wave signals arising from the mergers of binary BHs \cite{perna2018binary,cholis2017gravitational,zackay2021detecting,abbott2019gwtc,krolak2021recent,liu2020multiband,dominik2015double}. Furthermore, there have been groundbreaking observations of high-resolution images of BHs at the center of our Milky Way galaxy, known as Sagittarius A* (Sgr A*), which has indeed been a major focus of study and observation by the Event Horizon Telescope (EHT). The EHT captures high-resolution images of BHs. They also released the first-ever image of a BH, which is the supermassive BH in the center of the M87 galaxy \cite{collaboration2019eht}.\\\\
BHs are classified into distinct types based on their properties, specifically mass, angular momentum (rotation), and electric charge. Firstly, Schwarzschild BHs \cite{schwarzschild1979gravitational}, characterized by their  mass and spherically symmetric, non-rotating nature, represent the simplest and most fundamental type. These BHs are defined by the Schwarzschild solution and harbor an event horizon beyond which nothing, not even light, can escape. Second, Kerr BHs, known as rotating BHs, possess both mass and angular momentum, as described by the Kerr solution. They feature an event horizon and an ergosphere due to their rotation, introducing distinctive characteristics such as frame-dragging effects. Within the category of charged BHs, Reissner-Nordstr\"{o}m (RN) BHs emerge as the third type. These BHs, encapsulated by the RN solution, carry both mass and  electric charge, with this charge potentially being positive, negative, or zero.\\\\
Furthermore, dark matter (DM) is a fascinating and mysterious topic that continues to captivate researchers in various scientific fields, including astrophysics, cosmology, etc. There is a growing body of evidence supporting the existence of DM, which comes from sources like the unusual rotation patterns of spiral galaxies \cite{van1978kinematics,corbelli2000extended}, gravitational lensing effects \cite{massey2010dark}, the formation of large-scale structures in the Universe \cite{blumenthal1984formation}, observations of the cosmic microwave background radiation \cite{serpico2020cosmic,thomas2016constraining}, and the study of baryon acoustic oscillations \cite{BAO1,BA02,BAO3}.  When we combine the data from studies on the cosmic microwave background, it becomes evident that approximately 26.8\% of the Universe is composed by DM, while dark energy, generating the observed accelerated expansion of the cosmic flow,  makes up a substantial 68.3\% of the cosmic coonstituents \cite{sharma2022exploring}. The presence of DM has a significant and long-lasting impact on the evolution and ultimate destiny of our Universe, playing a crucial role in shaping the cosmos. The leading contenders for DM are theoretical particles proposed by theories that go beyond the Standard Model of particle physics. These potential candidates encompass weakly interacting massive particles (WIMPs) \cite{gelmini2017light}, axions \cite{duffy2009axions}, sterile neutrinos \cite{boyarsky2019sterile}, dark photons \cite{hambye2019dark}, and millicharged particles \cite{liu2019reviving}. However, beside several astrophysical and cosmological evidences, DM escapes, up to now,  any probe at fundamental level so no final answer on its particle nature is today available.
DM typically forms halo-like structures around galaxies \cite{salucci2003intriguing,lovell2012haloes}, intriguingly encompassing the supermassive BHs situated at their cores \cite{alfred2007our,posti2019mass,Capozziello:2023rfv, Capozziello:2023tbo}. Hence, delving into and grasping the implications of DM halos on supermassive BHs in galactic centers holds paramount importance. In recent studies, researchers have delved into the influence of DM on the behavior of BHs. These investigations have spanned a wide range of subjects, encompassing circular geodesics \cite{hod2011hairy,rindler1990rotating}, the formation of BH shadows \cite{Wek5,shadows2,shadows3}, gravitational lensing \cite{bartelmann2010gravitational}, the dynamics of accretion disks \cite{accretion1,accretion2,accretion3,accretion4}, thermodynamics \cite{hawking1976black,Thermo1}, quasi-normal modes \cite{QNM1,kokkotas1999quasi,leaver1985analytic}, the phenomenon of BH echoes \cite{d2020black}, and even the complex processes of BH mergers. It is possible also to investigate alternative theories of gravity considering the BH shadows. See, e.g. Refs. \cite{Jusufi:2022loj,Addazi:2023pfx,Addazi:2021pty}.
Understanding the properties of DM, particularly its halo structures within galaxies, can be approached through various observational methods. Gravitational lensing is one of the most useful techniques in this regard. It is an extraordinary phenomenon rooted in  Einstein's theory of general relativity, which has immense implications for our understanding of the Universe \cite{renn1997origin,bennett2005astrophysical}. It occurs when massive objects, like galaxies or groups of galaxies, influence the very fabric of space and time, causing it to curve or bend. As a result, when light travels through this curved space, its path is altered, creating an effect akin to a cosmic magnifying glass. This lensing effect allows us to see and study distant celestial objects that would otherwise remain hidden from our view \cite{hacking1989extragalactic,wambsganss1998gravitational,treu2010strong}. It aids us in the discovery of galaxies located far away from us, which would be nearly impossible to observe directly \cite{lawrence1984discovery}. Moreover, it assists in mapping the distribution of DM, the enigmatic and elusive substance that makes up a substantial portion of the Universe mass \cite{jain2000statistics,kaiser1993mapping}. Gravitational lensing also plays a crucial role in the precise measurement of the Hubble constant \cite{jee2019measurement}. It encompasses two main categories: weak lensing and strong lensing, each with its own distinct characteristics and applications. Weak lensing occurs when the gravitational influence of a massive object, like a galaxy or a  galaxy cluster, subtly distorts the light from background objects. While the distortion is typically quite small, it provides invaluable insights into the distribution of matter in the foreground object and the larger cosmic structure. On the other hand, strong lensing involves more pronounced distortions in the path of light due to highly massive objects. This can create dramatic and often multiple images of a background object, such as a distant galaxy, around the massive foreground object. Strong lensing allows astronomers to probe the detailed properties of both the lensing object and the background sources. It has led to remarkable discoveries, such as the detection of distant galaxies and the study of the innermost regions of quasars. Over the past few decades, it has evolved into a crucial tool in astronomy and cosmology, enabling the test of gravity theories and astrophysical models \cite{Capozziello:1999hi,Capozziello:2006dp,Wek1,Wek2,Wek3,Wek4,Wek5,Wek6,Wek7,Wek8,Len1,Len2,Len3,Len4}.\\\\
The concept of gravitational lensing was initially introduced within the weak field limit \cite{Refsdal:1964yk, Refsdal:1964nw, schneider1992gravitational, petters2001singularity}. Subsequently, it garnered extensive exploration in the strong field limit \cite{darwin1959gravity, Claudel:2000yi, Virbhadra:2022iiy,Virbhadra:2008ws,Virbhadra:2022iiy,Virbhadra:2022ybp,Virbhadra:2007kw,Bozza:2001xd, Bozza:2005tg}. Pioneering research by Darwin on the trajectories of photons in the vicinity of a BH revealed significant bending of light rays. The foundational work by \cite{Frittelli:1999yf} played a crucial role in establishing the groundwork for a precise lens equation. Subsequent numerical investigations, as discussed in \cite{Virbhadra:1998dy,Virbhadra:1999nm}, focused on the lensing phenomenon induced by static, spherically symmetric naked singularities. These studies provided a comprehensive examination of various lensing observables. Building on these contributions, later in \cite{Perlick:2003vg} explored gravitational lensing within spherically symmetric and static spacetime.  Significantly, this exploration utilized the light-like geodesic equation without relying on approximations. The results derived from these investigations emphasize the profound importance of gravitational lensing, especially concerning BHs, underscoring its invaluable utility as a tool in astrophysics. Delving into the substantial gravitational effects near massive celestial bodies offers valuable insights into the characteristics of distant and faint stars. The phenomenon of gravitational lensing by various BHs has become a focal point in recent decades, underscoring its crucial role in astrophysical studies. Scientists have employed both functional and analytical approaches, drawing inspiration from the methods introduced in \cite{Bozza:2001xd}. 
These approaches are based on the principle of the substantial bending of light rays. Through these methodologies, it has been discovered that as light rays approach the photon sphere of a Schwarzschild BH, the deflection angle undergoes a logarithmic divergence. This approach extends beyond Schwarzschild BHs and has been applied to various scenarios. For instance, it has been employed in the investigation of RN BHs \cite{Eiroa:2002mk} and implemented in metric systems characterized by staticity and spherical symmetry \cite{Bozza:2002zj}.\\\\
In depth exploration of gravitational lensing has emerged from the realization that analyzing the trajectories of light in the vicinity of BHs offers valuable perspectives on the intrinsic features of the surrounding spacetime. Researchers have delved into the phenomenon of gravitational lensing resulting from diverse sources, encompassing the distribution of cosmic structures \cite{Mellier:1998pk,Bartelmann:1999yn,Heymans:2013fya}, dark energy \cite{Biesiada:2006zf,DES:2020ahh,Huber:2021nhx}, DM \cite{Kaiser:1992ps,clowe2006direct,Atamurotov:2021hoq}, quasars \cite{SDSS:2000jpb,Peng:2006ew,Oguri:2010ns,yue2022revisiting}, gravitational waves \cite{Seljak:2003pn,Diego:2021fyd,finke2021probing}, and various other compact celestial objects \cite{Liao:2015uzb,Nascimento:2020ime,Junior:2021svb,Molla:2022mjl,Molla:2022izk}. A variety of investigations collectively enhance our comprehension of the complex interplay among matter, energy, and spacetime. Notably, the Event Horizon Telescope collaboration has achieved a significant advancement in angular resolution. This breakthrough enables the observation of the supermassive BH situated at the core of the M87 galaxy. This remarkable achievement inaugurates a novel phase in exploring gravitational lensing within environments characterized by strong gravitational fields. \cite{EventHorizonTelescope:2019uob,EventHorizonTelescope:2019jan,EventHorizonTelescope:2019ths,EventHorizonTelescope:2019pgp,EventHorizonTelescope:2019ggy}. Moreover, the utilization of relativistic images for investigating gravity under strong deflection conditions has found application in diverse BH metrics. This encompasses examinations in both the framework of general relativity and modified gravitational theories. Numerous scholars have made significant contributions to this realm, as evident in \cite{fernando2002gravitational,Bozza:2003cp,majumdar2005gravitational,Eiroa:2008ks,bozza2014gravitational,Sahu:2015dea,Islam:2021dyk}. In this study, our aim is to thoroughly examine two models of DM halos within the framework of strong gravitational lensing, specifically in the context of static and charged spacetime.\\\\
The structure of this paper is organized as follows: Section \ref{sec2} provides a brief overview of the structure of the RN BH spacetime in the presence of DM halos. Moving on to Section \ref{sec3}, we delve into the gravitational lensing phenomena induced by BHs, detailing aspects such as the lens equation, deflection angle, and strong lensing coefficients. This section focuses in particular on two supermassive BHs $M87^*$ and $SgrA^*$ within the framework of charged BHs accompanied by DM halos. We explore various strong lensing observables, including the angular position of the innermost image, image separation for the first and \(n^{th}\) relativistic images, brightness differences among these images, and the time delay between the first and second relativistic images. Section \ref{sec4} conducts a comparative analysis of strong lensing observables, contrasting the outcomes between standard Schwarzschild BH, standard RN BH, and charged BH spacetime coupled with a DM halo. In this comparative study two distinct models, the universal rotation curve (URC) and cold dark matter (CDM) halo with the Navarro-Frenk-White (NFW) profile are considered. Moving forward, Section \ref{sec5} is dedicated to constraining the charge parameter \(Q\) with DM halos based on observational data from the Event Horizon Telescope (EHT) for both \(M87^*\) and \(SgrA^*\). Finally, in Section \ref{sec6}, we encapsulate our key findings and draw conclusions from the study.
\section{Black Hole Geometry and Dark Matter Halo}\label{sec2}
To examine the geometry around the BH in this section, we initiate our analysis with a spherically symmetric static spacetime characterized by the following metric:
\begin{equation}\label{1}
ds^2  =  - f(r) dt^2 + \frac{1}{f(r)} dr^2 + r^2 d\theta^2 + r^2 \sin^2\theta d\varphi^2,
\end{equation}
The BH lapse function is defined as $f(r)$ in the above equation. This study will center around the examination of the lapse function described in: \cite{accretion4, toledo2018black}:
\begin{equation}\label{2}
f(r)=1-\frac{2 M G_N}{c^2 r}+\frac{Q^2}{r^{2}},
\end{equation}
where $Q^2=(q^2 G_N)/(4\pi\epsilon_0 c^4)$. In addition, $q$ and $M$ represent the charge and mass of BH. Now, we shall incorporate the URC profile and CDM halo with NFW profile into the above BH geometry to see how they affect the BHs solutions. It is worth noticing that charged BH solutions constitute, today,  a very active research area due to their interesting properties in standard GR and in alternative gravities \cite{Capozziello:2023vvr,Nashed:2019tuk,Nashed:2017fnd,Awad:2017tyz}.
\subsection{Universal Rotation Curve Profile}
The well-known DM halo-like URC profile will be briefly discussed here in this subsection. According to \cite{salucci2000dark}, the following equation characterizes the distribution of URC profile
\begin{equation}\label{3}
\rho(r) = \frac{\rho_0 r_0^3}{(r+r_0)(r^2+r_0^2)},
\end{equation}
where $\rho_0$ and $r_0$ express the central density and the characteristic radius of the URC DM halo. The best fit values for $\rho_0$ and $r_0$ was given in \cite{donato2009constant, salucci2019distribution} by using the the recent observations. For the  M87* galaxy in the framework of URC DM profile, the involved parameters are measured as: $\rho_0 = 6.9\times 10^6 \text{M}_{\odot}/{\rm kpc}^{3}$ and $r_0 = 91.2\; {\rm kpc}$. For the other case, i.e., Milky Way galaxy $\rho_0 = 5.2 \times 10^7 \text{M}_{\odot}/{\rm kpc}^{3}$ and $r_0 = 7.8\; {\rm kpc}$ \cite{lin2019dark} are measured. Within the scenario of URC DM profile, the lapse function $f(r)$ \cite{jusufi2019black} from the metric Eq  (\ref{2}) is revised as:
\begin{equation}\label{4}
\begin{aligned}
f(r) = & e^{-\frac{2 \pi ^2 \rho _0 r_{0}^2 G_N}{c^2}} \left(\frac{r^2}{r_{0}^2}+1\right)^{-\frac{\left(1-\frac{r}{r_{0}}\right) \left(2 \pi  \rho _0 r_{0}^3 G_N\right)}{c^2 r}} \\
& \times \left(\frac{r}{r_{0}}+1\right)^{-\frac{\left(\frac{r}{r_{0}}+1\right) \left(4 \pi  \rho _0 r_{0}^3 G_N\right)}{c^2 r}} \\
& \times e^{\frac{4 \pi  \rho _0 r_{0}^3 G_N \tan ^{-1}\left(\frac{r \left(\frac{r}{r_{0}}+1\right)}{r_{0}}\right)}{c^2 r}} -\frac{2 M G_N}{c^2 r} + \frac{Q^2}{r^{2}}.
\end{aligned}
\end{equation}

In the above relation, we shall consider $M=4.3 \times 10^6~\text{M}_{\odot}$ for the Sgr A* BH and $M=6.5 \times 10^9~\text{M}_{\odot}$ for the M87* central BH according to \cite{jusufi2019black}. In Eq.  (\ref{4}), the lapse function is very complicated. For the present study, we shall discuss strong gravitational lensing by using the series solution alone for the sake of simplicity, which may be computed directly as:
\begin{widetext}
\begin{equation}\label{5}
f(r) = \left(1-2 \pi ^2 A B\right) \left(1-4 \pi  A B \left(\frac{B r}{M }+1\right)\right) \left(4 \pi  A B \left(\frac{B r}{M }+1\right)+1\right)  \left(\frac{2 \pi  A B^2 r \left(B r-M \right)}{M^2 }+1\right)-\frac{2 M}{r}+\frac{Q^2}{r^{2}}.
\end{equation}
\end{widetext}
In the above Eq. \eqref{5}, we have two new paramters namely $A$ and $B$. For the current analysis, both are calculated as $A=\frac{\rho _0 r_{0}^{3}}{M}$ and $B=\frac{M G_N}{c^2 r_{0}}$. Parameters $A$ and $B$ are determined now using the observationally well-fit values for M87*, as follows:
$B=\frac{M G_N}{c^2 r_{0}}=3.40611*10^{-9}$ and $A=\frac{\rho _0 r_{0}^{3}}{M}=805.231;$.
For Sgr A*, $A$ and $B$ are measured as: $B=\frac{M G_N}{c^2 r_{0}}=2.6346*10^{-11}$,
$A=\frac{\rho _0 r_{0}^{3}}{M}=5738.77.$
\subsection{The Cold Dark Matter Halo with Navarro Frenk White Profile}
The CDM halo model with NFW profile was constructed by using $N$-body simulations, commonly known as the universal spherically averaged density profile \cite{navarro1997universal}, and is stated as:
\begin{equation}\label{6}
\rho(r) = \frac{\rho_0}{(r/r_0)(1+r/r_0)^2},
\end{equation}
where $\rho_0$ and $r_0$ define the density of the universe of halo and characteristic radius respectively at the stage of collapsing. As per the latest observational findings related to the Milky Way galaxy \cite{lin2019dark}, the optimal numerical values for the parameters $\rho_0$ and $r_0$ are determined when considering the influence of CDM halo characterized by the NFW profile are measured as $\rho_0 = 5.23 \times 10^7 \text{M}_{\odot}/{\rm kpc}^{3}$ and $r_0=8.1 \; {\rm kpc}$. For the M87* BH model, these values are measured as $\rho_0 = 0.008 \times 10^{7.5}~\text{M}_{\odot}/ \text{kpc}^3$ (see \cite{oldham2016galaxy} and ${r_0} = 130~\text{kpc}$ \cite{jusufi2019black}. Within the context of the CDM halo profile, \cite{xu2018black} provides the following expression for the function $f(r)$ in the metric defined by Eq. (\ref{2}) as
\begin{equation}\label{7}
f(r)=\left(1+\frac{r}{{r_0}}\right)^{-\frac{8 \pi G_N \rho_0 r_0^3}{c^2 r}} - \frac{2 G_N M}{c^2 r}+\frac{Q^2}{r^{2}},
\end{equation}
The central BH mass for M87*, denoted as \(M\), is \(6.5 \times 10^9~\text{M}_{\odot}\), and for the Sgr A* BH model, the specific mass is \(M = 4.3 \times 10^6~\text{M}_{\odot}\). When considering strong gravitational lensing, the expression in Eq.  \eqref{7} becomes notably intricate due to the exponential function in the lapse function as defined by Eq. \eqref{2}. To address this complexity, it is essential to employ a series solution for the present BH lapse function, taking into account the influence of the CDM halo model with a NFW) profile. The updated form of the lapse function is now given by
\begin{eqnarray}\label{8}
f(r)&=&1-\frac{2M}{r}+32 \pi ^2 A^2 B^2\nonumber\\&+&\frac{4 \pi  A B (B r+2 M)}{M}+\frac{Q^2}{r^{2}}.
\end{eqnarray}
For the above relation values of the parameters $A$ and $B$ for M87* are measured as: $A=\frac{\rho _0 r_{0}^{3}}{M}=85.508;$
$B=\frac{M G_N}{c^2 r_{0}}=2.38952*10^{-9}$. For the Sgr A* BH model values of $A$ and $B$ are calculated as: $A=\frac{\rho _0 r_{0}^{3}}{M}=6463.81;$
$B=\frac{M G_N}{c^2 r_{0}}=2.537*10^{-11}$.
\section{Strong Gravitational Lensing effects by charged Black Holes with Dark Matter Halos}\label{sec3}
Here, we shall study strong gravitational lensing by BHs using two different DM halo models. The first case involves charged BHs with the URC DM halo model, and the second case concerns charged BHs with the CDM halo model. To do this, we investigate the deflection of photon rays by charged BHs with the URC and CDM DM halo models in the equatorial plane ($\theta=\frac{\pi}{2}$), ensuring that both the observer and source lie in the same plane. To study the strong deflection angle of photon rays, we first rewrite Eq  \eqref{1} through the following transformations $t\rightarrow \frac{t}{2M}$, $r\rightarrow
\frac{r}{2M}$, $Q\rightarrow \frac{Q}{2M}$ as:
 \begin{equation}\label{9}
d\bar{s}^2=-P(r)dt^2+ R(r) dr^2 +S(r) d\phi^2.
\end{equation}
With the help of Eq  \eqref{5}, one can express the metric coefficient functions for a charged BH with the URC DM halo within the framework of Eq  \eqref{9} as follows:
\begin{widetext}
\begin{equation}\label{10}
P(r) = \left(1-2 \pi ^2 A B\right) \left(4 \pi  A B^2 r (2 B r-1) +1\right) (4 \pi  A B (2 B r+1)+1) (1-4 \pi  A B (2 B r+1))+\frac{Q^2}{r^2}-\frac{1}{r}.
\end{equation}
\end{widetext}
and
\begin{widetext}
\begin{equation}\label{11}
R(r) = \left[ \left(1-2\pi^2 AB\right) \left(4\pi AB^2 r (2Br-1) +1\right) \left(4\pi AB (2Br+1)+1\right) \left(1-4\pi AB (2Br+1)\right) + \frac{Q^2}{r^2} - \frac{1}{r} \right]^{-1}.
\end{equation}
\end{widetext}
\begin{equation}\label{12}
    S(r)=r^2.
\end{equation}
and using Eq  \eqref{8}, the metric coefficient functions for a charged BH with the CDM DM halo with NFW profile model within the framework of Eq  \eqref{9} as :
\begin{eqnarray}\label{13}
    P(r)&=&1-\frac{1}{r}+\frac{Q^2}{r^2}+32 \pi ^2 A^2 B^2+8 \pi  A B (B r+1)\nonumber\\&,
\end{eqnarray}
\begin{eqnarray}\label{14}
    R(r)&=&\bigg[1-\frac{1}{r}+\frac{Q^2}{r^2}+32 \pi ^2 A^2 B^2\nonumber\\ &+&8 \pi  A B (B r+1)\bigg]^{-1},
\end{eqnarray}
\begin{equation}\label{15}
    S(r)=r^2.
\end{equation}
By employing the Eq denoted as Eq.  \eqref{9}, it is possible to deduce null geodesics in relation to the affine parameter $\tau$ through the following set of Eqs.
   \begin{equation}\label{16}
   \dot{t}=\frac{dt}{d\tau}=\frac{E}{P(r)}
       \end{equation}

   \begin{equation}\label{17}
   \dot{\phi}= \frac{d\phi}{d\tau}=\frac{L}{r^2}
   \end{equation}

   \begin{equation}\label{18}
 \biggr( \frac{dr}{d\tau}\biggr)^2 = \dot{r}^2=E^2-\frac{L^2 }{r^2}P(r)
   \end{equation}
In the given context, $E$ represents the total energy of the test particle, while $L$ denotes the angular momentum of the particle. These parameters are associated with the $\partial_t$ and $\partial_{\phi}$ killing vectors, respectively. The function $P(r)$, a metric component, is given by the Eqs  \eqref{10} and  \eqref{13} for the URC and CDM halo models, respectively. Eq  \eqref{18} can be expressed as
\begin{equation}\label{19}
   \left(\frac{dr}{d\tau}\right)^2+V_{eff} =E^2
\end{equation}
where the effective potential of a photon is expressed by the following expression
\begin{equation}\label{20}
  V_{eff}=\frac{L^2 }{r^2}P(r)
\end{equation}
To determine the radius \(r_{ph}\) of the unstable circular photon orbit, certain conditions on the effective potentials must be satisfied. Specifically, the conditions are that the first derivative of the effective potential with respect to \(r\), evaluated at \(r_{ph}\), must be equal to zero (\(\frac{dV_{eff}}{dr}\Big|_{r_{ph}} = 0\)), and the second derivative of the effective potential with respect to \(r\), also evaluated at \(r_{ph}\), must be less than zero (\(\frac{d^2V_{eff}}{dr^2}\Big|_{r_{ph}} < 0\)). Therefore, the photon sphere radius \(r_{ph}\) is identified as the real root of the following expression \cite{Virbhadra:2002ju,Claudel:2000yi,Adler:2022qtb}:
\begin{equation}\label{21}
    2P(r_{ph})-r_{ph}P^{\prime}(r_{ph})=0
\end{equation}
It is noteworthy that at the photon sphere radius, denoted as $r_{\text{ph}}$, the condition $\frac{d^2V_{\text{eff}}}{dr^2}\big|_{r_{\text{ph}}} < 0$ is satisfied in the context of the charged BH spacetime incorporating URC and CDM models, as described in Eq  \eqref{9}. The fulfillment of this condition implies the instability of these orbits against small perturbations. In this scenario, photons originate from infinity, nearing the BH with a certain impact parameter \(u\) and subsequently recede to infinity after reaching their closest approach, denoted as \(r_0\). At this minimum distance, where \(\frac{dr}{d\tau} = 0\), we can express the ratio \(\frac{L}{E}\) as a function of the impact parameter ``u" in relation to the closest distance \(r_0\) as \cite{Bozza:2002zj}
\begin{equation}\label{22}
   u=\frac{L}{E}=\frac{r_0}{\sqrt{P(r_0)}}
\end{equation}
The strong deflection angle grows unboundedly as $r_0$ approaches $r_{ph}$, and it remains finite only when $r_0 > r_{ph}$. Therefore, the critical impact parameter, denoted as $u_{ph}$, is described  by
 \begin{equation}\label{23}
u_{ph} =\frac{r_{ph}}{\sqrt{P(r_{ph})}}
 \end{equation}
For impact parameters \(u\) less than \(u_{\text{ph}}\), photons experience gravitational attraction towards the BH. Conversely, when the impact parameter exceeds \(u_{\text{ph}}\), photons approach their minimum distance to the BH, denoted as \(r_0\). At the critical point where the impact parameter \(u\) equals \(u_{\text{ph}}\), photons initiate an unstable circular orbit around the BH, creating a photon sphere characterized by a radius of \(r_{\text{ph}}\). By solving the null geodesic equations, the expression for the strong deflection angle of a photon originating from infinity, within the framework of charged BH spacetime, as described by Eq.  \eqref{9}, can be formulated as follows \cite{Virbhadra:1998dy}.
 \begin{equation}\label{24}
\alpha_D(r_0)=I(r_0)- \pi
\end{equation}
where
\begin{equation}\label{25}
I(r_0)=2 \int _{r_0}^\infty  \frac{d\phi}{dr} dr
\end{equation}
or,
\begin{equation}\label{26}
 I(r_0)=  \int_{r_0}^\infty \frac{2\sqrt{R(r)}dr}{\sqrt{S(r) }\sqrt{ \frac{P(r_0)S(r)}{S(r)S(r_0)}-1} }dr
\end{equation}
In the given context, $r_0$ represents the closest approach distance of the photon's trajectory. It is important to highlight that the magnitude of the strong deflection angle, denoted as $\alpha_D(r_0)$, is influenced by the interplay between $r_{ph}$ and $r_0$. When $r_0$ approaches the vicinity of $r_{ph}$, a notable enhancement in the deflection angle is observed. To address this relationship more explicitly, we introduce a novel parameter, designated as "z," drawing inspiration from previous scholarly works \cite{Tsukamoto:2016qro,Tsukamoto:2016jzh}
\begin{equation}\label{27}
 z=1-\frac{r_0}{r}
\end{equation}
and obtain
\begin{equation}\label{28}
 I(r_0)=\int_{0}^1 F(z,r_0)H(z,r_0)dz
\end{equation}
where
\begin{equation}\label{29}
F(z,r_0)= \frac{2(1-P(r_0))\sqrt{S(r_0)}}{S(r)A^\prime(r)}\sqrt{P(r)R(r)},
\end{equation}
\begin{equation}\label{30}
 H(z,r_0)=\frac{\sqrt{S(r)}}{\sqrt{S(r)P(r_0)-P(r)S(r_0)}}
\end{equation}
The function $F(z, r_0)$ exhibits regular behavior for all values of $z$ and $r_0$, in contrast to the function $H(z, r_0)$, which diverges as $z$ approaches zero. The integral given by Eq  \eqref{28} can be appropriately defined in this context.
\begin{equation}\label{31}
I(r_0) =  I^{D}(r_0)+I^{R}(r_0)
\end{equation}
where  the regular part and the divergent part of the integral are
\begin{equation}\label{32}
 I^{R}(r_0)=\int_{0}^1 g(z,r_0) dz
\end{equation}

\begin{equation}\label{33}
 I^{D}(r_0)=\int_{0}^1 F(0,r_{ph})H_0(z,r_{0}) dz
\end{equation}
and $g(z, r_0) = F(z, r_0)H(z, r_0) - F(0, r_{ph}) H_0(z, r_0)$. To determine the order of divergence of the integrand in Eq  \eqref{33}, one can express the term inside the square root of $H_0(z, r_0)$ as follows:
\begin{equation}\label{34}
H_0(z,r_0)= \frac{1}{\sqrt{\eta(r_0) z +\zeta(r_0)  z^2 + \mathcal{O}(z^3) }}
\end{equation}
where
\begin{equation}\label{35}
\begin{split}
\eta(r_0) =  \frac{(1 - P(r_0))}{P^\prime(r_0) S(r_0)} \bigg( P(r_0) S^\prime(r_0) \\
- P^\prime(r_0) S(r_0) \bigg)
\end{split}
\end{equation}
\begin{equation}\label{36}
 \begin{split}
&\zeta (r_0) = \frac{(1-P(r_0))^2}{{P^\prime }^3(r_0)
S^2(r_0)}\biggr(2S(r_0)S^\prime(r_0) {P^\prime(r_0)}^2
+ (S(r_0) S^{\prime\prime}(r_0)\\
&-
2{S^\prime}^2(r_0))P(r_0)P^\prime(r_0) - S(r_0)S^\prime(r_0)P(r_0)P^{\prime\prime}(r_0)
 \biggr)\\
\end{split}
\end{equation}
In the given context, the prime symbol signifies differentiation concerning the variable $r$. When $r_0$ is approximately equal to $r_{ph}$, the coefficient $\eta(r_0)$ becomes zero, resulting in a divergence order of $z^{-1}$. This divergence leads to a logarithmic divergence of the integral. When $r_0$ is close to $r_{ph}$, the expression for the deflection angle can be articulated as follows \cite{Tsukamoto:2016qro,Iyer:2006cn,Tsukamoto:2016jzh,Tsukamoto:2022uoz,Tsukamoto:2022tmm}
\begin{equation}\label{37}
\alpha(u)= -\bar{a}~ log\left(\frac{u}{u_{ph}}-1\right) +\bar{b} +\mathcal{O}((u -u_{ph})log(u -u_{ph}))
\end{equation}
where
\begin{equation}\label{38}
\bar{a}=\frac{F(0,r_{ph})}{2\sqrt{\zeta(r_{ph})}}
=\sqrt{\frac{2P(r_{ph}) R(r_{ph})}{P(r_{ph})S^{\prime\prime}(r_{ph})-P^{\prime\prime}(r_{ph}) S(r_{ph})}}
\end{equation}
and
\begin{equation}\label{39}
\bar{b}=-\pi + a_R + \bar{a} log \biggr({\frac{4\zeta(r_{ph})S(r_{ph})}{u_{ph} |P(r_{ph})| ( 2u_{ph} P(r_{ph}))}}\biggr),
\end{equation}
$ a_R=I^{R}(r_{ph})= \int_{0}^{1} g(z,r_{ph} ) dz  $ which is numerically obtained. Now, we explore the diverse astrophysical implications through the observation of strong gravitational lensing. In this investigation, we focus on a scenario characterized by a high alignment of the source, the lens (BH), and the observer. Additionally, both the observer and the source are positioned at considerable distances from the BH. Consequently, the lens equation governing strong gravitational lensing can be formulated as follows \cite{Bozza:2001xd}
\begin{equation}\label{40}
 \beta=\tilde{\alpha}-\frac{D_{ls}}{D_{os}}\Delta \alpha_{n}
\end{equation}
The symbols $D_{ls}$ and $D_{os}$ denote the lens-source, and observer-source distance, respectively, where $D_{os} = D_{ol} + D_{ls}$. The parameter $\beta$ represents the angular separation between the lens (BH), while $\tilde{\alpha}$ represents the angular separation between the source of the image and the lens relative to the optical axis. The offset of the deflection angle is given by $\Delta\alpha_{n} = \tilde{\alpha} - 2n\pi$, where $n$ is an integer. Utilizing Eqs  \eqref{37} and  \eqref{40}, the angular position of the $n^{th}$ relativistic image can be approximated as follows:
\begin{equation}\label{41}
\theta _n =  \theta^0 _n - \frac{u_{ph} e_n  D_{os}(\theta_n^0-\beta)}{\bar{a}D_{ls}D_{ol}}
\end{equation}
where $$ e_n=e^{\frac{\bar{b}-2n\pi}{\bar{a}}},$$
$$\theta^0_n=\frac{ u_{ph}(1+e_n)}{D_{ol}},$$ $\theta^0_n$ is  the
image position corresponding to $\alpha=2n\pi$.The magnification of $n$-th relativistic image, is expressed as \cite{Bozza:2002zj}
\begin{equation}\label{42}
\mu_n=\biggr(\frac{\beta}{\tilde{\alpha}}\frac{d\beta}{d\tilde{\alpha}}\biggr)^{-1}\biggr|_{\theta_n^0}=\frac{e_n u^2_{ph}(1+e_n)D_{os}}{\bar{a}\beta D_{ls}D^2_{ol}}
\end{equation}
It is important to highlight that the initial relativistic image exhibits the highest brightness, and its magnification undergoes an exponential decline as \(n\) increases. Particularly, when \(\beta \rightarrow 0\), Eq  \eqref{42} exhibits divergence. Therefore, achieving precise alignment significantly improves the probability of detecting relativistic images. If we denote \(\theta_n\) as the asymptotic position where a cluster of images converges as \(n\) approaches infinity, then the brightest image, specifically the outermost image labeled as \(\theta_1\), can be uniquely identified. Simultaneously, all inner images converge closely together at the position \(\theta_{\infty}\). Utilizing the strong deflection angle, as elaborated in Eq  \eqref{37}, and the lens equation for strong gravitational lensing given in Eq  \eqref{40}, we have derived three observable quantities: the angular position of the set of asymptotic relativistic images, denoted as \(\theta_{\infty}\); the angular separation \(S\) between the outermost image (\(\theta_1\)) and the closely packed inner images (\(\theta_{\infty}\)); and the relative magnification of the outermost relativistic image compared to the remaining relativistic images. \cite{Kumar:2022fqo,Bozza:2002zj}
\begin{equation}\label{43}
\theta_{\infty}=\frac{u_{ph}}{D_{ol}}
\end{equation}
\begin{equation}\label{44}
S= \theta_1-\theta_{\infty}\approx\theta_{\infty}e^\frac{(\bar{b} -2\pi)}{\bar{a}}
\end{equation}
\begin{equation}\label{45}
r_{mag}=\frac{\mu_1}{\Sigma^\infty_{n=2}\mu_{n}}\approx \frac{5\pi}{\bar{a}log(10)}
\end{equation}
Upon assessing the measurable parameters $\theta_{\infty}$, $S$, and $r_{\text{mag}}$, the lensing coefficients $\bar{a}$, $\bar{b}$, and $u_{\text{ph}}$ can be deduced utilizing Eqs  \eqref{43},  \eqref{44}, and  \eqref{45}. The obtained values can then be juxtaposed with the predictions from theoretical models. This comparison allows for the characterization of charged BHs with DM halos surrounding other astrophysical BHs. In scenarios where the source, BH (lens), and observer are in alignment, denoted by $\beta=0$ in Eq.  \eqref{41}, the BH (lens) alters the trajectory of light rays in all directions, resulting in the formation of a ring-shaped image known as an Einstein ring. \cite{Einstein:1936llh,Liebes:1964zz,Mellier:1998pk,Bartelmann:1999yn,Schmidt:2008hc,Guzik:2009cm}. By simplifying the Eq \eqref{25} for the case where $\beta=0$, the resulting formula provides the angular radius of the $n^{th}$ relativistic image.
\begin{equation}\label{30}
     \theta _n =  \theta^0 _n \biggr(1 - \frac{u_{ph} e_n  D_{os}}{\bar{a}D_{ls}D_{ol}}\biggr)
 \end{equation}
In the scenario where the BH (acting as a lens) is positioned at a midpoint between the source and receiver, denoted as $D_{os}=2D_{ol}$, and assuming that $D_{ol}$ significantly exceeds the photon impact parameter $u_{ph}$, the angular radius of the $n^{th}$ relativistic Einstein ring within the framework of a modified Bardeen BH can be expressed as follows:
\begin{equation}\label{31}
\theta^E_n=\frac{ u_{ph}(1+e_n)}{D_{ol}}
\end{equation}
Time delay, denoted as $\Delta T_{2,1}$, emerges as a crucial and significant observable in the context of strong lensing. It characterizes the time lag between the formation of two relativistic images, accounting for fluctuations in the source's luminosity. The varying travel times for photons along their distinct paths create temporal disparities among different relativistic images. This temporal offset becomes a pivotal parameter, rooted in the fluctuations in the times of image formation. The ability to differentiate between the temporal signals of primary and secondary images during observational studies allows for the computation of the time delay between these signals \cite{Bozza:2003cp}. While a photon makes its journey from the source to an observer, the duration it requires to complete an orbit around the BH can be calculated using the formula proposed by \cite{Bozza:2003cp}.
\begin{equation}\label{46}
\tilde{T}=\tilde{a}log\biggr(\frac{u}{u_{ph}}-1\biggr)+\tilde{b}+\mathcal{O}(u-u_{ph}
\end{equation}
By employing the Eq  \eqref{46}, one can calculate the time delay between two relativistic images. In the context of a spherically symmetric spacetime, the time delay between two images, when both images occur on the same side of the BH, can be formulated as per \cite{Bozza:2003cp}.
\begin{equation}\label{47}
\Delta T_{2,1}=2\pi u_{ph}=2\pi D_{ol} \theta_{\infty}
\end{equation}
By utilizing Eq  \eqref{47}, if we make precise estimations of the time delay and critical impact parameter, it becomes possible to derive the distance of the BH with negligible error. This study focuses on the assessment of the time delay between images produced through strong field lensing caused by a charged BH, considering the influence of the URC and CDM DM halo. The investigation is specifically conducted in the context of two supermassive BHs, denoted as $M87^*$ and $SgrA^{*}$.
\subsection{URC profile}
\begin{figure*}[htbp]
\centering
\includegraphics[width=.45\textwidth]{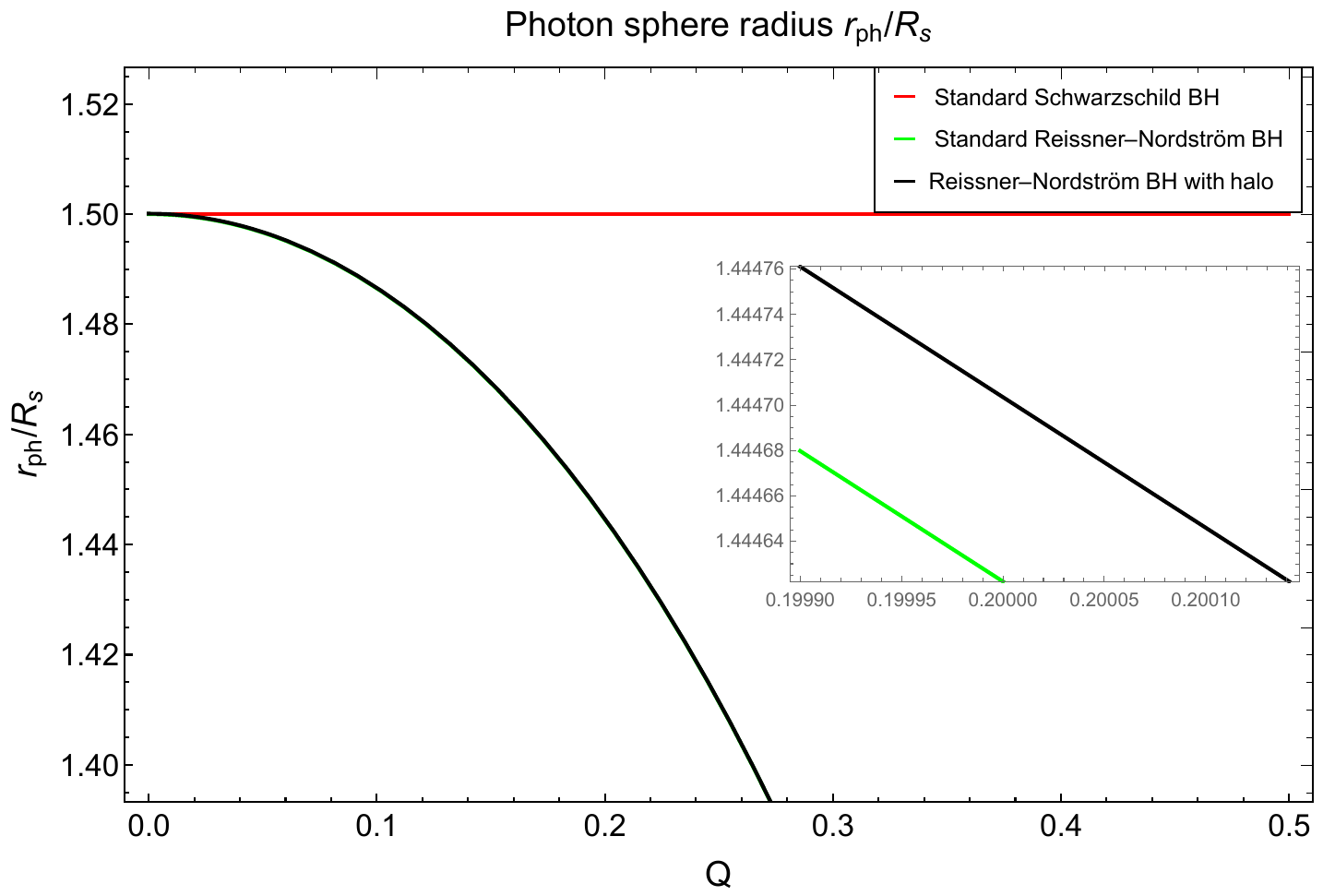}(a)
\qquad
\includegraphics[width=.45\textwidth]{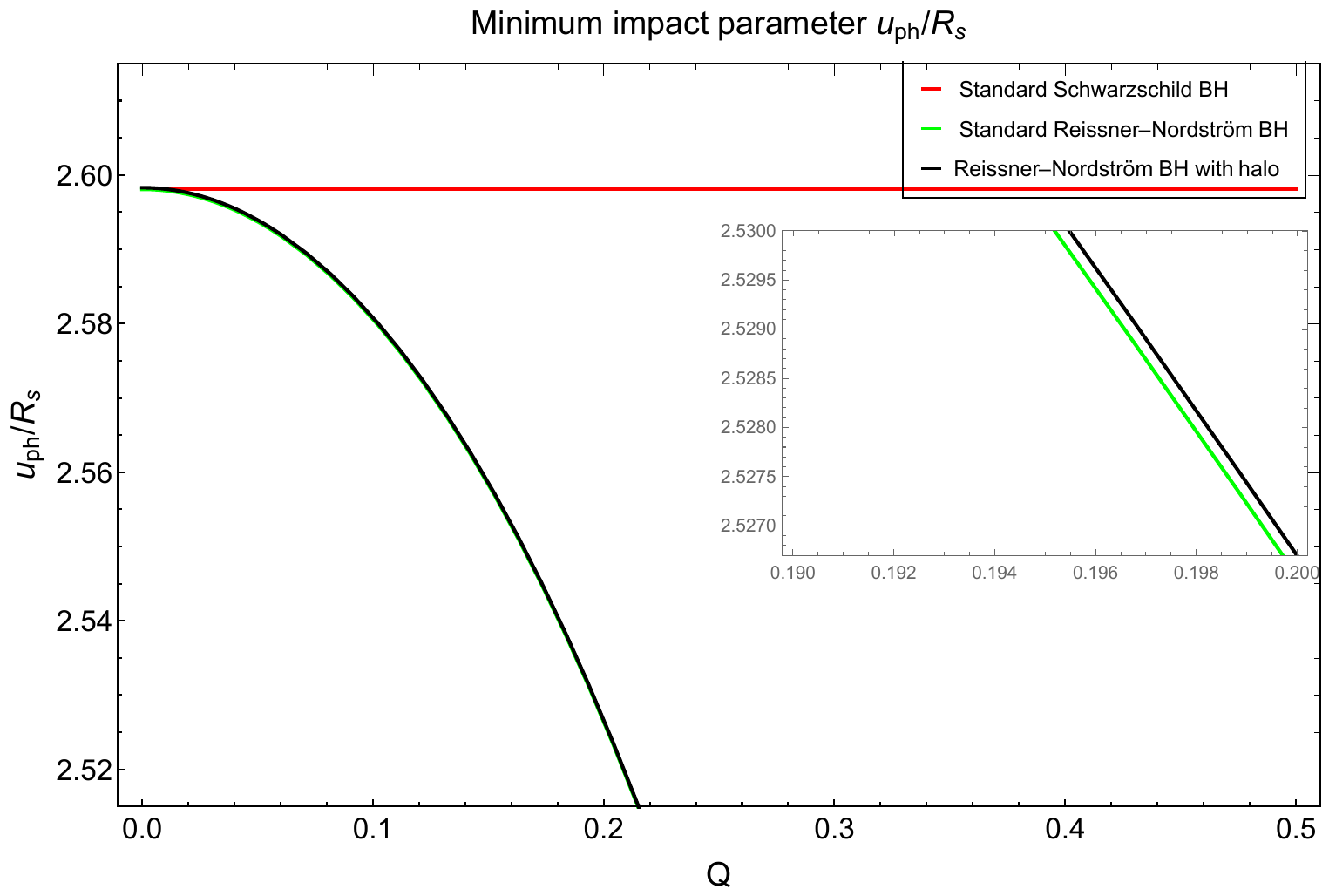}(b)
 \caption{The left panel depicts the photon sphere radius (\(r_{\text{ph}}/R_s\) vs Q) and the right panel illustrates the minimum impact parameter (\(u_{\text{ph}}/R_s\) vs Q) for RN BHs with a URC DM halo (black line), contrasted with standard RN BHs (green line) and standard Schwarzschild BHs (red line). It is noteworthy that both the photon sphere radius (\(r_{\text{ph}}/R_s\)) and the minimum impact parameter (\(u_{\text{ph}}/R_s\)) exhibit nearly identical characteristics for the $M87^{*}$ and $Sgr A^{*}$ scenarios.}
 \label{fig:1}
\end{figure*}

\begin{figure*}[htbp]
\centering
\includegraphics[width=.45\textwidth]{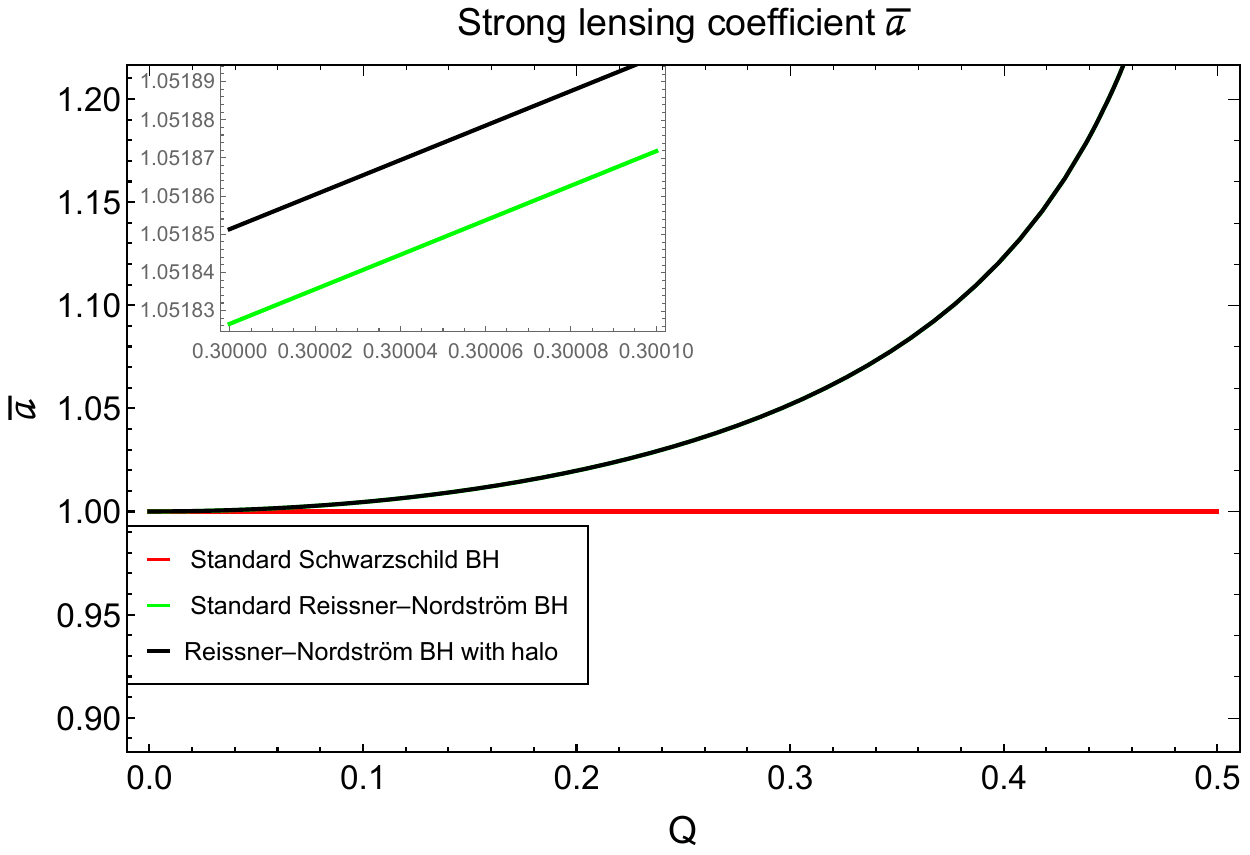}(a)
\qquad
\includegraphics[width=.45\textwidth]{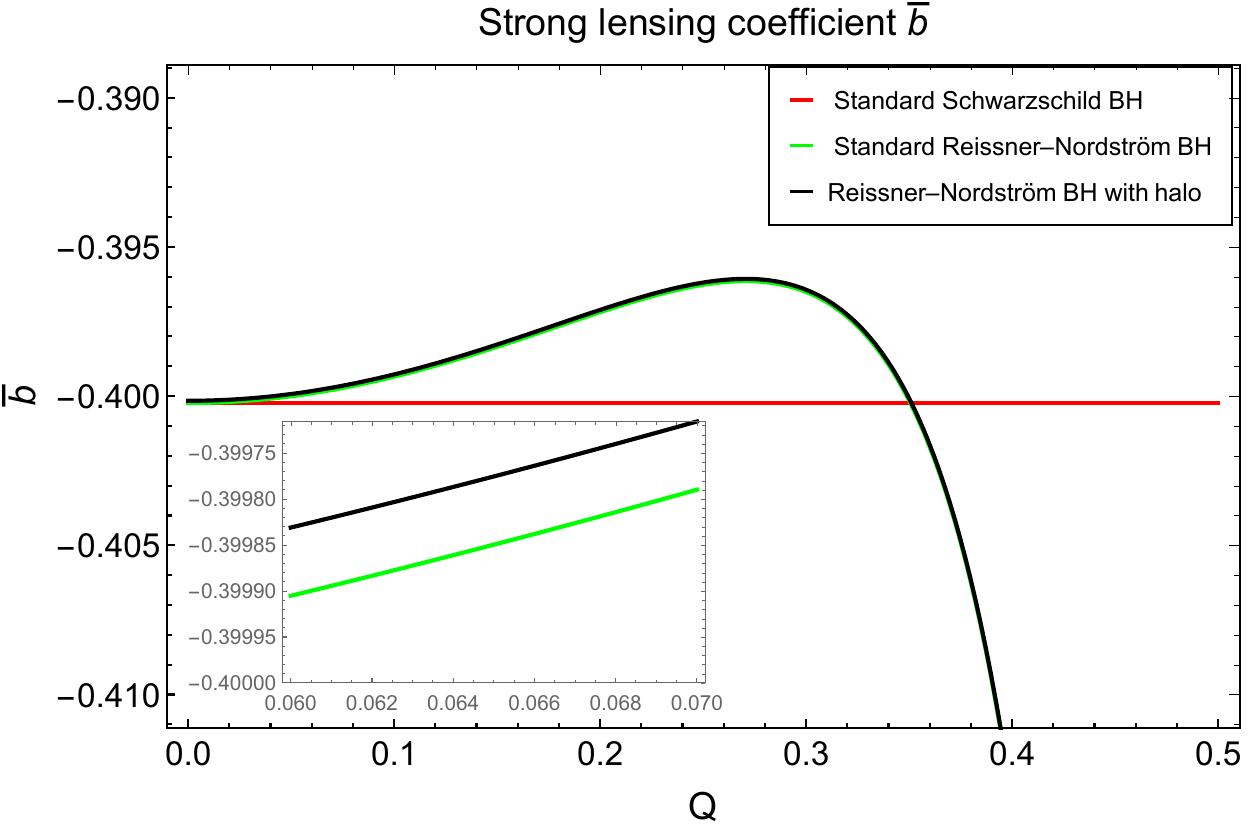}(b)
\caption{The left panel illustrates the deflection limit coefficients ($\mathit{\bar{a}}$ vs Q), while the right panel depicts ($\mathit{\bar{b}}$ vs Q) for the RN BH with a URC DM halo (black line), contrasted with standard RN BHs (green line) and standard Schwarzschild BHs (red line). It is noteworthy that the deflection limit coefficients $\mathit{\bar{a}}$ and $\mathit{\bar{b}}$ demonstrate nearly identical characteristics for the $M87^{*}$ and $Sgr A^{*}$ scenarios.}
\label{fig:2}
\end{figure*}

\begin{figure*}[htbp]
\centering
\includegraphics[width=.45\textwidth]{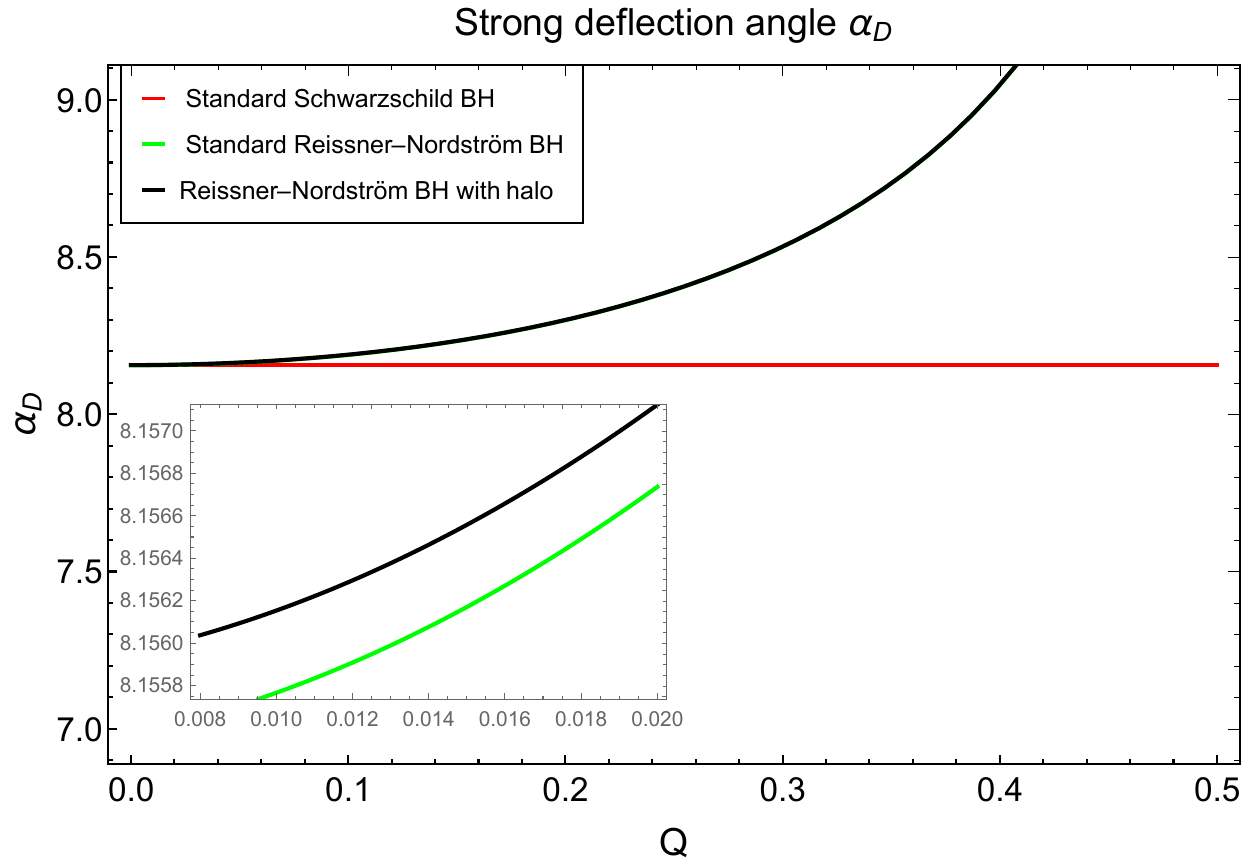}(a)
\qquad
\includegraphics[width=.45\textwidth]{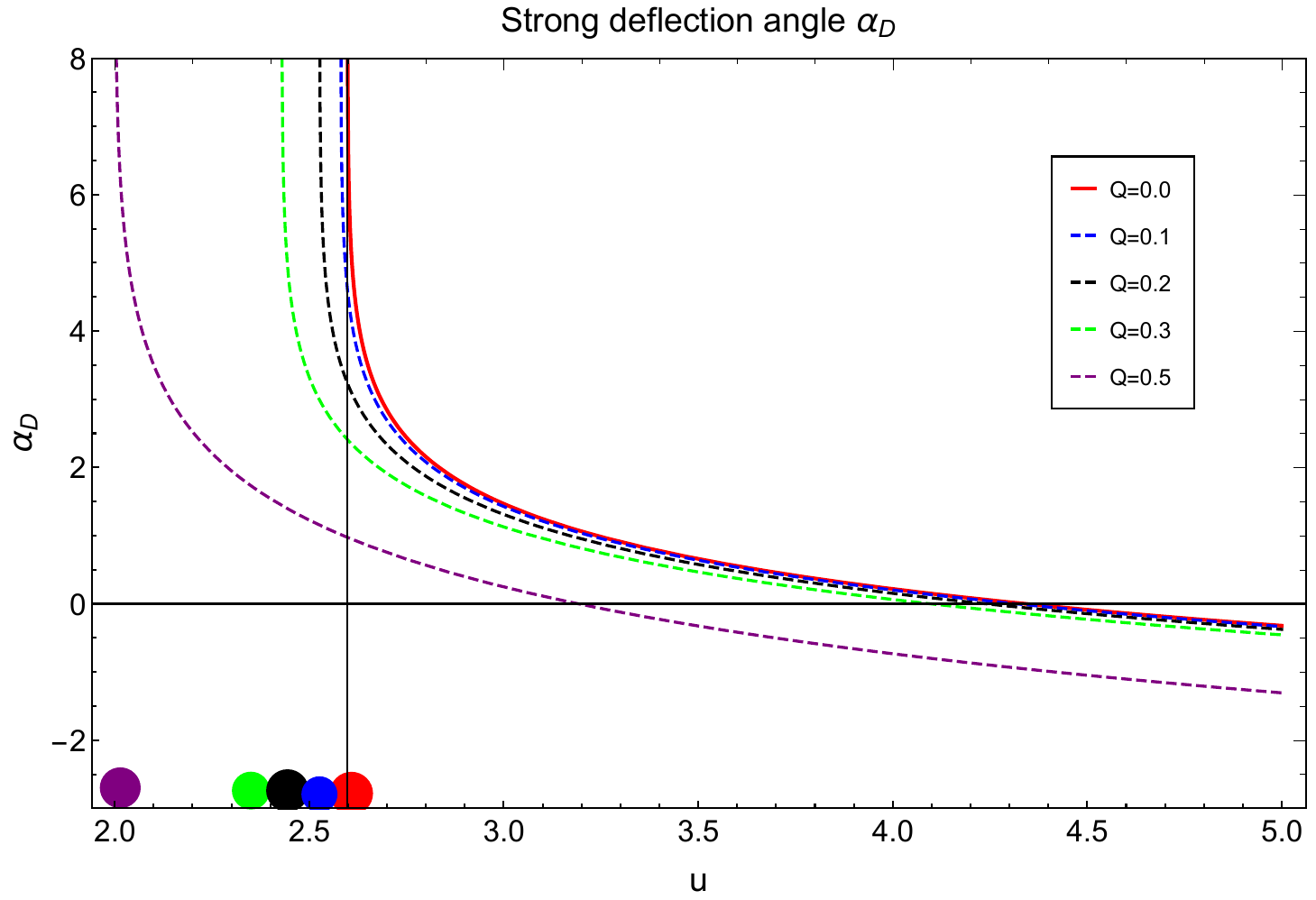}(b)
\caption{The left panel depicts the strong deflection angle ($\mathit{\alpha_{D}}$) as a function of charge parameter $Q$ at $u=u_{ph}+0.0002$ for the RN BH with a URC DM halo (black line), contrasting it with the standard RN case (green line) and the standard Schwarzschild BH (red line). In the right panel, the strong deflection angle ($\mathit{\alpha_{D}}$) is shown against the impact parameter $u$ for various values of charge parameter $Q$. The dotted lines indicated in Fig(b) are the value of the impact parameter$u=u_{ph}$, where the deflection angle diverges. It is noteworthy that the strong deflection angle ($\mathit{\alpha_{D}}$) exhibits remarkably similar characteristics for both the $M87^{*}$ and $Sgr A^{*}$ scenarios.}\label{fig:3}
\end{figure*}

\begin{figure*}[htbp]
\centering
\includegraphics[width=.45\textwidth]{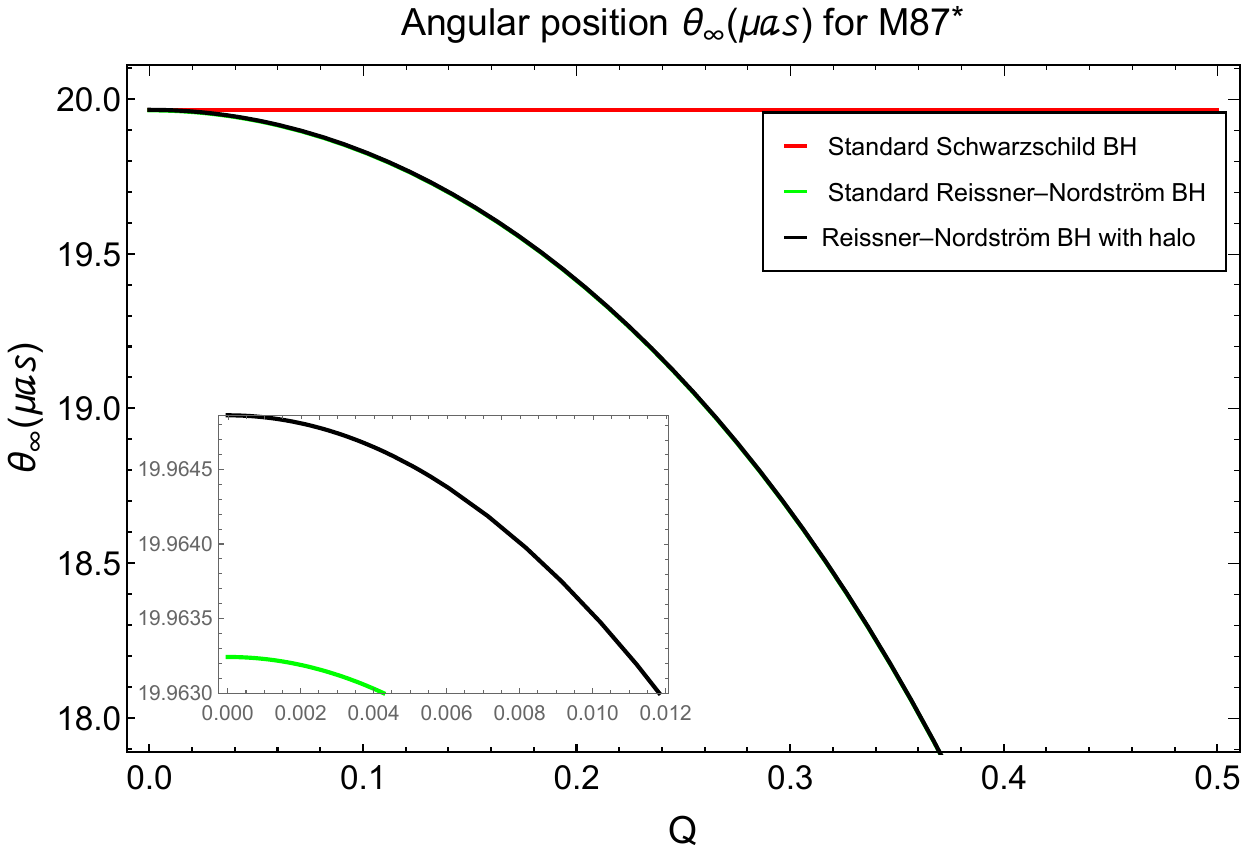}(a)
\qquad
\includegraphics[width=.45\textwidth]{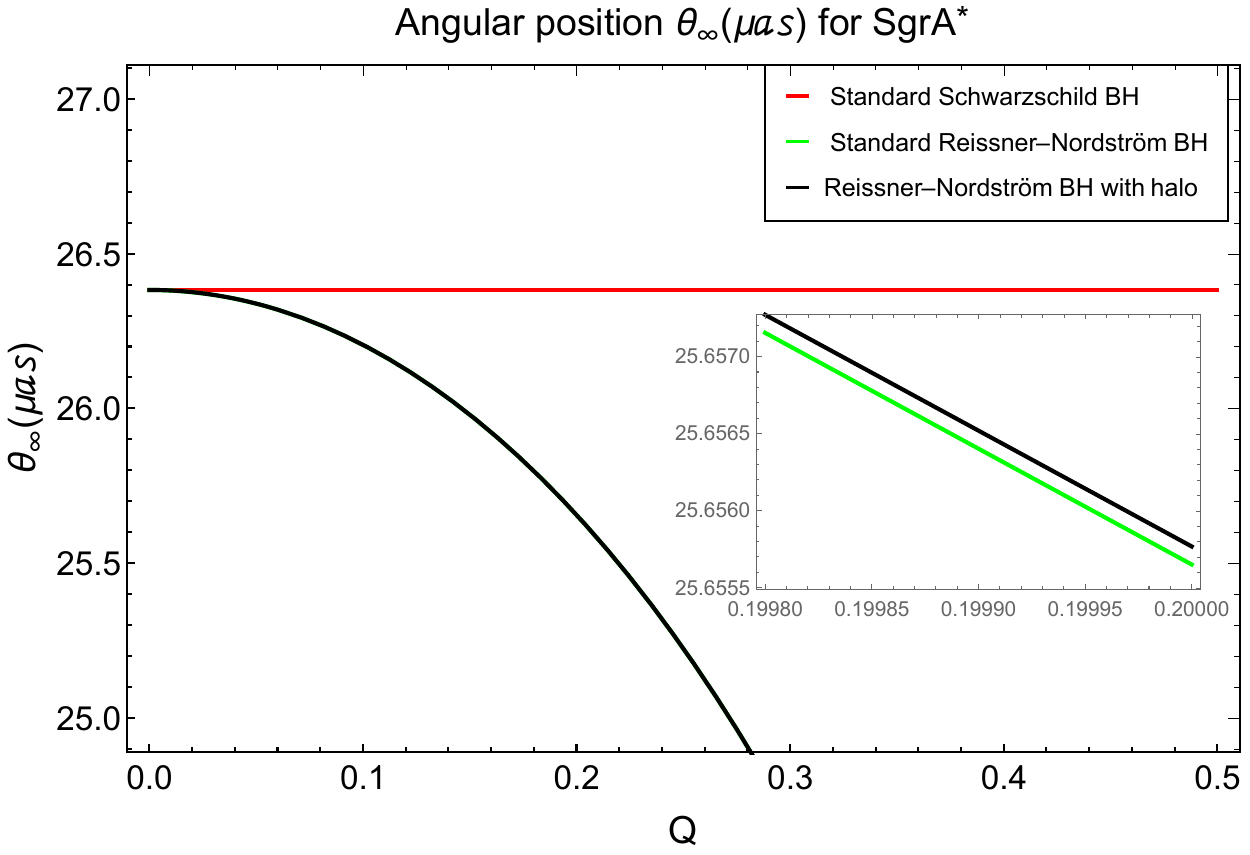}(b)
\caption{Comparison of the observable quantity, angular position ($\mathit{\theta_{\infty}}$), versus the parameter Q for both $M87^{*}$ (left panel) and $Sgr A^*$ (right panel). The black line represents the RN solution with a URC DM halo, while the green line corresponds to the standard RN BH and the red line represents the standard Schwarzschild BH.}\label{fig:4}
\end{figure*}

\begin{figure*}[htbp]
\centering
\includegraphics[width=.45\textwidth]{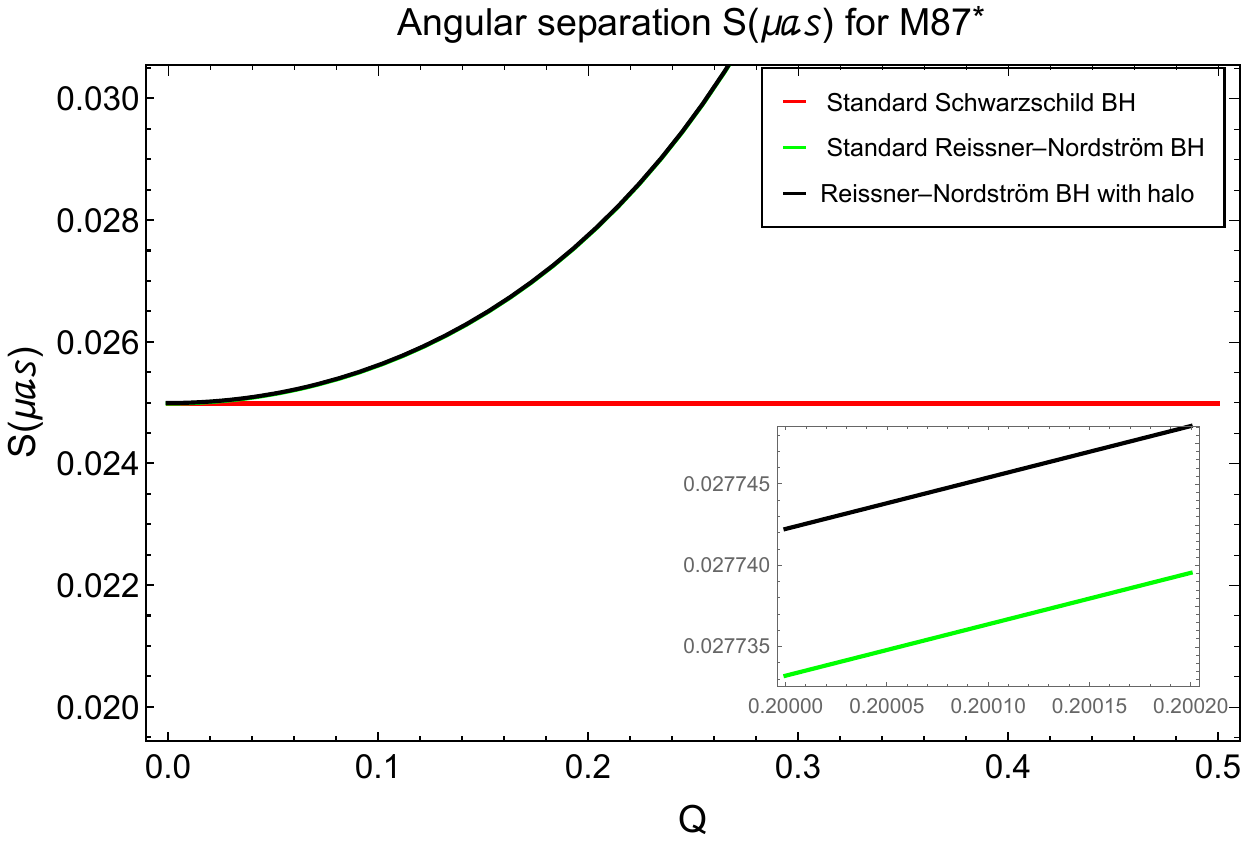}(a)
\qquad
\includegraphics[width=.45\textwidth]{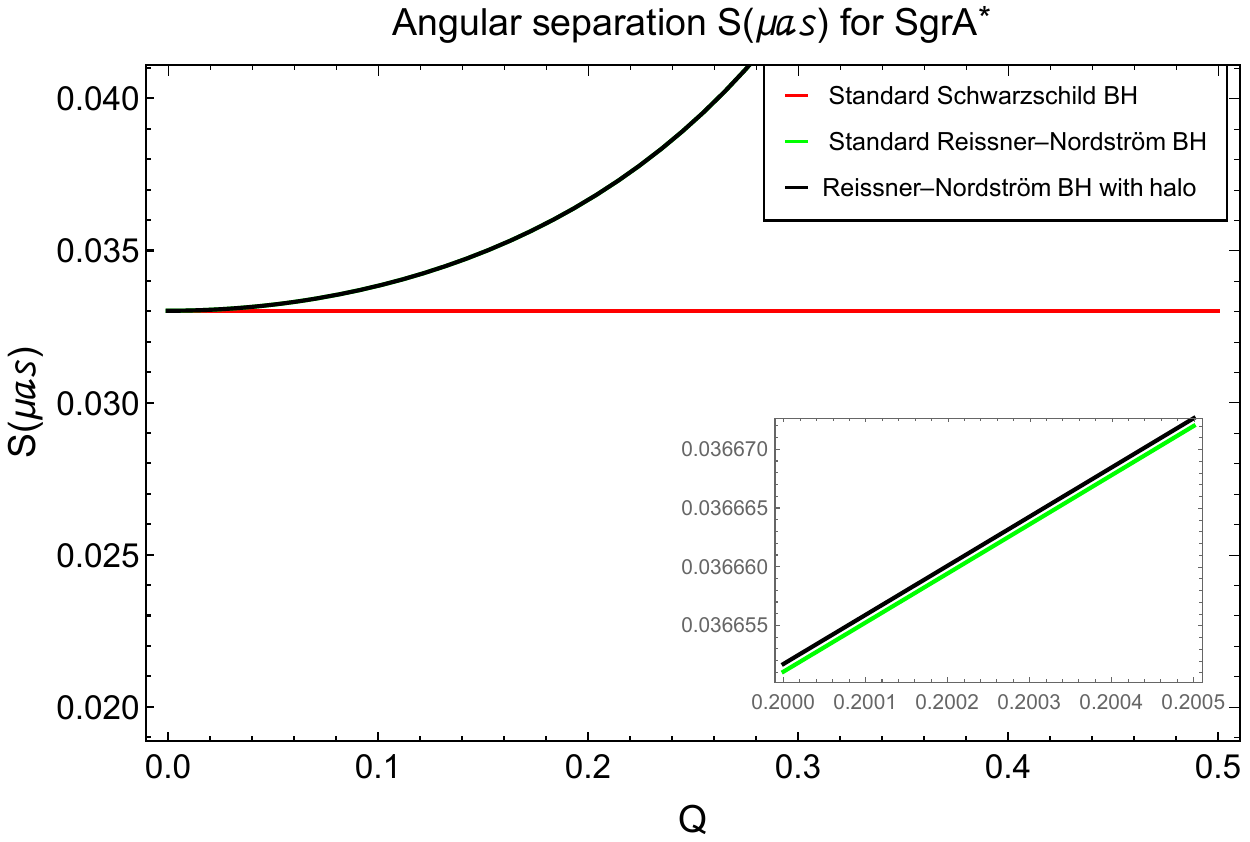}(b)
\caption{The figure illustrates the observable angular separation ($S$ )  vs $Q$ for $M87^{*}$ (left panel) and $Sgr A^*$ (right panel) within the framework of the RN metric, incorporating a URC DM halo, represented by the black line. This is compared with the predictions of the standard RN BH (green line) and standard Schwarzschild BH (red line).}\label{fig:5}
\end{figure*}

\begin{figure*}[htbp]
\centering
\includegraphics[width=.45\textwidth]{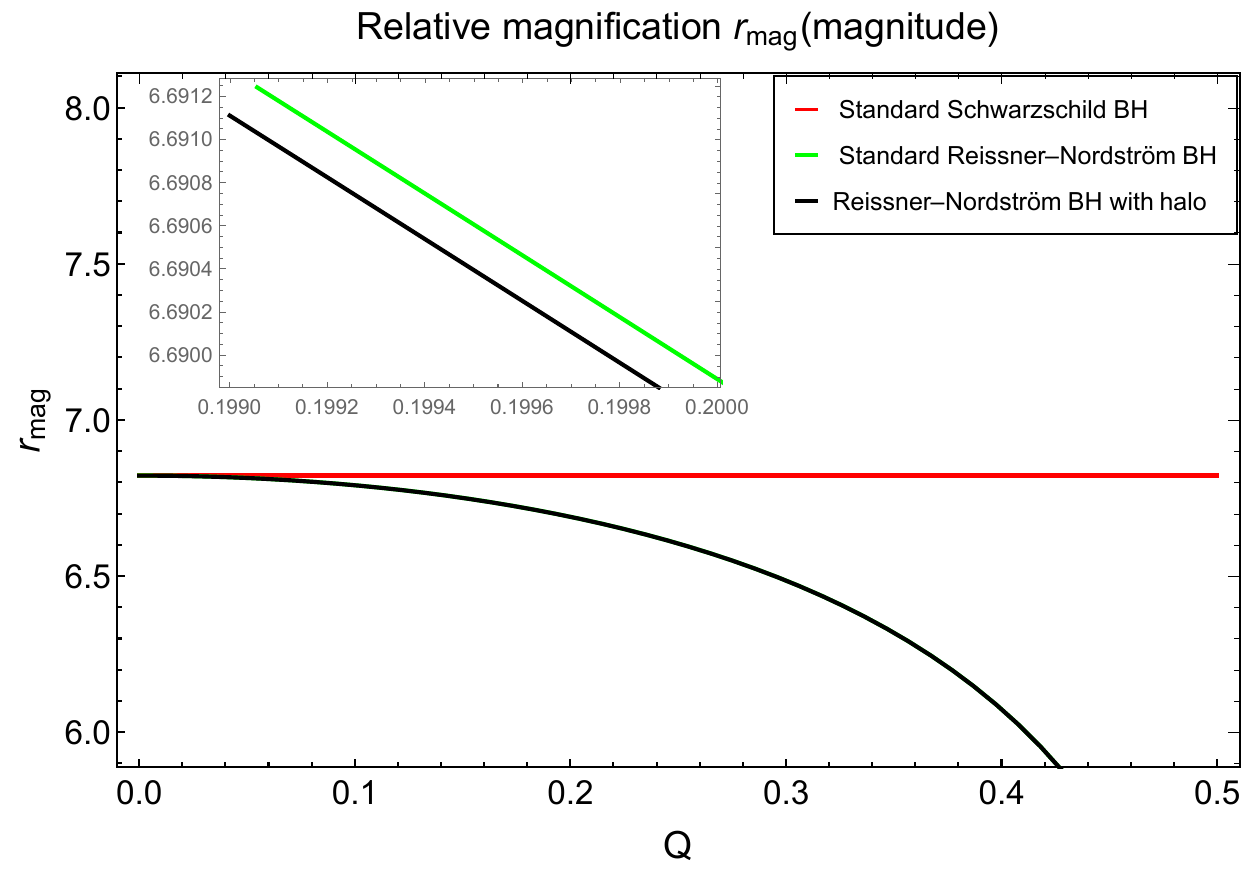}
 \caption{The observable quantity, relative magnification ($\mathit{r_{\text{mag}}}$) versus Q, is depicted for the RN BH with a URC dark DM halo (black line), in comparison to the standard RN BH (green line) and the standard Schwarzschild BH (red line). It is crucial to emphasize that the relative magnification $\mathit{r_{\text{mag}}}$ exhibits nearly identical characteristics for the $M87^{*}$ and $Sgr A^{*}$ scenarios.}\label{fig:6}
\end{figure*}

\begin{figure*}[htbp]
\centering
\includegraphics[width=.45\textwidth]{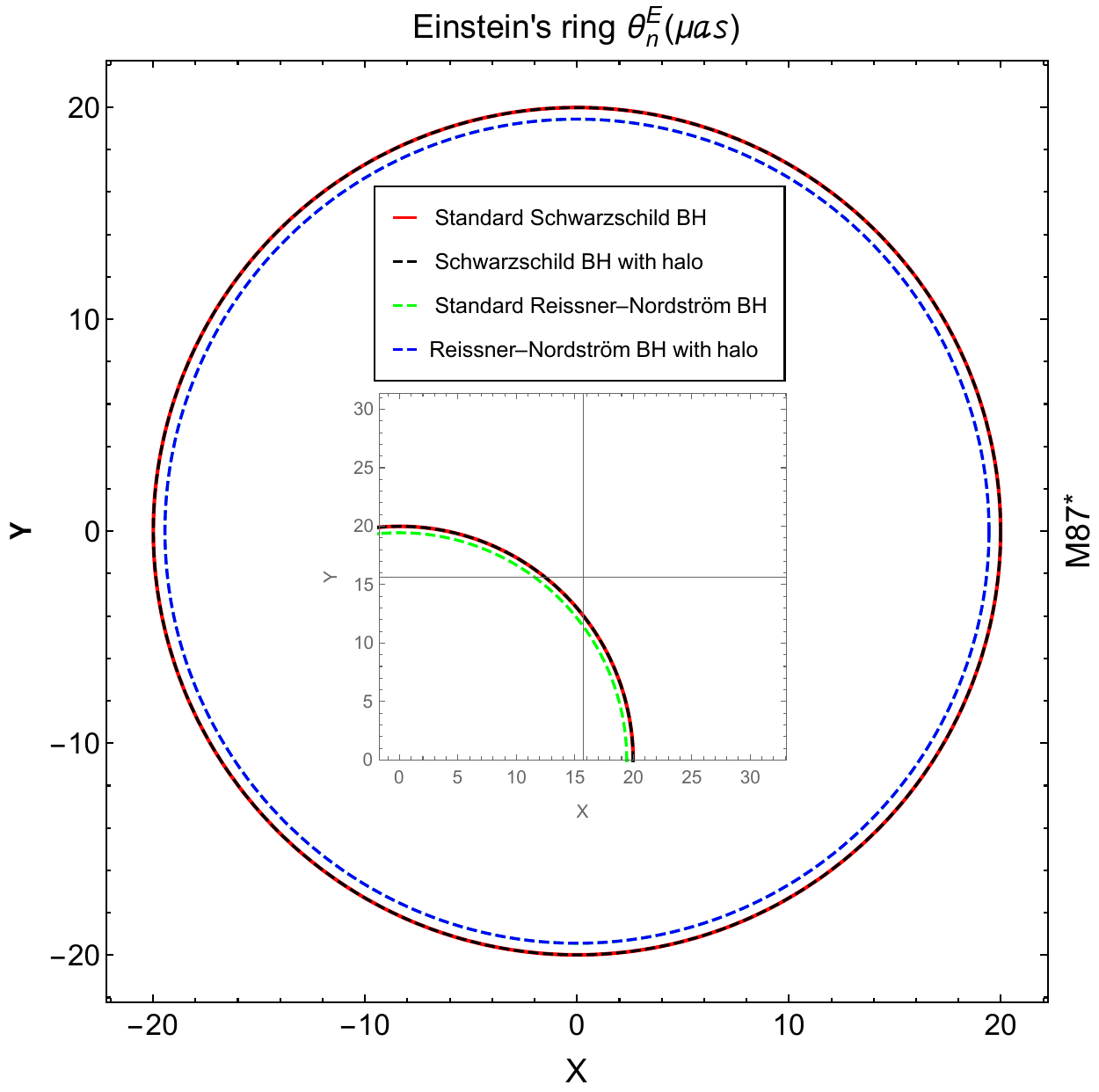}(a)
\qquad
\includegraphics[width=.45\textwidth]{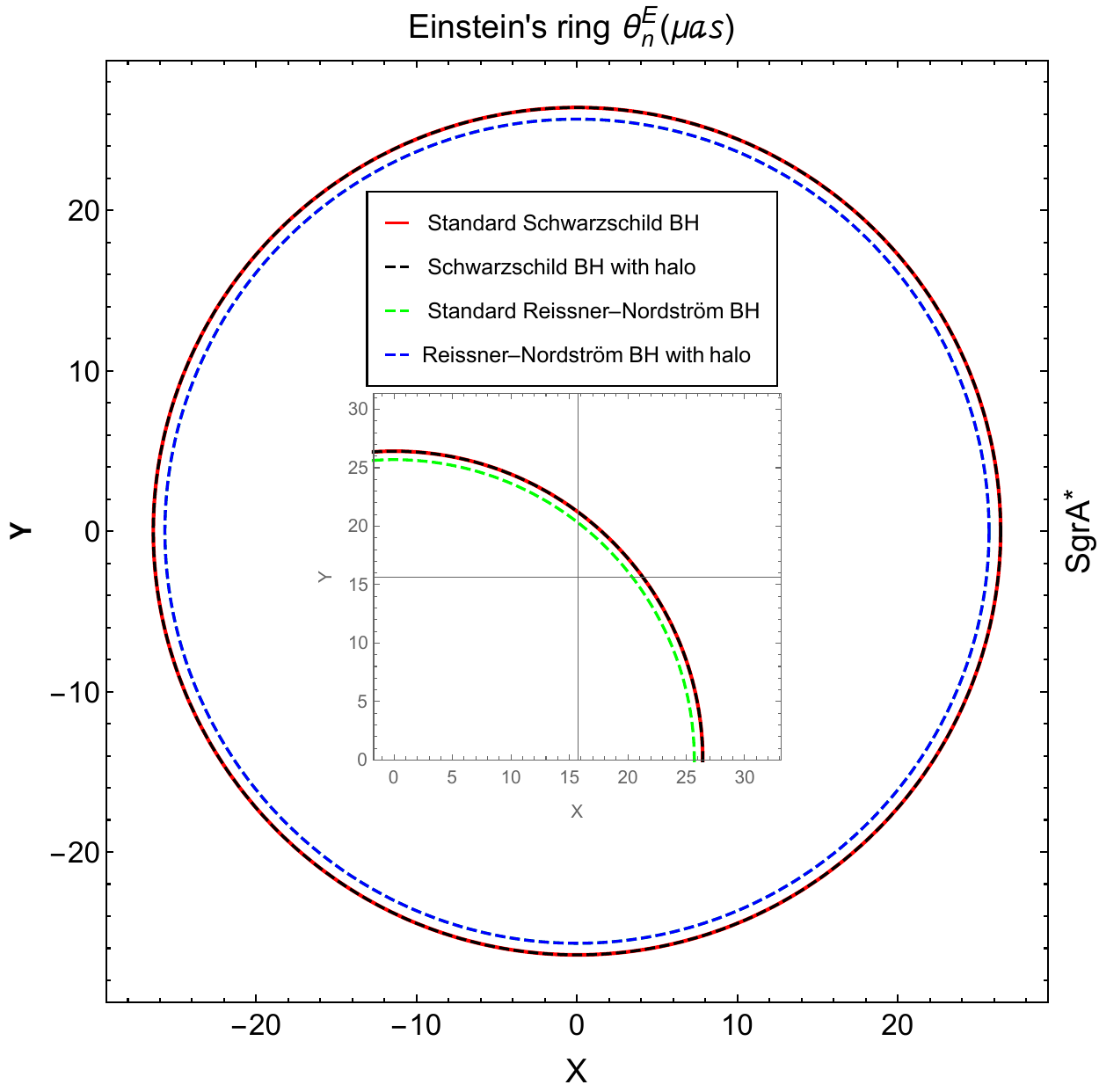}(b)
\caption{The observable quantity, Einstein's ring ($\theta^{E}_n$) plotted against Q, is presented for both $M87^{*}$ (left panel) and $Sgr A^*$ (right panel). The blue dotted line represents the RN BH with a URC DM halo, while the black dotted line corresponds to the Schwarzschild BH with a URC DM halo. Additionally, the green dotted line represents the RN BH without the URC DM  halo, and the red solid line corresponds to the standard Schwarzschild BH. This comparison provides insights into the gravitational lensing behavior of these BHs under different configurations.}\label{fig:7}
\end{figure*}

\begin{figure*}[htbp]
\centering
\includegraphics[width=.45\textwidth]{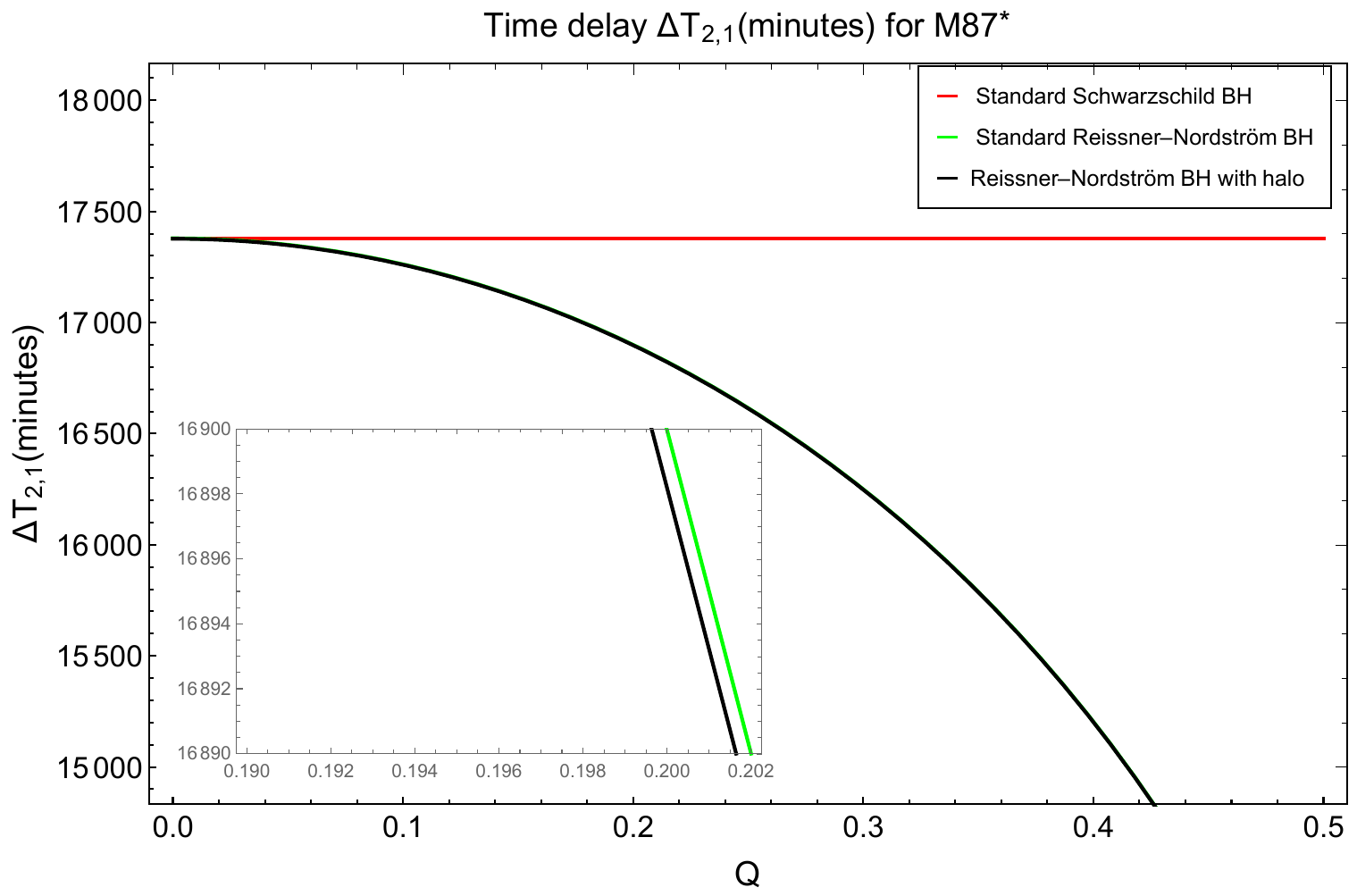}(a)
\qquad
\includegraphics[width=.45\textwidth]{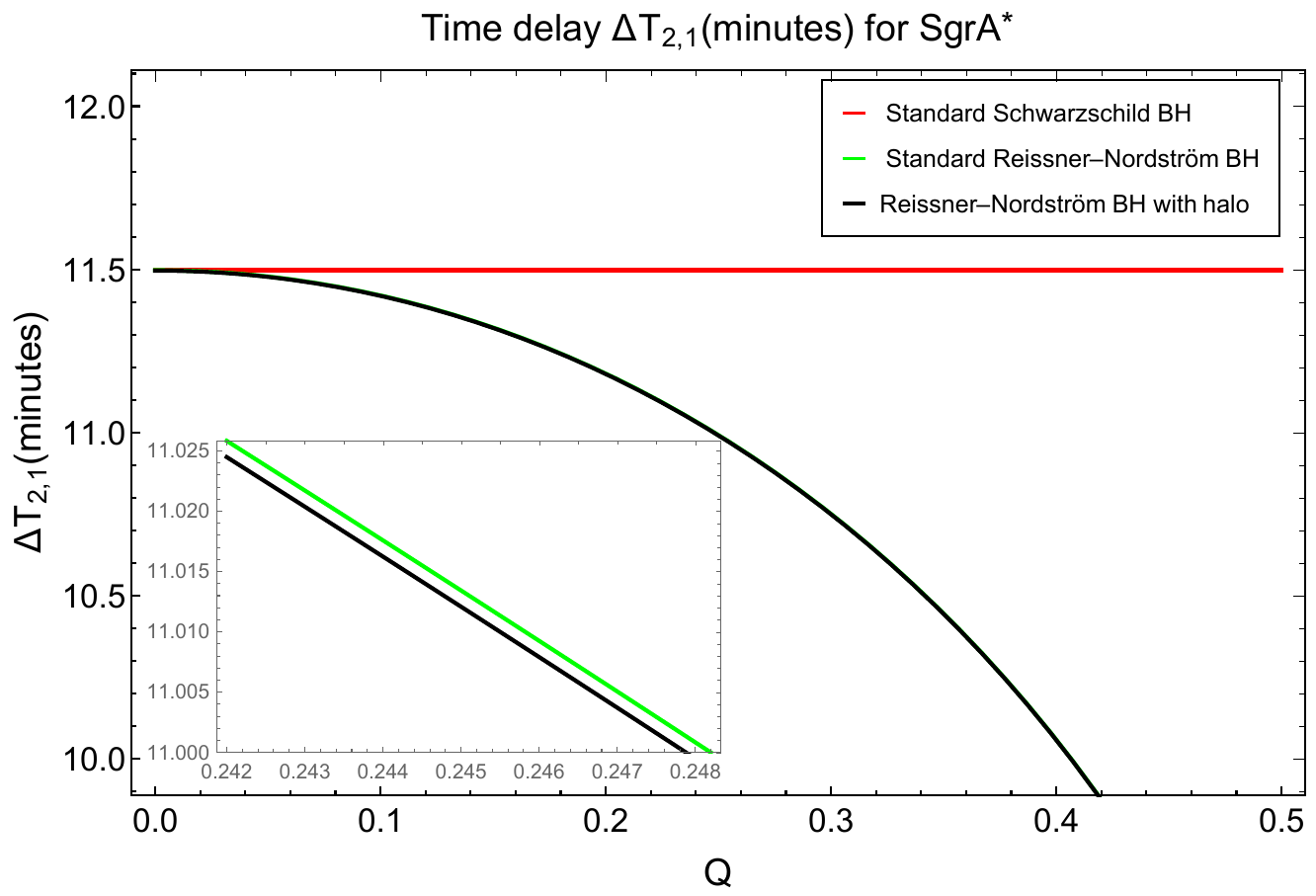}(b)
\caption{The observable quantity, time delays ($\Delta T_{2,1}$), is depicted for two relativistic images corresponding to $M87^{*}$ (left panel) and $Sgr A^*$ (right panel). The black line represents the RN solution with a URC DM halo, while the green line corresponds to the standard RN BH, and the red line represents the standard Schwarzschild BH.}\label{fig:8}
\end{figure*}

In this subsection, we investigate the strong gravitational lensing by charged BHs with a URC DM halo and its various astrophysical consequences in the context of two supermassive BHs $M87^*$ and $SgrA^{*}$. Furthermore, we compare the strong gravitational lensing phenomena by charged BH with a URC DM halo with the standard RN, as well as standard Schwarzschild BHs. The radius of the photon sphere $\mathit{r_{ph}}/R_s$ in Fig. \ref{fig:1}(a) and the impact parameter $\mathit{u_{ph}/R_s}$ in Fig. \ref{fig:1}(b) are depicted as functions of the charge parameters $Q$ for the URC model. From Figs. \ref{fig:1}(a) \& \ref{fig:1}(b), it is found that the radius of the photon sphere $\mathit{r_{ph}}$ and the impact parameter $\mathit{u_{ph}/R_s}$ decrease with charge parameters $Q$. Furthermore, the impact parameter $\mathit{u_{ph}/R_s}$ for the case of a charged BH with a URC DM halo is slightly greater than the case of the standard RN ($Q=0.3$) and smaller than the case of standard Schwarzschild BHs see Fig. \ref{fig:1}(b) and Table \ref{table:1} \& \ref{table:2}. The deflection limit coefficients $\mathit{\bar{a}}$ and $\mathit{\bar{b}}$ for the case of the URC DM model as functions of the charge parameter $Q$ are respectively displayed in Figs. \ref{fig:2}(a) and \ref{fig:2}(b). It is observed that the deflection limit coefficient $\bar{a}$ increases with Q, while the deflection limit coefficient $\bar{b}$ first increases with Q, then reaches its maximum value, and then decreases with Q. The strong deflection angle $\alpha_D$ for the case of the URC DM model as a function of the charge parameter $Q$ is displayed in Fig. \ref{fig:3}(a), and as a function of the impact parameter u with different values of the charge parameter $Q$ is displayed in Fig. \ref{fig:3}(b). In Fig. \ref{fig:3}(a), it is seen that the deflection angle $\alpha_D$ increases with the magnitude of the charge parameter $Q$. The deflection angle $\alpha_D$ for the case of a charged BH with a URC DM halo (black solid line) is slightly greater than the case of the standard RN (green solid line) as well as standard Schwarzschild (red solid line) BHs. Moreover, the strong deflection angle $\alpha_D$ for the URC DM model decreases with the minimum impact parameter $u$, and the strong deflection angle $\alpha_D$ diverges at the critical impact parameter $u=u_{ph}$ at the photon radius $r=r_{ph}$ see Fig.\ref{fig:3}(b). From Table \ref{table:3} and Figs. \ref{fig:4} \& \ref{fig:5}, it is found that the angular image position $\theta_{\infty}$ and relative magnification $r_{mag}$ decrease with the parameter $Q$. In contrast, the observable quantity, the angular image separation $S$ for the case of a charged BH with a URC DM halo is slightly greater than the cases of standard RN as well as for the cases of Schwarzschild BH with a halo and standard Schwarzschild BHs see Fig. \ref{fig:5}(a) \& \ref{fig:5}(b) and Table \ref{table:3} . Still, the angular separation $S$ increases with the parameter $Q$. Furthermore, the observable quantity, the angular image position $ \theta_{\infty}$ for the case of a charged BH with a URC DM halo is slightly greater than that for the standard RN BH ($Q=0.3$) and smaller than those for the cases of Schwarzschild BH with a halo, as well as standard Schwarzschild BHs see Fig. \ref{fig:4}(a) \& \ref{fig:4}(b) and Table \ref{table:3}. Meanwhile, the observable quantity, the angular image separation $ S$ for the case of a charged BH with a URC DM halo is slightly greater than the cases of standard RN, as well as for the cases of Schwarzschild BH with a halo and standard Schwarzschild BHs see Fig. \ref{fig:5}(a) \& \ref{fig:5}(b) and Table \ref{table:3} . However, the observable quantity, the relative magnification $r_{\text{mag}}$ for the case of a charged BH with a URC DM halo is slightly smaller than the cases of standard RN, as well as for the cases of Schwarzschild BH with a halo and standard Schwarzschild BHs see Fig. \ref{fig:6} and Table \ref{table:3}. It is important to emphasize that the photon sphere radius $\mathit{r_{\text{ph}}/R_s}$, the minimum impact parameter $\mathit{u_{\text{ph}}/R_s}$, strong deflection limit coefficients $\mathit{\bar{a}}$ and $\mathit{\bar{b}}$, strong deflection angle $\mathit{\alpha_D}$, and the lensing observable quantity, the relative magnification $r_{\text{mag}}$ for the case of a charged BH with a URC DM halo exhibit nearly identical characteristics for the $M87^{*}$ and $Sgr A^{*}$ scenarios. In Figs. \ref{fig:7}(a) \& \ref{fig:7}(b), the outermost relativistic Einstein rings ($\theta^E_n$) of $M87^{*}$ and $Sgr A^{*}$ have been depicted for the case of a charged BH with a URC DM halo, standard RN, Schwarzschild BH with a halo, and standard Schwarzschild BH. It has been observed that the Einstein ring radius is slightly smaller in the cases of the standard RN BH (with Q=0.2) and in the cases of Schwarzschild BHs with a halo compared to standard Schwarzschild BHs. In Figs. \ref{fig:8}(a) \& \ref{fig:8}(b), We plot the time delays $\Delta T_{2,1}$ between two different relativistic images. From these figures and Table. \ref{table:3}, it is evident that the time delays $\Delta T_{2,1}$ between two different relativistic images decrease with the parameter $Q$ in the context of the $M87^{*}$ and $Sgr A^{*}$ BHs with a URC DM halo. Furthermore, it is observed that the time delays $\Delta T_{2,1}$ are smaller than those in the cases of the standard RN BH (Q=0.2), as well as in the cases of Schwarzschild BHs with a halo and standard Schwarzschild BH.
\begin{table*}
\centering
\begin{tabular}{|p{2.5cm}|p{4cm}|p{3.5cm}|p{3cm}|p{1cm}|}
\hline
\multicolumn{4}{|c|}{Strong Lensing Coefficients }\\
\hline
 $Q$& $\bar{a}$  & $\bar{b} $ &$ u_{ph}/R_{s}$\\
\hline
Standard Schwarzschild BH&  1.00&-0.40023&2.59808\\
\hline
Standard RN BH(0.3)&  1.05183&-0.396509&2.42935\\
\hline
0& 1.00003&-0.400156&2.59829\\
0.1& 1.00458&-0.399274&2.58083\\
0.2&  1.01976&-0.39711&2.5267\\
0.3&  1.05185&-0.396434&2.429555\\
0.4&  1.12319&-0.413559&2.2732\\
0.5&  1.41414&-0.732937&2.00022\\
\hline
\end{tabular}
\caption{Estimation of strong lensing coefficients $\bar{a}$, $\bar{b}$, and $u_{ph}/R_{s}$ for various values of the charge parameter $Q$ in the context of a charged BH with a URC DM halo, specifically for the M87* BH. The constants used in the calculations are $A=805.231$ and $B=3.40611*10^{-9}$.}\label{table:1}
\end{table*}

\begin{table*}
\centering
\begin{tabular}{|p{2.5cm}|p{4cm}|p{3.5cm}|p{3cm}|p{1cm}|}
\hline
\multicolumn{4}{|c|}{Strong Lensing Coefficients }\\
\hline
 $Q$& $\bar{a}$  & $\bar{b} $ &$ u_{ph}/R_{s}$\\
\hline
Standard Schwarzschild BH&  1.00&-0.40023&2.59808\\
\hline
Standard RN BH(0.3)&  1.05183&-0.396509&2.42935\\
\hline
0& 1.0000&-0.400226&2.59809\\
0.1& 1.00456&-0.399344&2.58063\\
0.2&  1.01974&-0.39718&2.5265\\
0.3&  1.05183&-0.396505&2.42936\\
0.4&  1.12317&-0.413634&2.273\\
0.5&  1.41421&-0.733188&2.00001\\
\hline
\end{tabular}
\caption{Estimation of strong lensing coefficients $\bar{a}$, $\bar{b}$, and $u_{ph}/R_{s}$ for various charge parameters $Q$ in  presence of a URC DM Halo around the Supermassive BH Sgr* with parameters $A=5738.77$ and $B=2.6346\times10^{-11}$.}\label{table:2}
\end{table*}

\begin{table*}
\begin{center}
\begin{tabular}{|c|cccc|cccc|}
\hline
Parameters & & & $M87^*$ & & & &$SgrA^*$& \\
\hline
$Q$ & $ \theta_{\infty} (\mu as)$&$S(\mu as)$&
$r_{mag}$&$\Delta T_{2,1}(minutes)$& $ \theta_{\infty} (\mu as)$&$S(\mu as)$&$r_{mag}$ & $\Delta T_{2,1}$ (minutes)\\
\hline
Standard & & & & & & & &\\
Schwazschild BH&19.9633&0.024984&6.82188&17378.8&26.38&0.0330177&6.82188&11.4973\\
\hline

Standard Reissner & & & & & & & &\\
Nordstr\"om BH(0.3)&18.6668&0.0325876&6.48575&16250.2&24.6692&0.0430665&6.48575&10.7507\\
\hline

0&19.9649&0.0249923&6.8217&17377&26.3827&0.0330183&6.82187&11.4973\\
\hline

0.1&19.8307&0.0256108&6.79076&17260.3&26.2054&0.0338356&6.79093&11.4200\\

0.2&19.4148&0.0277422&6.68968&16898.2&25.6558&0.0366517&6.68984&11.1805\\

0.3&18.6683&0.0325976&6.48559&16248.4&24.6693&0.0430672&6.48574&10.7506\\

0.4&17.4669&0.0449611&6.07368&15202.6&23.0816&0.0594043&6.07377&10.6587\\
0.5&15.3694&0.107632&4.82406&13376.4&20.3095&0.142238&4.8238&8.85059\\
\hline
\end{tabular}
\caption{Estimating various strong lensing observables for a URC DM halo in the presence of supermassive BHs, specifically $ M87^*$, $Sgr A^*$, with different values of the charge parameter $Q$.}\label{table:3}
\end{center}
\end{table*}
\subsection{CDM halo with NFW profile}
\begin{figure*}[htbp]
\centering
\includegraphics[width=.45\textwidth]{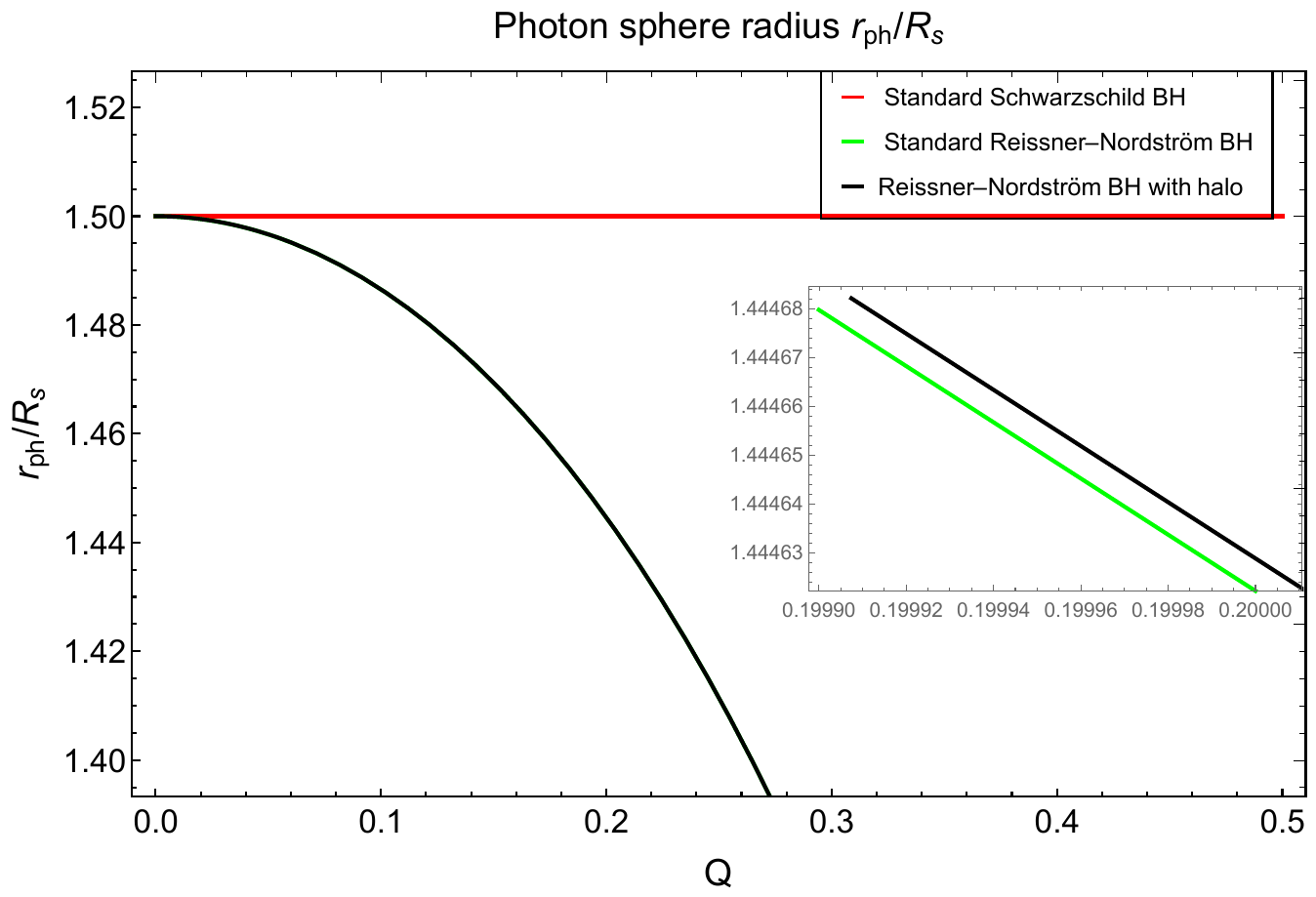}(a)
\qquad
\includegraphics[width=.45\textwidth]{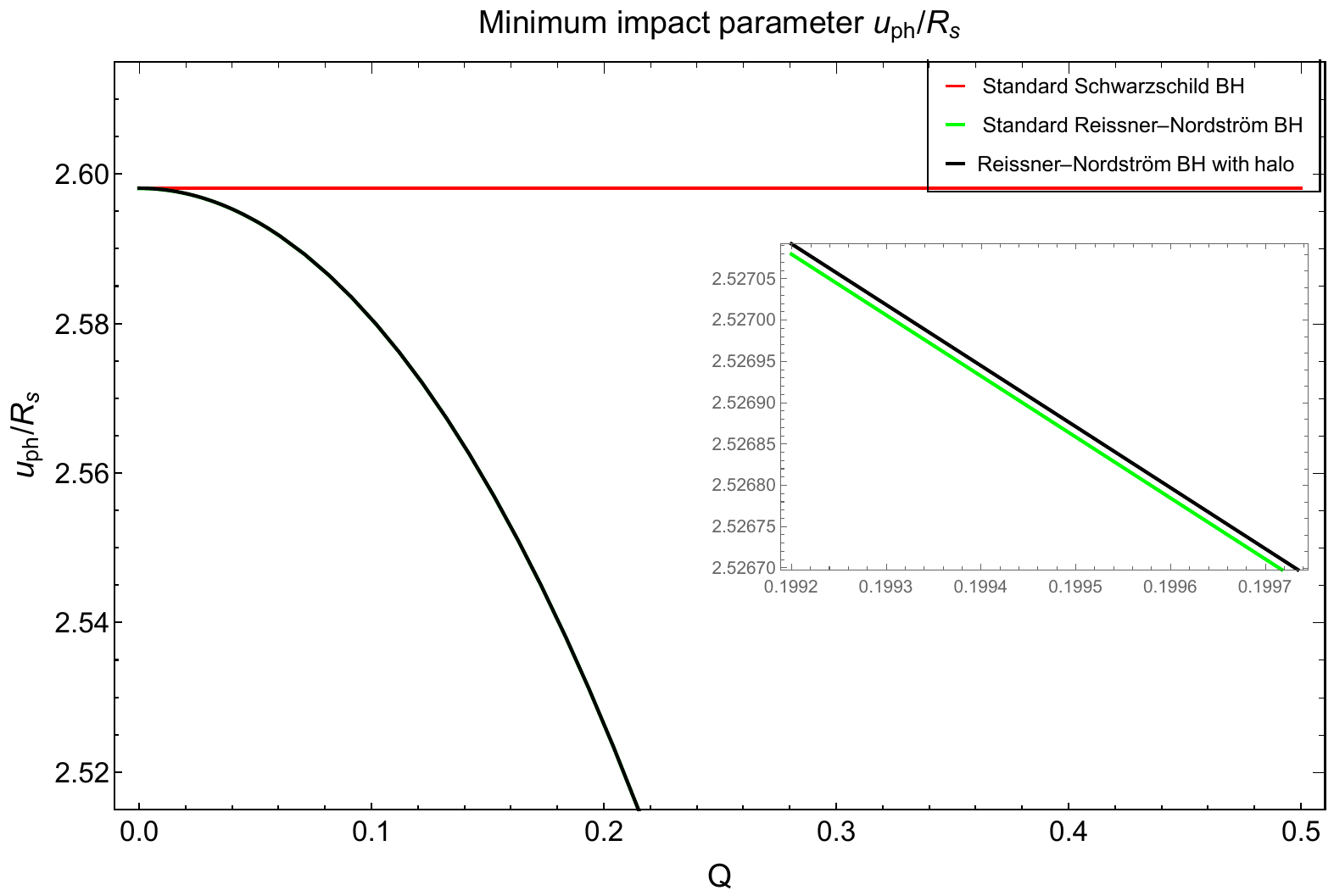}(b)
\caption{The left panel depicts the photon sphere radius (\(r_{\text{ph}}/R_s\) vs \(Q\)), while the right panel illustrates the minimum impact parameter (\(u_{\text{ph}}/R_s\) vs \(Q\)) for BHs with a CDM halo (depicted by the black line), in comparison with standard RN BHs (depicted by the green line) and standard Schwarzschild BHs (depicted by the red line). It is crucial to emphasize that both the photon sphere radius (\(r_{\text{ph}}/R_s\)) and the minimum impact parameter (\(u_{\text{ph}}/R_s\)) exhibit nearly identical characteristics for the \(M87^{*}\) and \(Sgr A^{*}\) scenarios.}\label{fig:9}
\end{figure*}

\begin{figure*}[htbp]
\centering
\includegraphics[width=.45\textwidth]{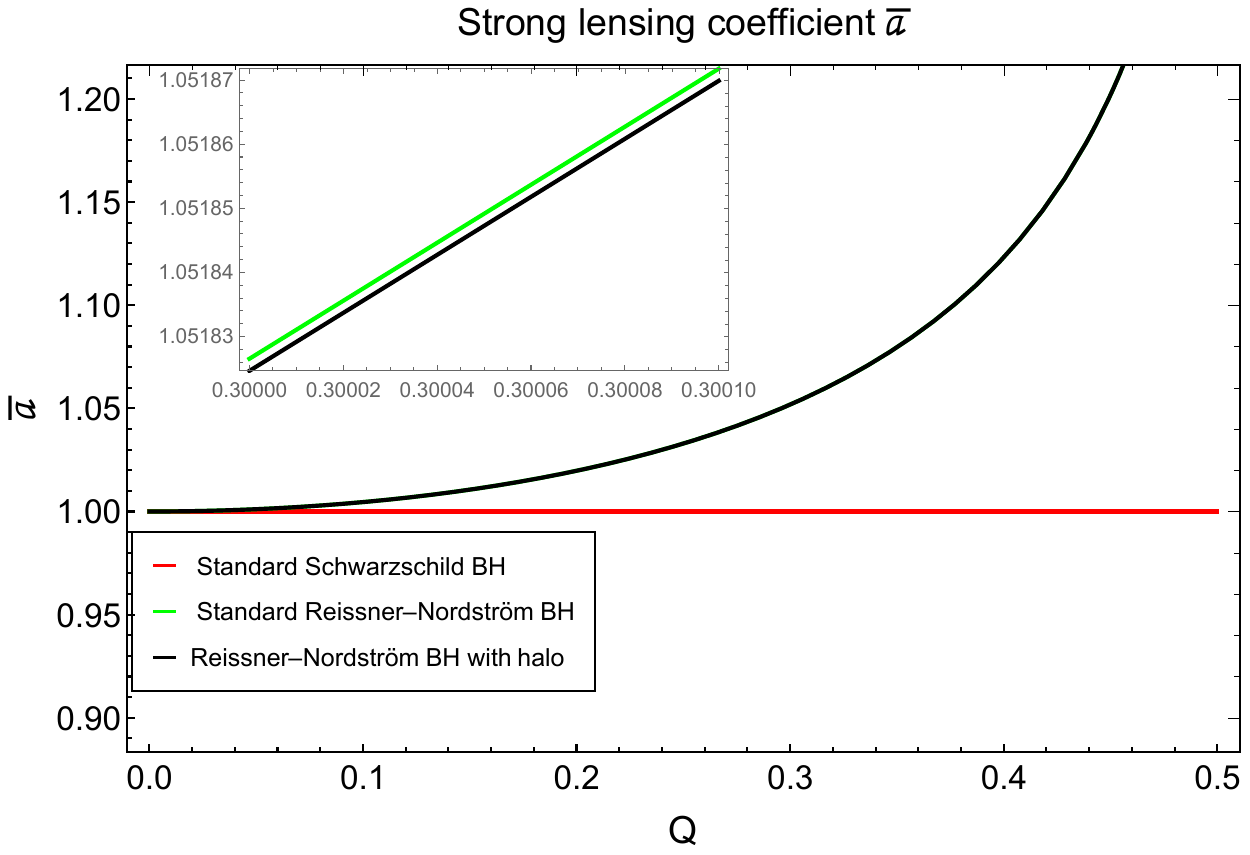}(a)
\qquad
\includegraphics[width=.45\textwidth]{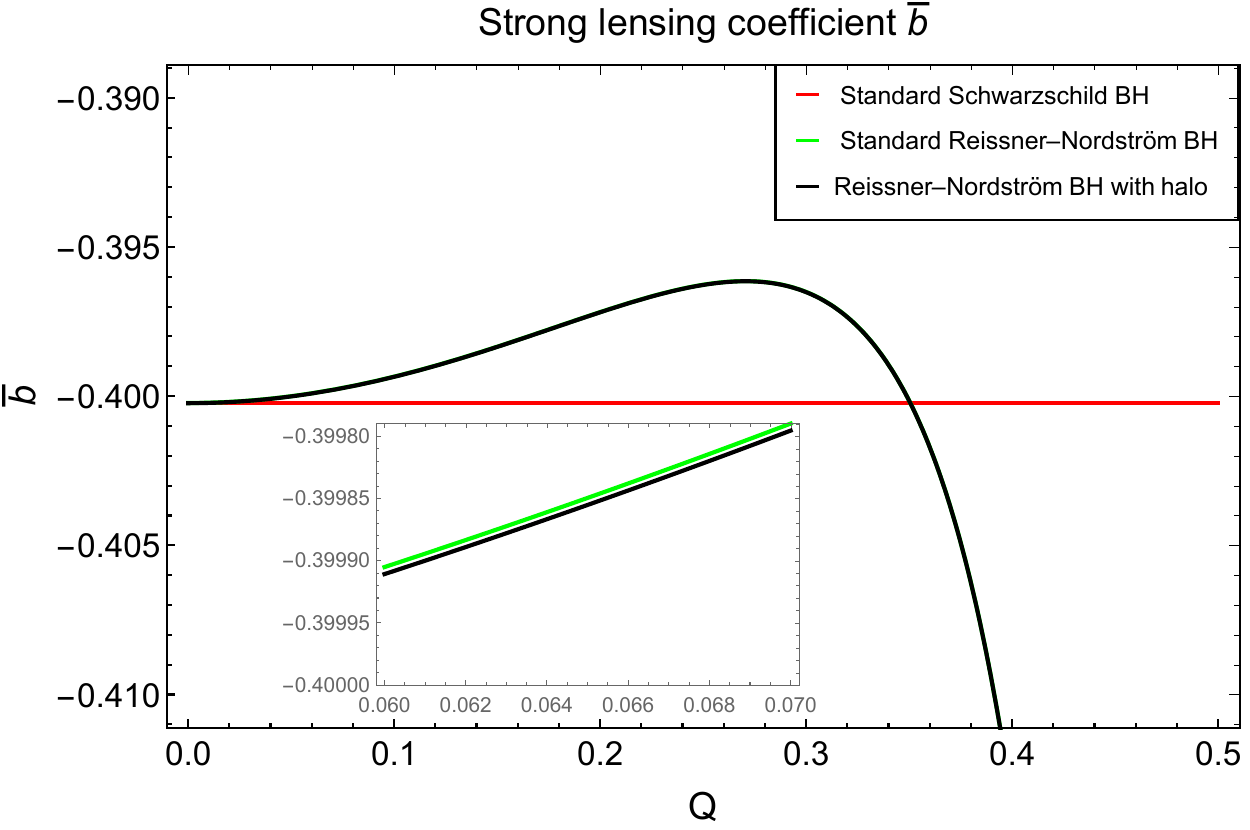}(b)
\caption{The left panel displays the deflection limit coefficients ($\mathit{\bar{a}}$ vs Q), while the right panel shows ($\mathit{\bar{b}}$ vs Q) for the RN BH with a CDM halo (depicted by the black line), contrasted with standard RN BHs (green line) and standard Schwarzschild BHs (red line). Notably, it is crucial to underscore that the deflection limit coefficients $\mathit{\bar{a}}$ and $\mathit{\bar{b}}$ exhibit nearly identical characteristics in both the $M87^{*}$ and $Sgr A^{*}$ scenarios.}\label{fig:10}
\end{figure*}

\begin{figure*}[htbp]
\centering
\includegraphics[width=.45\textwidth]{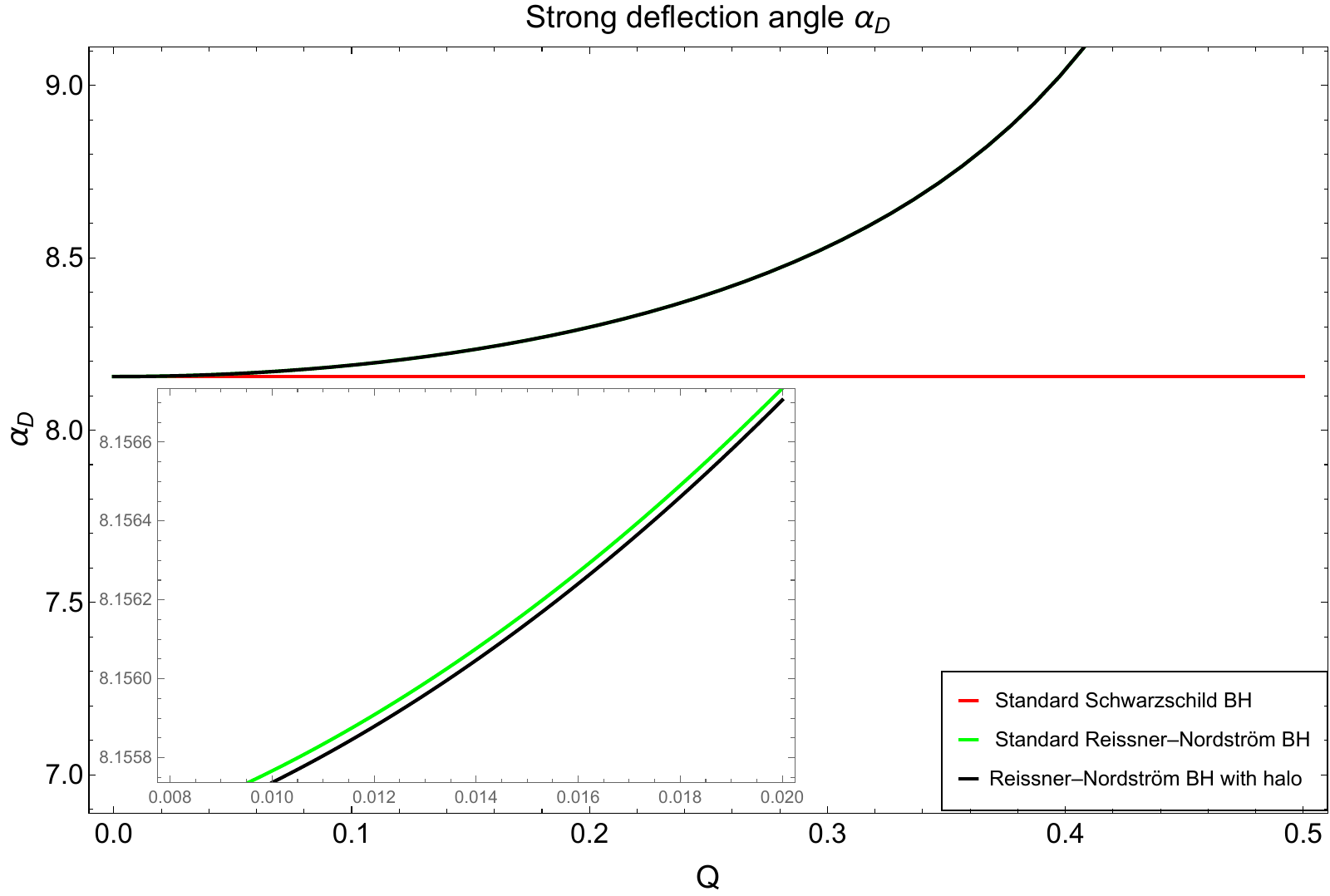}(a)
\qquad
\includegraphics[width=.45\textwidth]{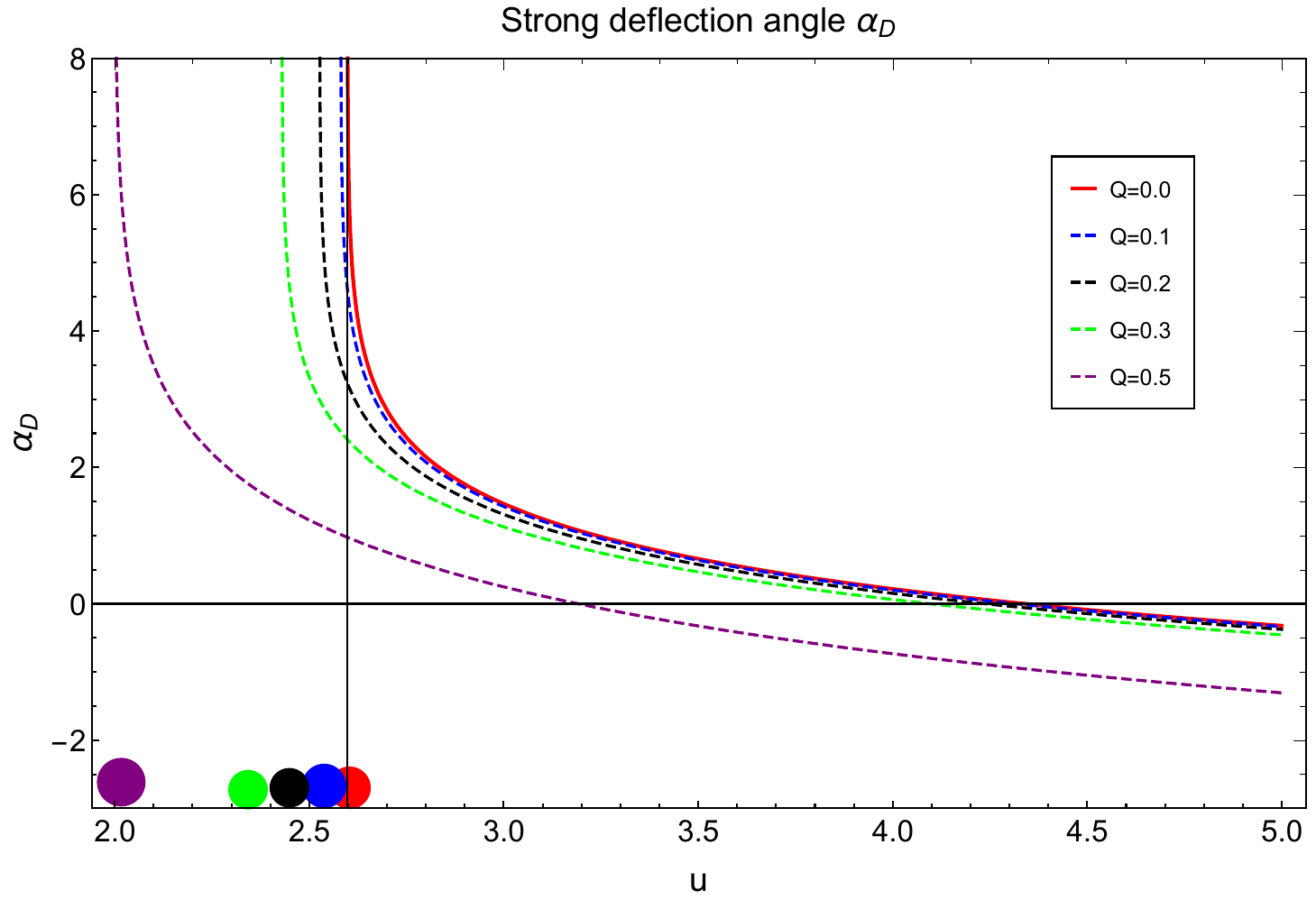}(b)
\caption{The left panel illustrates the strong deflection angle ($\mathit{\alpha_{D}}$) versus charge parameter ($Q$) at $u=u_{ph}+0.0002$ for the RN BH with a CDM halo (depicted by the black line), contrasted with the standard RN case (green line) and the standard Schwarzschild BH (red line). In the right panel, the strong deflection angle ($\mathit{\alpha_{D}}$) is plotted with  the impact parameter ($u$) for various values of charge ($Q$).The dotted lines indicated in Fig(b) are the value of the impact parameter$u=u_{ph}$, where the deflection angle diverges. Notably, the strong deflection angle ($\mathit{\alpha_{D}}$) exhibits nearly identical characteristics for both the $M87^{*}$ and $Sgr A^{*}$ scenarios.}\label{fig:11}
\end{figure*}

\begin{figure*}[htbp]
\centering
\includegraphics[width=.45\textwidth]{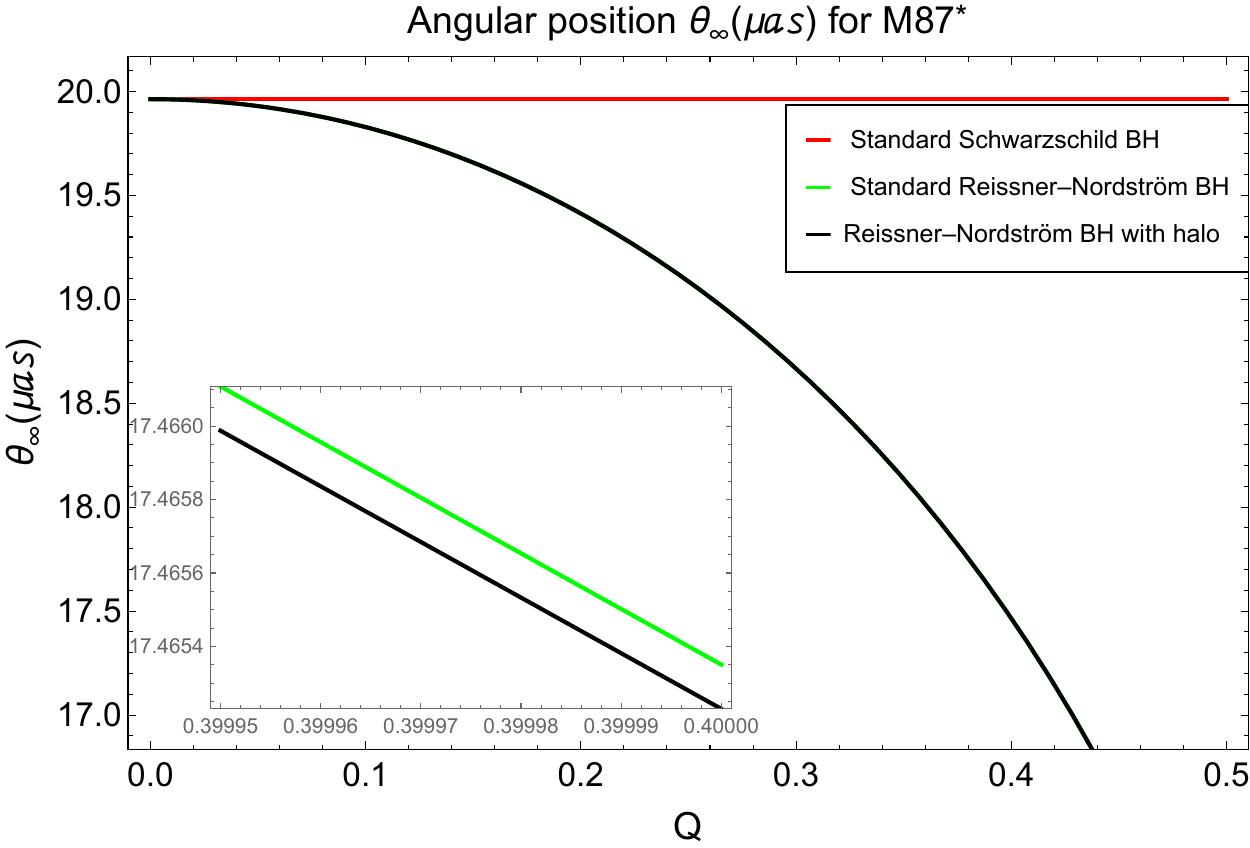}(a)
\qquad
\includegraphics[width=.45\textwidth]{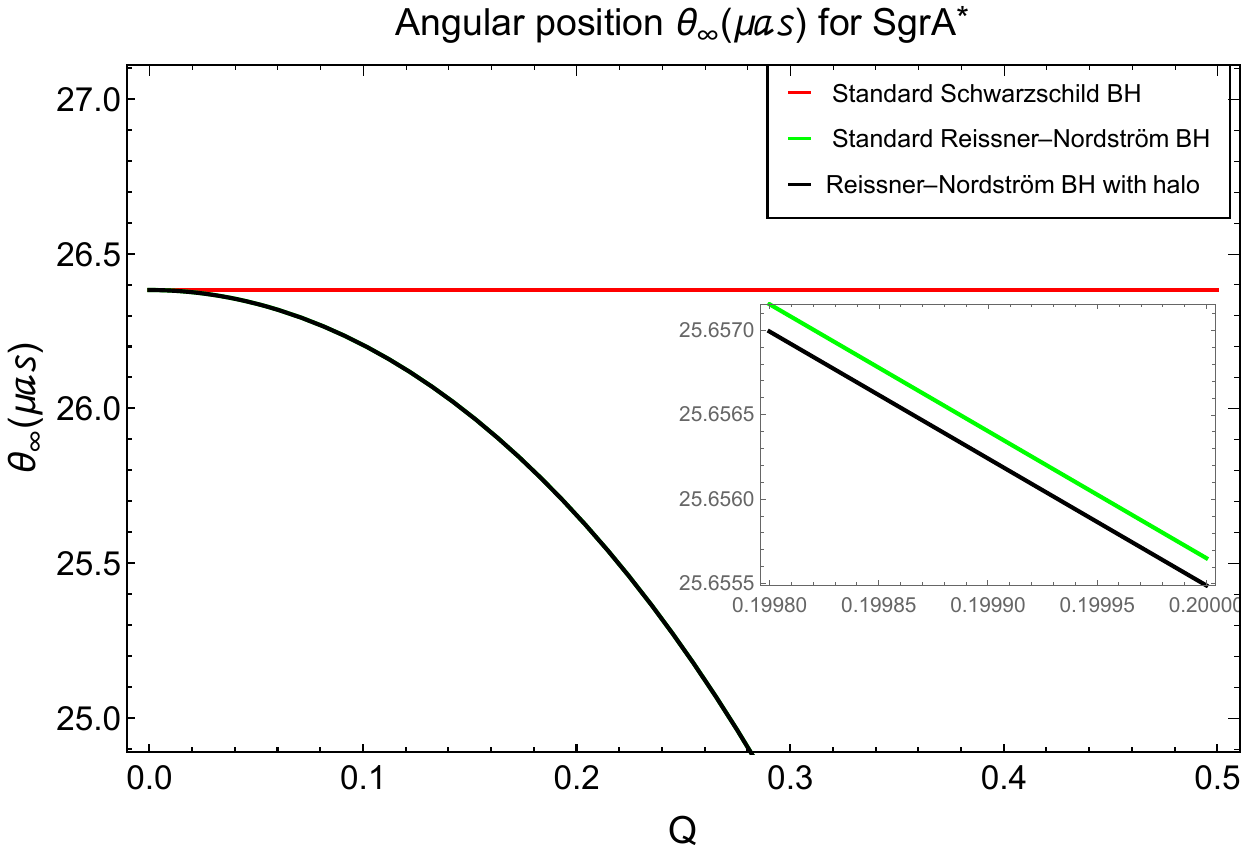}(b)
\caption{The observable quantity, angular position ($\mathit{\theta_{\infty}}$) plotted against Q for $M87^{*}$ (left panel) and $Sgr A^*$ (right panel) in the RN BH with a CDM halo (depicted by the black line), as compared to the standard RN solution (green line) and the standard Schwarzschild BH (red line).}\label{fig:12}
\end{figure*}

\begin{figure*}[htbp]
\centering
\includegraphics[width=.45\textwidth]{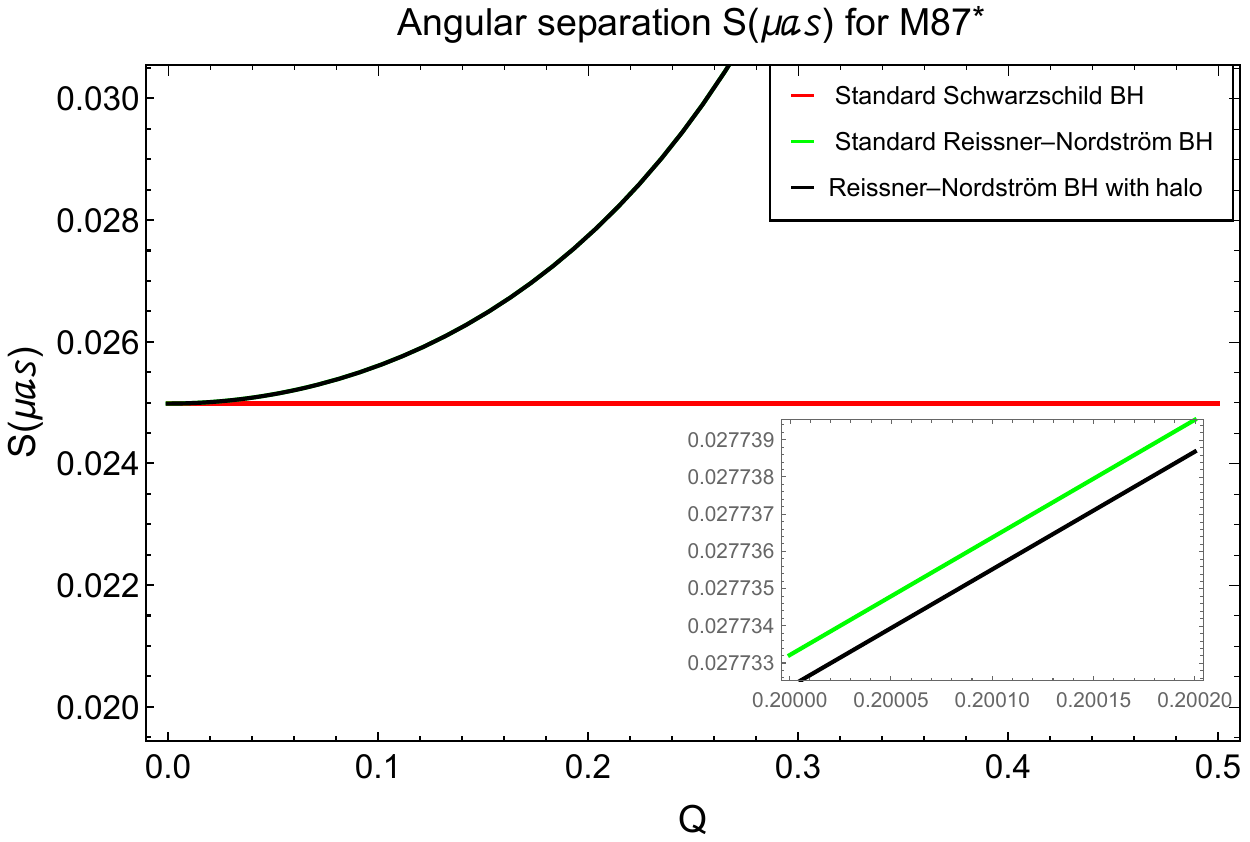}(a)
\qquad
\includegraphics[width=.45\textwidth]{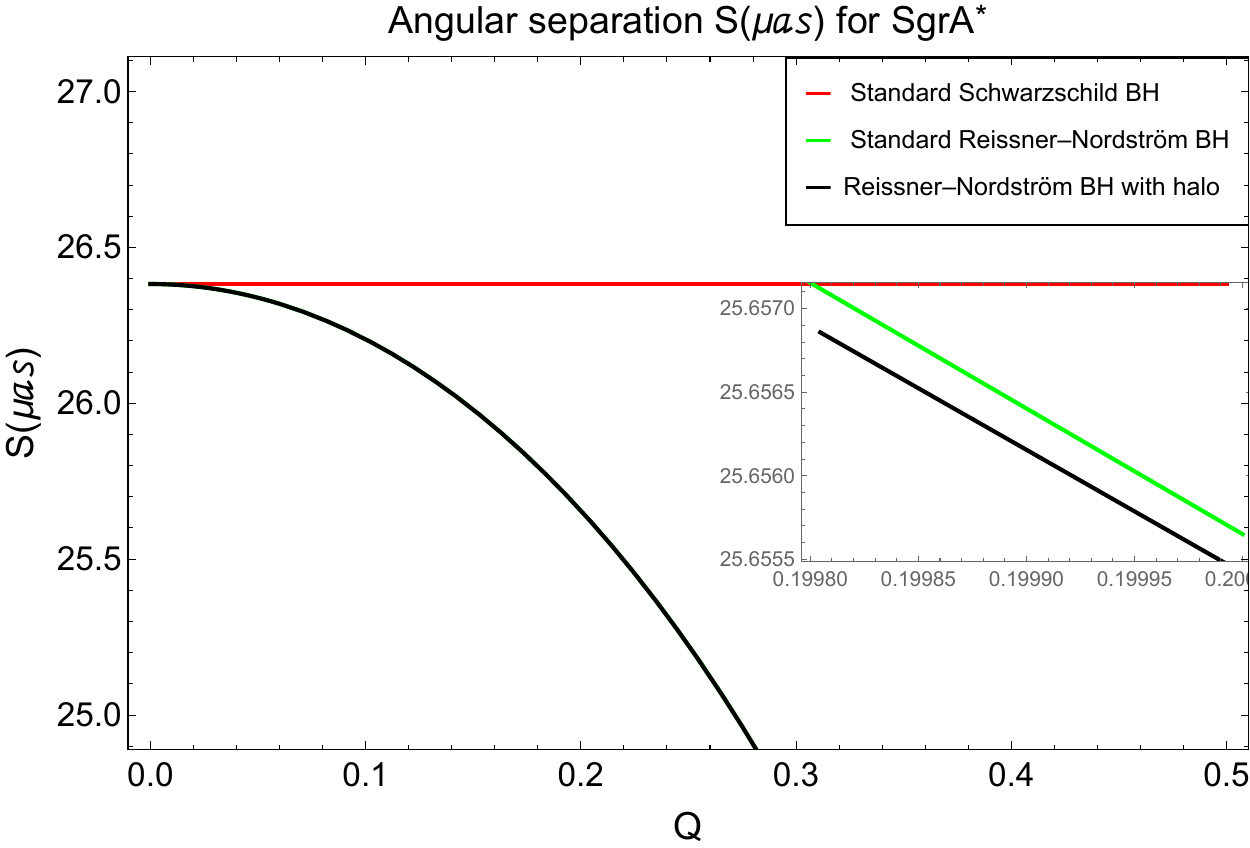}(b)
\caption{The observable angular separation ($S$ )  vs $Q$ for $M87^*$ (left panel) and $Sgr A^*$ (right panel) is depicted. The black line represents the RN solution with a CDM halo, while the green line corresponds to the standard RN BH, and the red line represents the standard Schwarzschild BH.}\label{fig:13}
\end{figure*}

\begin{figure*}[htbp]
\centering
\includegraphics[width=.45\textwidth]{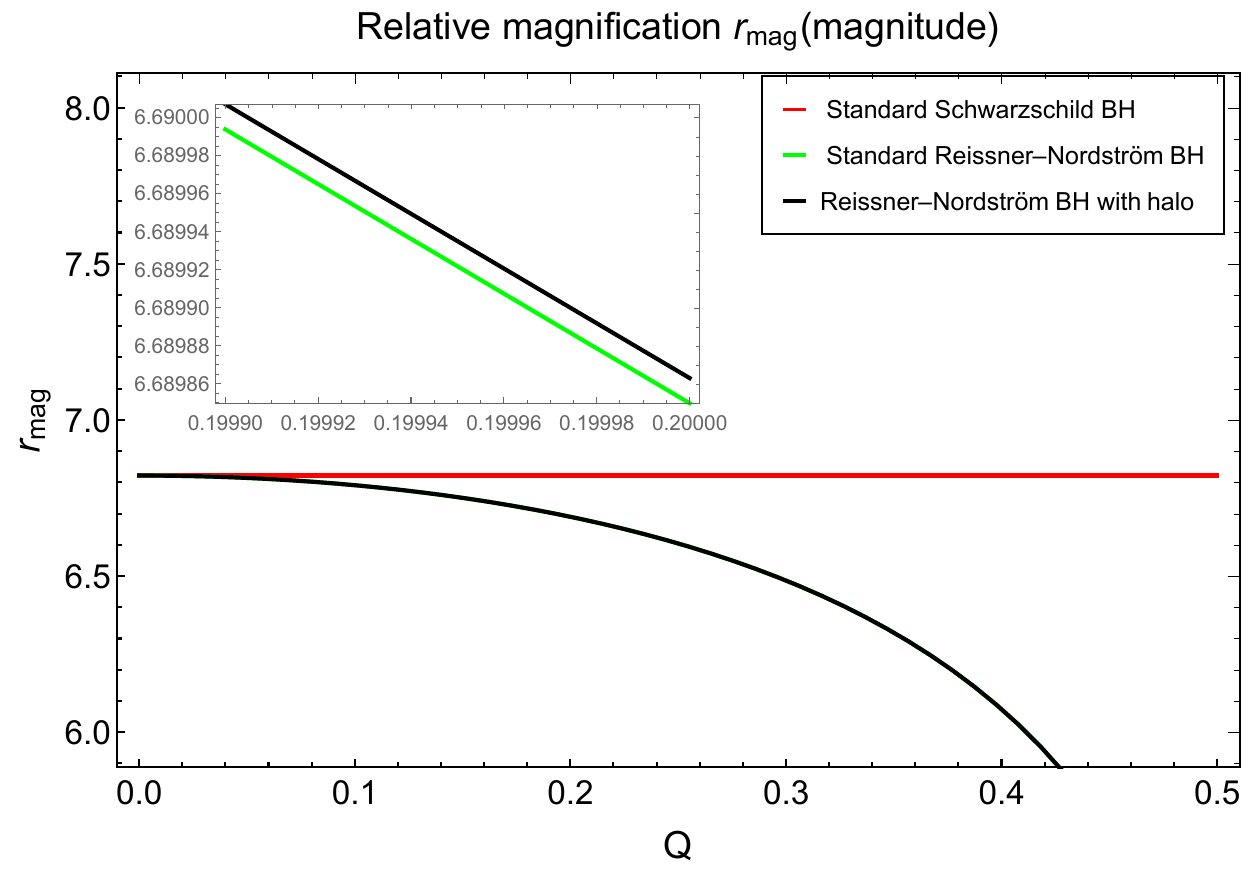}
\caption{Comparison of the observable quantity, relative magnification ($\mathit{r_{mag}}$) versus Q, for the RN BH with a CDM halo (depicted by the black line) in contrast to the standard RN BH (green line) and the standard Schwarzschild BH (red line). Notably, it is crucial to emphasize that the relative magnification $\mathit{r_{mag}}$ demonstrates nearly identical characteristics for both the $M87^{*}$ and $Sgr A^{*}$ scenarios.}\label{fig:14}
\end{figure*}

\begin{figure*}[htbp]
\centering
\includegraphics[width=.45\textwidth]{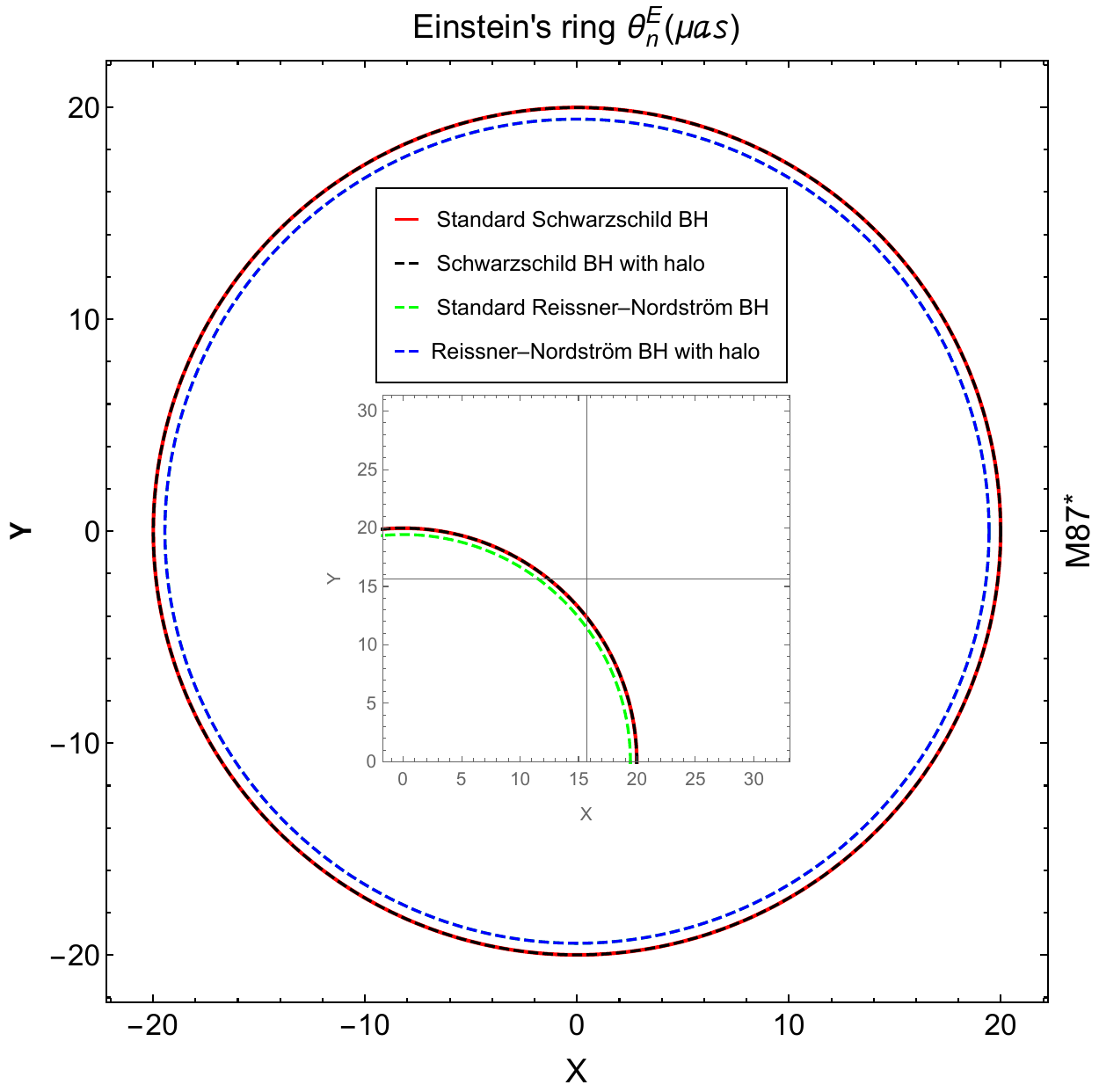}(a)
\qquad
\includegraphics[width=.45\textwidth]{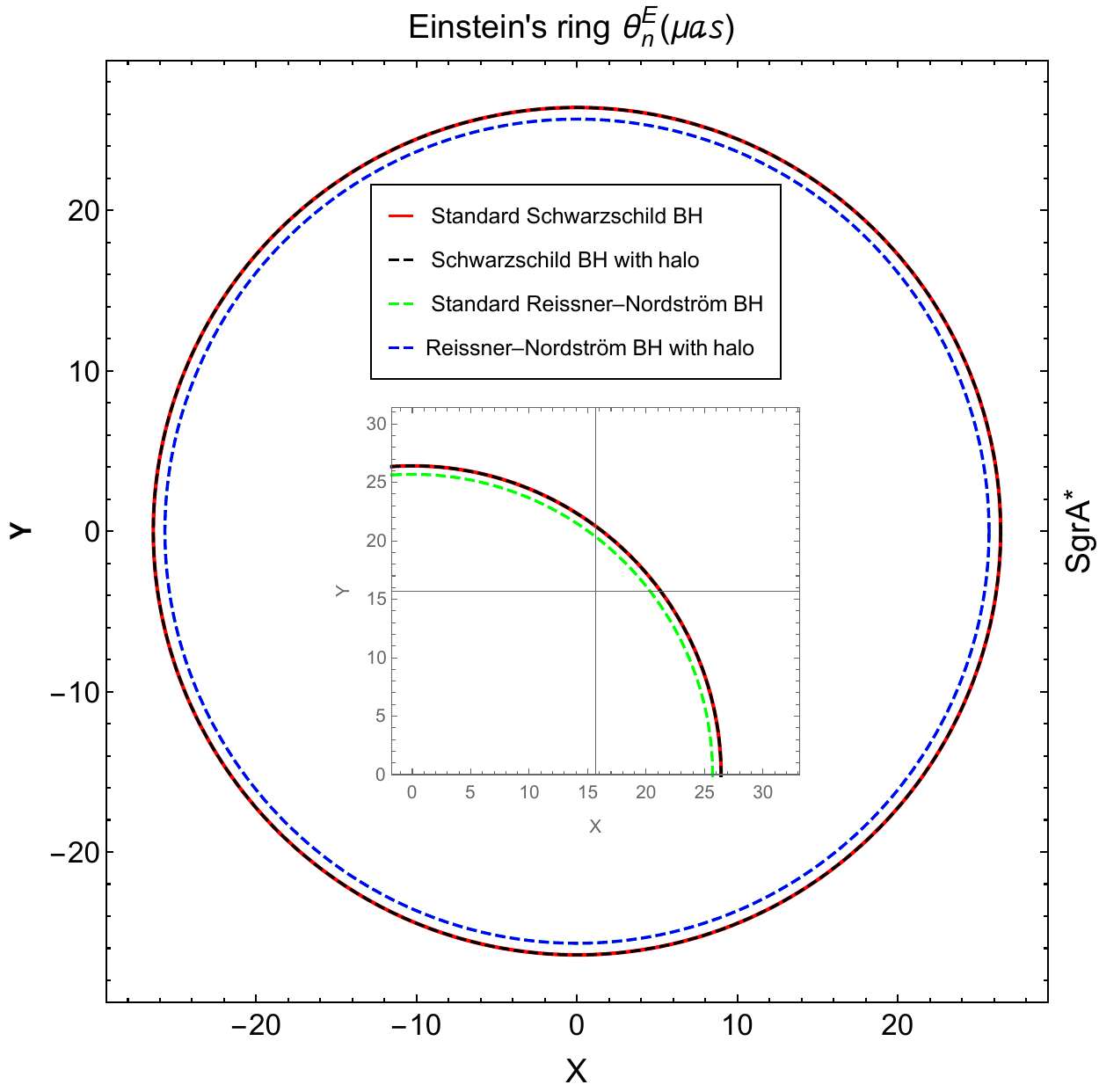}(b)
\caption{The observable quantity the Einstein's ring ($\theta^{E}_n$ vs Q) for $M87^{*}$ (left panel) and for $Sgr A^*$(right panel) for the RN with CDM halo (blue dotted line) in comparison with Schwarzschild with URC DM halo (black dotted line), standard RN (green dotted line) and standard Schwarzschild (red  solid line) BHs.}
    \label{fig:15}
\end{figure*}

\begin{figure*}[htbp]
\centering
\includegraphics[width=.45\textwidth]{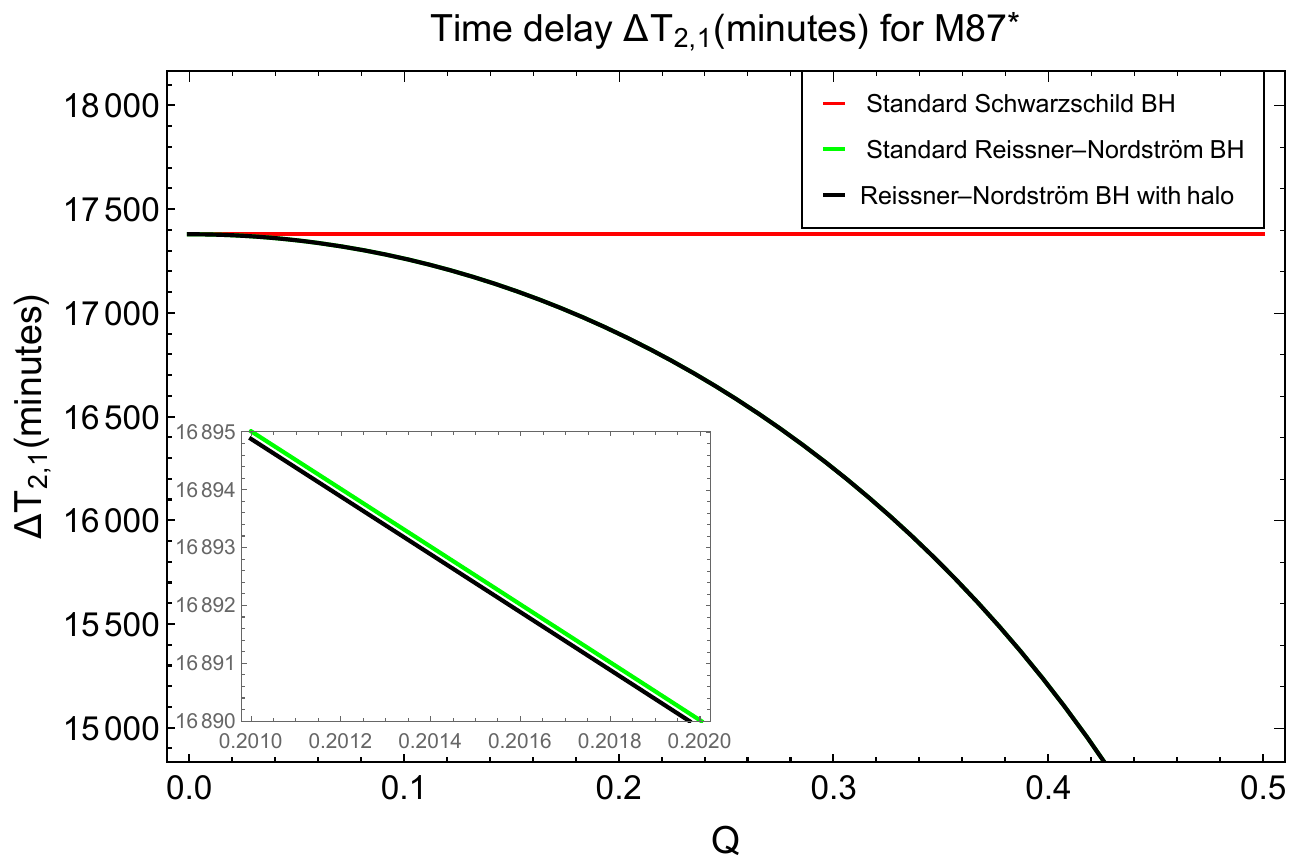}(a)
\qquad
\includegraphics[width=.45\textwidth]{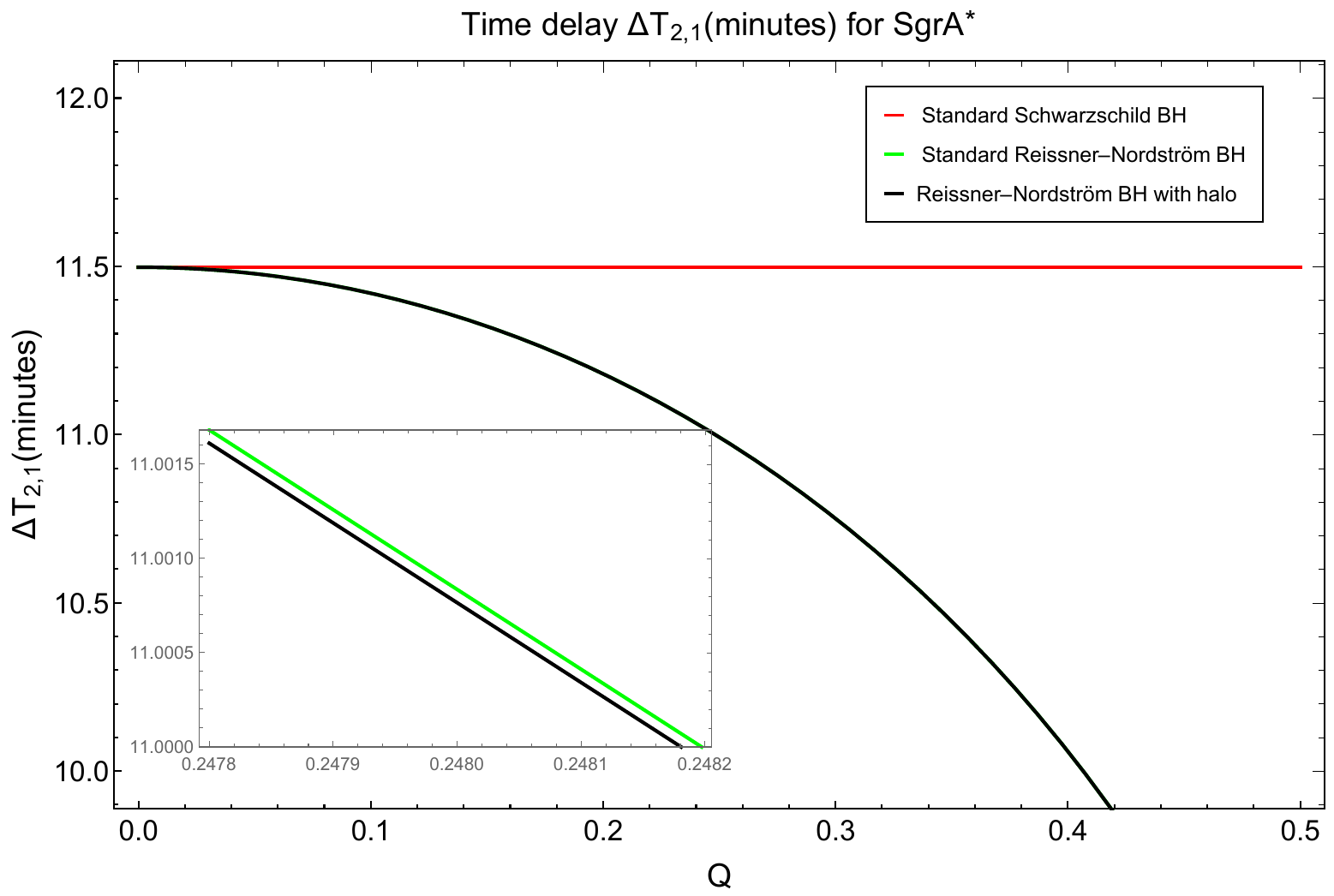}(b)
\caption{The observable quantity the  time delays $\Delta T_{2,1}$ between two relativistic images for $M87^{*}$ (left panel) and  for $Sgr A^*$(right panel) for the RN with URC DM halo (black line) in comparison with standard RN (green line) and standard Schwarzschild (red line) BHs.}
    \label{fig:16}
\end{figure*}

In this subsection, we investigate strong gravitational lensing by a charged BH  with a CDM halo and explore its various astrophysical consequences in the context of two supermassive BHs, $M87^*$ and $SgrA^{*}$. Additionally, we compare the strong gravitational lensing phenomena by a charged BH with a CDM halo with standard RN, as well as standard Schwarzschild BHs. The radius of the photon sphere $\mathit{r_{ph}}/R_s$ in Fig. \ref{fig:9}(a) and the impact parameter $\mathit{u_{ph}/R_s}$ in Fig. \ref{fig:9}(b) are depicted as functions of charge parameters $Q$ for the URC model. From Figs. \ref{fig:9}(a) \& \ref{fig:9}(b), we observe that the radius of the photon sphere $\mathit{r_{ph}}$ and the impact parameter $\mathit{u_{ph}/R_s}$ decrease with charge parameters $Q$. Furthermore, the impact parameter $\mathit{u_{ph}/R_s}$ for the charged BH with URC DM halo is slightly greater than that for the standard RN ($Q=0.3$) and smaller than that for standard Schwarzschild BHs see Fig. \ref{fig:9}(b) and Table.\ref{table:4} \& \ref{table:5}. The deflection limit coefficients $\mathit{\bar{a}}$ and $\mathit{\bar{b}}$ for the CDM model as functions of the charge parameter $Q$ are respectively displayed in Fig. \ref{fig:10}(a)and Fig.\ref{fig:10}(b) . It is observed that the deflection limit coefficient $\mathit{\bar{a}}$ increases with $Q$, while the deflection limit coefficient $\mathit{\bar{b}}$ first increases with $Q$, then reaches its maximum value and finally decreases with $Q$. The strong deflection angle, $\alpha_D$, for the CDM model is depicted in Fig. \ref{fig:11} and as a function of the charge parameter $Q$. In Fig. \ref{fig:11}(b), it is shown as a function of the impact parameter u with different values of the charge parameter $Q$. In Fig. \ref{fig:11}(a), it is observed that the deflection angle $\alpha_D$ increases with the magnitude of the charge parameter $Q$. The deflection angle for the case of a charged BH with a CDM halo (black solid line) is slightly greater than the case of the standard RN solution (green solid line) as well as the standard Schwarzschild solution (red solid line). Additionally, the strong deflection angle $\alpha_D$ for the CDM model decreases with the minimum impact parameter u, and it diverges at the critical impact parameter $u=u_{ph}$ at the photon radius $r=r_{ph}$ see Fig. \ref{fig:11}(b). From Table \ref{table:6} and Figs. \ref{fig:12}, \ref{fig:13}, and \ref{fig:14}, it is evident that the angular image position $\theta_{\infty}$ and relative magnification $r_{mag}$ decrease with the parameter $Q$, while the angular separation S increases with the parameter $Q$. Furthermore, the observable quantity, the angular image position $\theta_{\infty}$ for the case of a charged BH with a CDM halo, is slightly greater than the case of the standard RN solution ($Q=0.3$) and smaller than the cases of Schwarzschild BHs with a halo, as well as standard Schwarzschild BHs see Fig. \ref{fig:12}(a) \& \ref{fig:12}(b) and Table \ref{table:6}. On the other hand, the observable quantity, the angular image separation S for the case of a charged BH with a CDM halo, is slightly greater than the cases of the standard RN solution, as well as for the cases of Schwarzschild BHs with a halo and standard Schwarzschild BHs see Fig. \ref{fig:13}(a) \& \ref{fig:13}(b) and Table \ref{table:6}. However, the observable quantity, the relative magnification $r_{mag}$ for the case of a charged BH with a CDM halo, is slightly smaller than the cases of the standard RN solution, as well as for the cases of Schwarzschild BHs with a halo and standard Schwarzschild BHs see Fig. \ref{fig:14} and Table \ref{table:6} . It is crucial to underscore that the photon sphere radius \(\mathit{r_{\text{ph}}/R_s}\) and the minimum impact parameter \(\mathit{u_{\text{ph}}/R_s}\), strong deflection limit coefficients \(\mathit{\bar{a}}\) and \(\mathit{\bar{b}}\), strong deflection angle \(\mathit{\alpha_D}\), and the lensing observable quantity the relative magnification \(r_{\text{mag}}\) exhibit nearly identical characteristics for the \(M87^{*}\) and \(Sgr A^{*}\) scenarios. In Figs. \ref{fig:15}(a) \& \ref{fig:15}(b), the outermost relativistic Einstein rings (\(\theta^E_n\)) of \(M87^{*}\) and \(Sgr A^{*}\) have been depicted for the case of a charged BH with a CDM halo, standard RN, Schwarzschild BH with a halo, and standard Schwarzschild BH. It is observed that the Einstein ring radius is slightly smaller than the cases of standard RN (Q=0.2) as well as for the cases of Schwarzschild BH with a halo and standard Schwarzschild BHs. Time delays \(\Delta T_{2,1}\) between two different relativistic images with the charge parameter \(Q\) are plotted in Figs. \ref{fig:16}(a) \& \ref{fig:16}(b). From these two figures and Table \ref{table:6}, it is found that the time delays \(\Delta T_{2,1}\) between two different relativistic images decrease with the parameter \(Q\) in the context of \(M87^{*}\) and \(Sgr A^{*}\) BHs with a CDM halo. It is further observed that the time delays \(\Delta T_{2,1}\) are smaller than the cases of standard RN (Q=0.2) as well as for the cases of Schwarzschild BH with a halo and standard Schwarzschild BH.
\begin{table*}
\centering
\begin{tabular}{|p{2.5cm}|p{4cm}|p{3.5cm}|p{3cm}|p{1cm}|}
\hline
\multicolumn{4}{|c|}{Strong Lensing Coefficients }\\
\hline
 $Q$& $\bar{a}$  & $\bar{b} $ &$ u_{ph}/R_{s}$\\
\hline
Standard Schwarzschild BH&  1.00&-0.40023&2.59808\\
\hline
Standard RN BH(0.3)&  1.05183&-0.396509&2.42935\\
\hline
0& 0.999997&-0.400237&2.59806\\
0.1& 1.00455&-0.399355&2.5806\\
0.2&  1.01973&-0.397191&2.52647\\
0.3&  1.05182&-0.396516&2.42933\\
0.4&  1.12317&-0.413646&2.27297\\
0.5&  1.41422&-0.733228&1.99998\\
\hline
\end{tabular}
\caption{Estimation  for strong lensing coefficients $\bar{a}$, $\bar{b}$, and $u_{\text{ph}}/R_{\text{s}}$ for various values of the charge parameter $Q$ in the context of a charged BH with a CDM halo around the supermassive BH $M87^*$. The constants are given as $A = 85.508$ and $B = 2.38952 \times 10^{-9}$ }\label{table:4}
\end{table*}
\begin{table*}
\centering
\begin{tabular}{|p{2.5cm}|p{4cm}|p{3.5cm}|p{3cm}|p{1cm}|}
\hline
\multicolumn{4}{|c|}{Strong Lensing Coefficients }\\
\hline
 $Q$& $\bar{a}$  & $\bar{b} $ &$ u_{ph}/R_{s}$\\
\hline
Standard Schwarzschild BH&  1.00&-0.40023&2.59808\\
\hline
Standard RN BH(0.3)&  1.05183&-0.396509&2.42935\\
\hline
0& 0.999998&-0.400236&2.59806\\
0.1& 1.00455&-0.399353&2.5806\\
0.2&  1.01973&-0.39719&2.52647\\
0.3&  1.05182&-0.396514&2.42933\\
0.4&  1.12317&-0.413644&2.27298\\
0.5&  1.41422&-0.733223&1.99998\\
\hline
\end{tabular}
\caption{Estimations for the strong lensing coefficients, denoted as $\bar{a}$ and $\bar{b}$, as well as the ratio $u_{ph}/R_{s}$ are sought for varying values of the charge parameter $Q$ in the context of a charged BH with a CDM halo, particularly focusing on the supermassive BH SgrA*. The parameters $A$ and $B$ are specified as $6463.81$ and $2.537 \times 10^{-11}$, respectively.}\label{table:5}
\end{table*}
\begin{table*}
\begin{center}
\begin{tabular}{|c|cccc|cccc|}
\hline
parameters &  & $M87^*$ & &&  &$SgrA^*$&\\
\hline
$Q$ & $ \theta_{\infty} (\mu as)$&$S(\mu as)$&
$r_{mag}$&$\Delta T_{2,1}(minutes)$& $ \theta_{\infty} (\mu as)$&$S(\mu as)$&$r_{mag}$& $\Delta T_{2,1}$ (minutes)\\
\hline
Standard & & & & & & \\
Schwazschild BH&19.9633&0.024984&6.82188&17378.8&26.38&0.0330177&6.82188&11.4973\\
\hline
 Standard Reissner & & & & & & & & \\
 Nordstr\"{o}m BH(0.3)&18.6668&0.0325876&6.48575&16250.2&24.6692&0.0430665&6.48575&10.7507\\
\hline
0&19.96431&0.0249923&6.8219&17378.7&26.3824&0.0330169&6.8219&11.4973\\
\hline
0.1&19.8289&0.0256108&6.79096&17261.9&26.2052&0.0338341&6.79095&11.4200\\
0.2&19.413&0.0277422&6.68987&16899.8&25.6555&0.0366502&6.68986&11.1805\\
0.3&18.666&0.0325976&6.48576&16250.1&24.66931&0.0430655&6.48576&10.7506\\
0.4&17.4652&0.0449611&6.07379&15204.2&23.0813&0.0594023&6.07379&10.6587\\
0.5&15.3676&0.107632&4.82377&13378.1&20.3092&0.142237&4.82378&8.85059\\
\hline
\end{tabular}
 \caption{Estimate various strong lensing observables for CDM halos in the context of supermassive BHs, specifically $M87^*$ and $Sgr A^*$, while considering different values of the charge parameter $Q$.}\label{table:6}
\end{center}
\end{table*}
\section{Comparison with observations}\label{sec4}
In this study, we employ the gravitational lensing method developed by Bozza \cite{Bozza:2002zj} to investigate strong field limit phenomena. This method enables the differentiation of various types of static spherically symmetric BHs and explores their potential astrophysical implications. We specifically compare observational outcomes for a charged BH with a URC, DM and CDM halo. Additionally, we contrast these results with those of standard RN BHs and Schwarzschild BHs, both with and without halos. To facilitate this comparison, we focus on a supermassive BH, denoted as $SgrA^{*}$, positioned at the centers of nearby galaxies. We calculate lensing coefficients, namely $\bar{a}$, $\bar{b}$, and $u_{Ph}/R_s$, as presented in Tables \ref{table:1}, \ref{table:2}, \ref{table:4}, \ref{table:5}, and \ref{table:7}. Furthermore, we estimate various observable quantities such as $\theta_{\infty}$, $S$, and $r_{mag}$, as outlined in Tables \ref{table:3}, \ref{table:6}, and \ref{table:7}. The charged BH scenarios include specific charge values ($Q=0.2,0.3$) in conjunction with URC DM and CDM halos. The comparison also encompasses standard RN BHs with a charge of $Q=0.2$, as well as cases involving Schwarzschild BHs with halos ($Q=0$) and standard Schwarzschild BHs ($A=0, B=0, \& Q=0$). Refer to Tables \ref{table:3}, \ref{table:6}, and \ref{table:7} for detailed numerical values.
\begin{table*}
\centering
\begin{tabular}{ |p{1.1cm}|p{1.2cm}|p{1.4cm}| p{5.7cm}|p{5.6cm}|}
\hline
& {Standard Schwarz\vfill -schild BH}& {Standard\vfill Reissner Nordstrom\vfill BH\vfill($Q=0.2$)}& { Charged BH for URC model \vfill  $A=5738.77$
$B=2.6346*10^{-11} $  }\vfill {\begin{tabular}{cccc}
    \hline
    \multicolumn{1}{c}{}\\
    $Q=0.0$ ~~~ &$Q=0.2$~~&~~~$Q=0.3$\\
\end{tabular}} & {Charged BH for CDM model \vfill $A=6463.81$
$B=2.537*10^{-11} $ }\vfill{\begin{tabular}{cccc}
\hline
    \multicolumn{1}{c}{}\\
$Q=0.0$ ~~~ &$Q=0.2$~~&~~~$Q=0.3$\\
\end{tabular}}\\
\hline
\textbf {$\theta_{\infty}$\vfill($\mu$as)} & 26.3826& 25.6557 &{\begin{tabular}{cccc}
    \hline
    \multicolumn{1}{c}{}\\
    26.3827~~~& ~~~25.6558~&~ ~~~24.6693 \\
\end{tabular}} &{\begin{tabular}{cccc}
    \hline
    \multicolumn{2}{c}{}\\
     ~ 26.3824~~~&~ 25.6555~&~~~ 24.6691\\
\end{tabular}}\\
\hline
\textbf {$S$ \vfill($\mu$as) }&0.0330177& 0.0366511  &{\begin{tabular}{cccc}
    \hline
    \multicolumn{1}{c}{}\\
     0.0330183&~0.0366517&~ 0.043067\\
\end{tabular}} &{\begin{tabular}{cccc}
    \hline
    \multicolumn{1}{c}{}\\
     0.0330169&0.0366502&0.043066\\
\end{tabular}}\\
\hline
$r_{mag}$ &6.82188& 6.68985&{\begin{tabular}{cccc}
    \hline
    \multicolumn{1}{c}{}\\
     6.82187~&~~~~~~6.68984&~~~~~~6.48574\\
\end{tabular}} &{\begin{tabular}{cccc}
    \hline
    \multicolumn{1}{c}{}\\
      ~~ 6.8219~~&~~~6.68986~&~~~~ ~6.48576\\
\end{tabular}}\\
\hline

$\Delta T_{2,1}$\vfill{(min)} &11.4973& 11.1805 &{\begin{tabular}{cccc}
    \hline
    \multicolumn{1}{c}{}\\
     11.4974~&~~~~~~11.1806~&~~~~~~10.7507\\
\end{tabular}} &{\begin{tabular}{cccc}
    \hline
    \multicolumn{1}{c}{}\\
       11.4973~~~~&~~~11.1805~~~&~~~~10.7506\\
\end{tabular}}\\
\hline
{$u_c/R_{sh}$} & 2.59808& 2.52649  &{\begin{tabular}{cccc}
    \hline
    \multicolumn{1}{c}{}\\
    ~ 2.59809~&~~~~~~2.5265~&~~~~~~2.42936\\
\end{tabular}}&{\begin{tabular}{cccc}
    \hline
    \multicolumn{1}{c}{}\\
    ~ 2.59806~&~~~~~2.52647~&~~~~~2.42933\\
\end{tabular}} \\
\hline
 {$\bar{a}$ }& 1 & 1.01974 &{\begin{tabular}{cccc}
    \hline
    \multicolumn{1}{c}{}\\
     1.0000015~&~~~~1.01974~&~~~~1.05183\\
\end{tabular}}&{\begin{tabular}{cccc}
    \hline
    \multicolumn{1}{c}{}\\
     ~ 0.999998~&~~~~1.01973~&~~~~1.05182\\
\end{tabular}}\\
\hline
{$\bar{b}$ }& -0.40023& -0.397184 &{\begin{tabular}{cccc}
    \hline
    \multicolumn{1}{c}{}\\
     -0.400226&~~~ -0.39718&~~~-0.396505\\
\end{tabular}} &{\begin{tabular}{cccc}
    \hline
    \multicolumn{1}{c}{}\\
      -0.400236&~~~-0.39719&~~~-0.396514\\
\end{tabular}}\\
\hline
\end{tabular}
 \caption{Estimating observables involves considering the supermassive BH (SgrA*), characterized by a mass of $4.3\times 10^6 M_{\odot}$ and a distance of $8.35$ kpc. This analysis is conducted within the frameworks of the Schwarzschild BH model, the BH URC model, and the BH CDM  model.}\label{table:7}
\end{table*}
In our analysis, we have observed that, considering equal mass and distance for BHs as outlined in Table \ref{table:7}, the angular position of the innermost images, $\theta_{\infty}$, in the background of a charged BH ($Q=0.2$) with a URC, DM halo is slightly greater by approximately $0.0001\mu as$, and with a CDM halo, it is slightly smaller by about $0.0002\mu as$ compared to the cases of a standard RN BH ($A=B=0, Q=0.2$). Similarly, the angular separation of images, denoted as $S$, in the background of a charged BH ($Q=0.2$) with URC DM halo is slightly greater by approximately $0.0000006\mu as$, and with CDM halo, it is slightly smaller by about $0.0000009\mu as$ compared to the cases of a standard RN BH ($A=B=0, Q=0.2$). However, the relative magnification, denoted as $r_{mag}$, in the background of a charged BH ($Q=0.2$) with URC DM halo is slightly greater by approximately $0.00001$ magnitude, and with CDM halo, it is slightly smaller by about $0.00001$ magnitude compared to the cases of a standard RN BH ($A=B=0, Q=0.2$). It means that the outermost images of the BH with a DM halo are very close to the remaining innermost packed images, making them distinguishable from the other BH images. Furthermore, the observable quantities $\theta_{\infty}$, $S$, and $r_{mag}$ in the background of a charged BH ($Q=0.2$) with URC DM halo are smaller than the cases of a Schwarzschild BH with a halo ($Q=0$) and a standard Schwarzschild BH ($A=0, B=0, Q=0$). From Figs \ref{fig:7}(a) \& \ref{fig:7}(b) and Figs. \ref{fig:15}(a) \& \ref{fig:15}(b), it is also evident that the Einstein ring radius $\theta^E_{n}$ in the background of a charged BH ($Q=0.2$) with both the URC DM halo and CDM halo is smaller than the cases of a standard RN BH ($A=B=0, Q=0.2$), a Schwarzschild BH with a halo ($Q=0$), and a standard Schwarzschild BH ($A=0, B=0, Q=0$). However, one can speculate about how the gravitational field surrounding a charged BH is affected by DM halos. The gravitational field in the vicinity of a BH is enhanced by the existence of DM. The findings from our observations lead to the conclusion that charged BHs, when accompanied by a DM halo (URC \& CDM), may be more easily detectable in strong gravitational lensing observations using existing technology. Moreover, if the outermost image can be resolved, it will facilitate the differentiation of a BH with a DM halo from a standard RN BH, a Schwarzschild BH with a halo, and a standard Schwarzschild BH. It is noteworthy that the gravitational lensing scenario provides a viable method for detecting the audible boundary of sound waves near the charged BH horizon, especially when influenced by DM halos (URC \& CDM). The astrophysical implications of these DM halos associated with BHs could offer valuable insights into the connections between the sonic fluid and the shape of BHs. We anticipate that our analytical discoveries will be instrumental in advancing research on BHs and their properties. These findings are expected to support further investigations into the characteristics of the geometry near the event horizon of typical astrophysical BHs.
\section{Constraints with dark matter halo by EHT observations data of $M87^*$ and $SgrA^*$}\label{sec5}
\begin{figure*}[htbp]
\centering
\includegraphics[width=.45\textwidth]{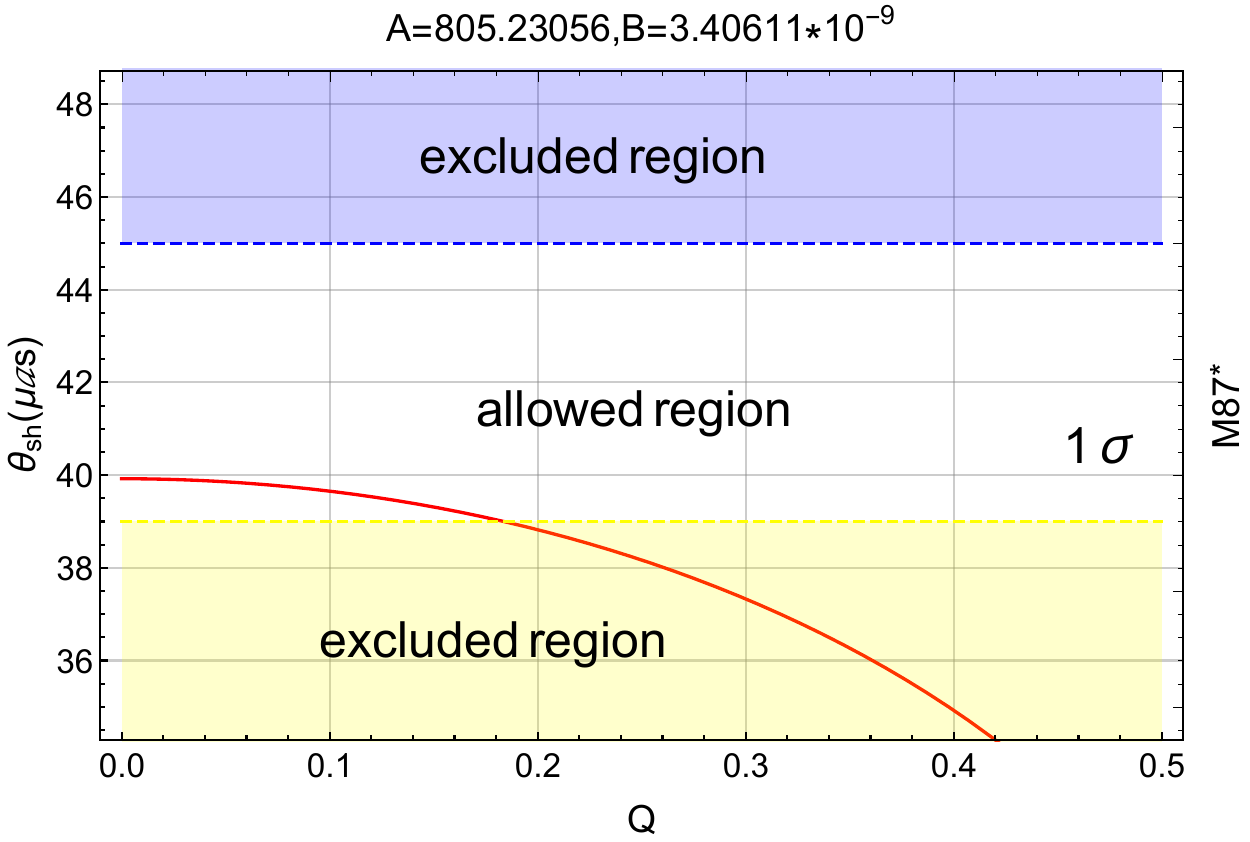}(a)
\qquad
\includegraphics[width=.45\textwidth]{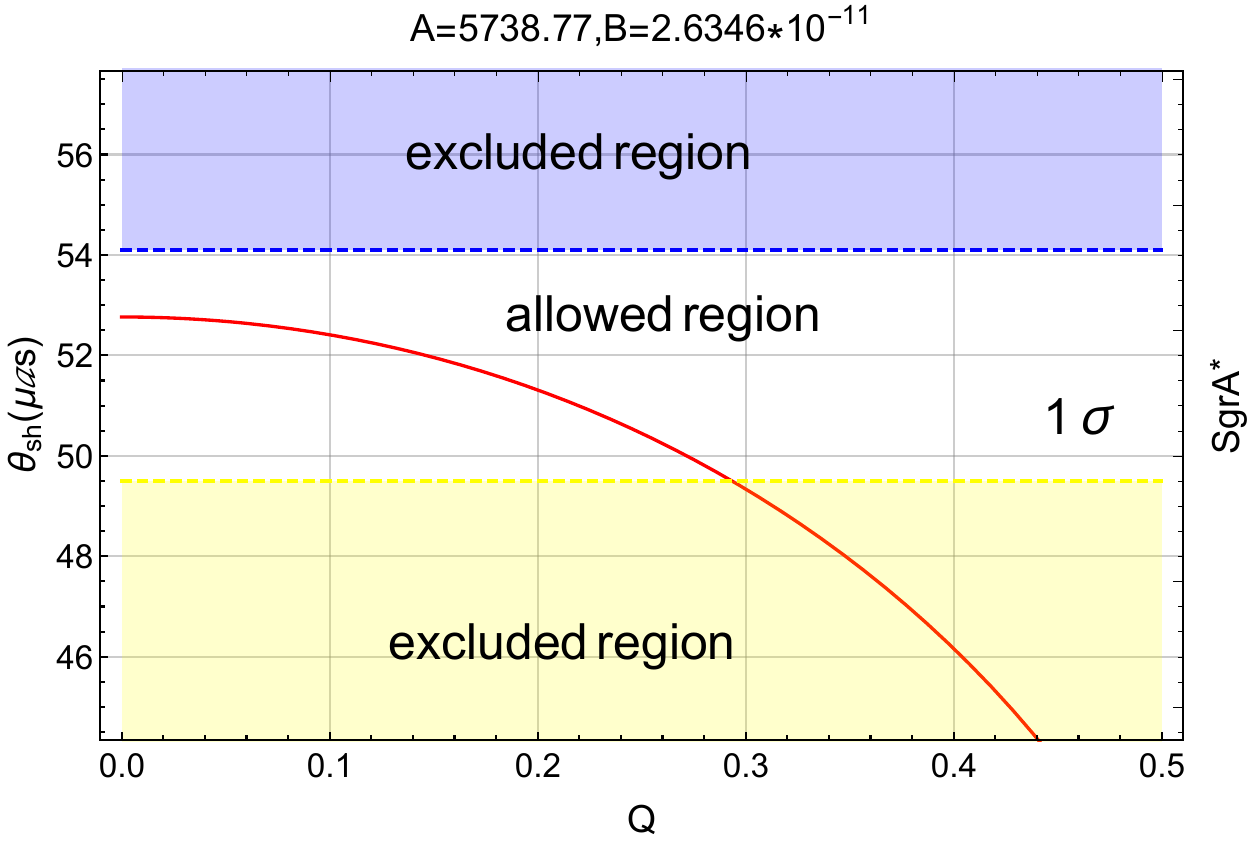}(b)
\caption{The angular diameter of shadow $\theta_{sh}(=\mathit{2\theta_{\infty}}$) vs Q for $M87^{*}$ (left panel) and for $Sgr A^*$(right panel)  for the RN with URC DM halo.The allowed (white) region represents the area that is $1\sigma$ consistent and the excluded (blue and yellow) region represents the areas that are $1\sigma$ inconsistent with EHT observation , suggesting the constraints of the BH charge parameter Q with  URC DM halo.}\label{fig:17}
\end{figure*}

\begin{figure*}[htbp]
\centering
\includegraphics[width=.45\textwidth]{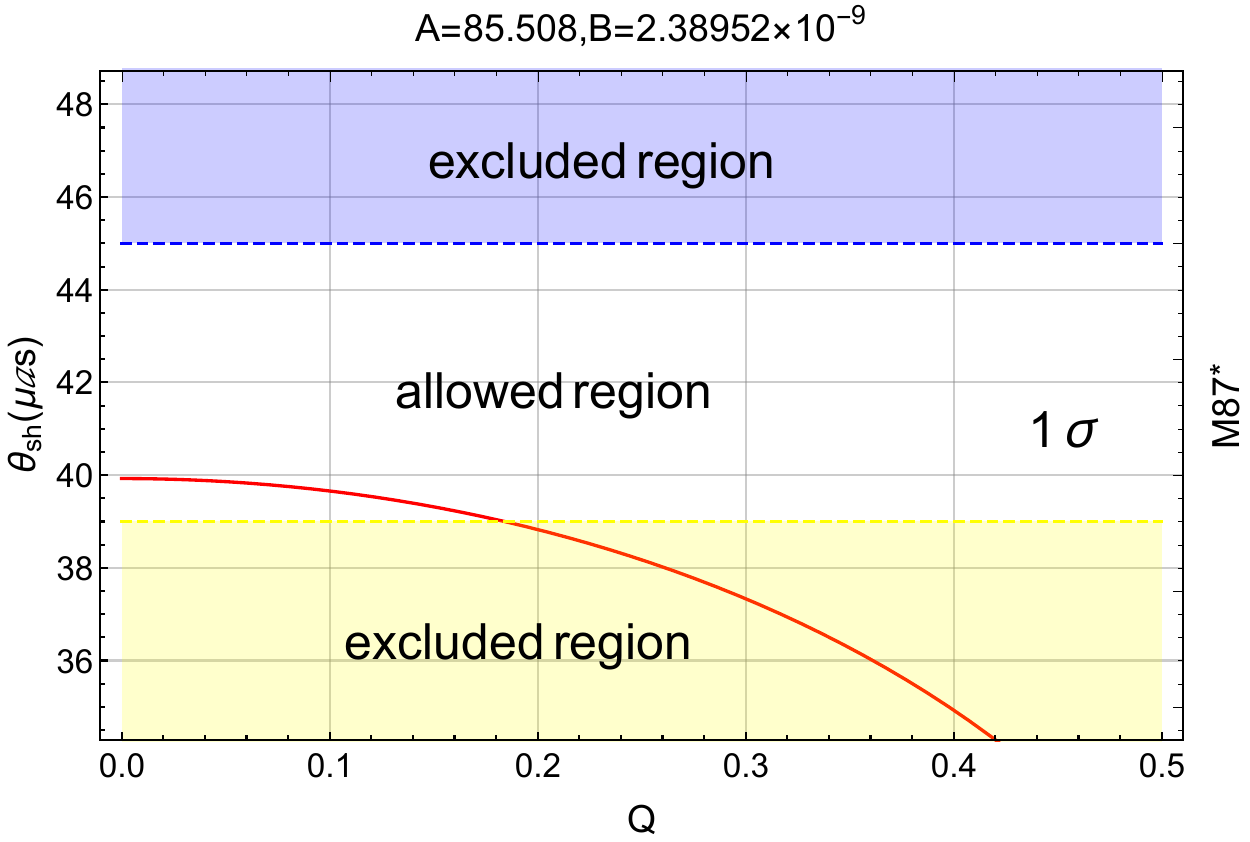}(a)
\qquad
\includegraphics[width=.45\textwidth]{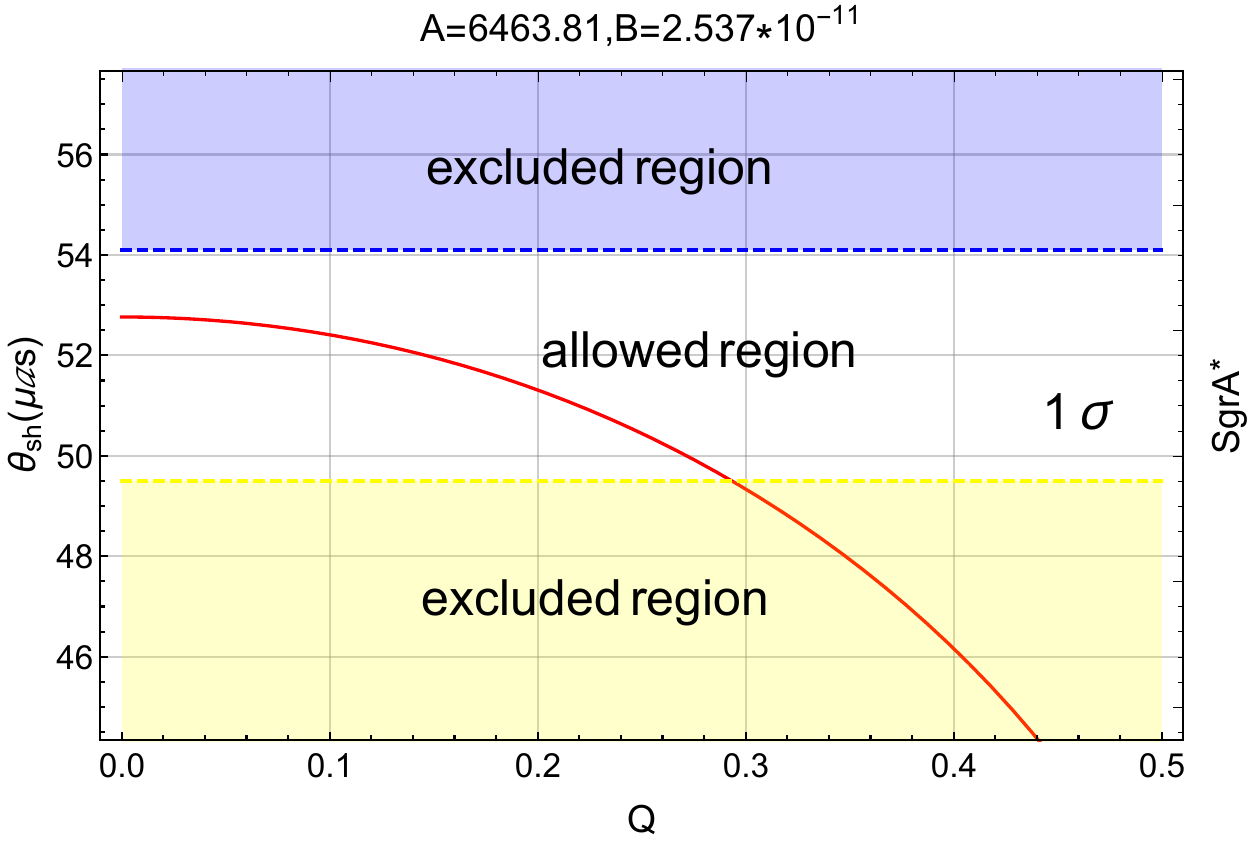}(b)
\caption{The angular diameter of shadow $\theta_{sh}(=\mathit{2\theta_{\infty}}$) vs Q for $M87^{*}$ (left panel) and for $Sgr A^*$(right panel) for the RN with CDM halo.The allowed (white) region represents the area that is $1\sigma$ consistent and the excluded (blue and yellow) region represents the areas that are $1\sigma$ inconsistent With EHT observation , suggesting the constraints of the BH charge parameter Q with CDM halo.}\label{fig:18}
\end{figure*}

In this section, our objective is to ascertain the constraints on the charge parameter of the BH in conjunction with the DM halo, utilizing the observational results from the Event Horizon Telescope (EHT) focused on the $ M87^*$ and $ SgrA^*$ celestial bodies. The EHT collaboration successfully unveiled the initial images of the BH at the center of the neighboring galaxy, $ M87^*$. \cite{E2019dse,EventHorizonTelescope:2019uob,EventHorizonTelescope:2019jan,EventHorizonTelescope:2019ths,EventHorizonTelescope:2019pgp,EventHorizonTelescope:2019ggy}. The discoveries pertaining to the shadow of $M87^*$ offer a captivating exploration into the implications of strong gravitational lensing, providing a robust testing ground for theories of gravity. With a distance of $16.8$ Mpc, the mass of $M87^*$ is determined to be $(6.5\pm 0.7)\times 10^9M_{\odot}$. The observations by the Event Horizon Telescope (EHT) collaboration reveal an angular diameter of the shadow, denoted as $\theta_d$, to be $42\pm 3 \mu as$. In a more recent development in 2022, the EHT collaboration unveiled images of the supermassive BH $SgrA^*$ in our Milky Way galaxy. The corresponding angular diameter of its shadow, $\theta_d$, was determined to be $51.8\pm 2.3 \mu as$, with a mass of $M=(4.3\pm 0.7)\times 10^6M_{\odot}$ and a distance of $d=8.35$ kpc \cite{EventHorizonTelescope:2022wkp}. Utilizing the observable quantity $\theta_{\infty}$, which serves as the angular radius of the BH shadow due to strong lensing, we can constrain the charge parameter $Q$ within the $1\sigma$ level. Analyzing the EHT results, we find that for $M87^*$ as illustrated in Fig. \ref{fig:17}(a)  , the charge parameter $Q$ for a BH with a URC DM halo is constrained to $0\leq |Q|\leq 0.366M$. Similarly, for $SgrA^*$ depicted in Fig. \ref{fig:17}(b), the constraints are $0\leq |Q|\leq 0.586M$. In the case of a BH with a CDM halo, the constraints for $M87^*$ shown in Fig. \ref{fig:18}(a) are $0\leq |Q|\leq 0.364M$, while for $SgrA^*$ illustrated in Fig.\ref{fig:18}(b), the constraints are $0\leq |Q|\leq 0.584M$. These findings suggest that charged BHs with a DM halo, whether URC or CDM, conform to the constraints set by the EHT observations. Moreover, the study implies the potential detectability of charged BHs with a DM halo in future observations.

\section{ Discussion and Conclusions}\label{sec6}
In this study, we investigated the gravitational lensing effects around charged BHs in the presence of DM halos. Specifically, we focused on supermassive BHs, namely \(M87^*\) and \(Sgr A^*\). Two distinct DM halo models, the URC and CDM models, were considered. The lapse functions corresponding to these models are provided in Eqs  \eqref{5} and  \eqref{8}. The determination of the parameters involved, such as \(\rho_0\) and \(r_0\), was achieved through the utilization of observational data pertaining to \(M87^*\) and \(Sgr A^*\).\\\\
We have conducted an investigation into the influence of the URC and CDM models, along with the charge parameter \(Q\), on the strong gravitational lensing phenomenon. Our study encompasses various observable quantities, including the angular position \(\theta_{\infty}\), separation \(S\), relative magnification \(r_{\text{mag}}\), Einstein rings for the relativistic images, and time delays \(\Delta T_{2,1}\) for relativistic images. We specifically examined these effects in the context of supermassive BHs \(M87^{*}\) and \(Sgr A^{*}\). To provide a comprehensive comparison, we contrasted these outcomes with scenarios involving a standard RN BH, a Schwarzschild BH with a halo, and a standard Schwarzschild BH. Firstly, we derive the null geodesic equations for the charged BH spacetime with DM halos using the
Hamilton-Jacobi action. Subsequently, we utilize these equations to determine the photon sphere radius, denoted as $r_{ph}$.
It is seen that the radius of photon sphere  $\mathit{r_{ph}}$ decreases with the magnitude of the charge parameter $Q$; and its value is greater than the corresponding case of standard RN BH and lesser than the corresponding cases of  Schwarzschild BH with halo and standard Schwarzschild BH for both the cases of URC DM and CDM halo models.\\\\
We numerically and graphically obtained the strong lensing
coefficients   $\mathit{\bar{a}}$ and $\mathit{\bar{b}}$
,$\mathit{u_{ph}/R_s}$ for both URC and CDM models. It is found
that the strong lensing coefficient $\mathit{u_{ph}/R_s}$ are decreased with charge parameters $Q$ and $\mathit{u_{ph}/R_s}$
for the case of charged BH with URC as well as CDM DM halo is a
little bit greater than the case of standard RN (
$Q=0.3$ ) and smaller than the case of standard Schwarzschild BHs. The coefficients for strong lensing, denoted as \(\bar{a}\) and \(\bar{b}\), in both the scenarios of URC and CDM models vary with the charge parameter \(Q\). Notably, the deflection limit coefficient \(\bar{a}\) tends to increase with the magnitude of \(Q\). On the other hand, \(\bar{b}\) initially increases with \(Q\), reaches a maximum, and then decreases with further increments in \(Q\). Analyzing the strong deflection angle \(\alpha_D\) for both URC and CDM models, it is evident that \(\alpha_D\) rises with the magnitude of the charge parameter \(Q\). Moreover, the deflection angle for the case of a charged BH with URC and CDM halo surpasses that of the standard RN (green line) and standard Schwarzschild (red line) BHs. This suggests that the presence of a charged BH parameter with URC and CDM halo enhances gravitational bending effects. Additionally, the strong deflection angle decreases with the minimum impact parameter \(u\) and diverges at the critical impact parameter \(u=u_{ph}\) corresponding to the photon radius \(r=r_{ph}\). This implies that charged BHs, in the presence of a DM halo and sound waves, may exhibit more detectable gravitational bending effects.\\\\
The strong lensing observable quantities, the angular image position$ \theta_{\infty}$, and relative magnification $r_{mag}$  decreases with the parameter $Q$ while angular separation $S $  increases with the parameter $Q$. Furthermore, the observable quantity, the angular image position$ \theta_{\infty}$  for the cases of URC \& CDM  halo models is a little bit greater than the case of standard RN ( $Q=0.3$ )and smaller than the cases of Schwarzschild BH with halo as well as standard Schwarzschild BHs while the observable quantity, the angular image separation $ S$  for the case of charged BH with URC \& CDM halo is a little bit greater than the cases of standard RN as well as for the cases of Schwarzschild BH with halo and standard Schwarzschild BHs, but the observable quantity, the relative magnification $r_{mag}$ for the case of charged BH with URC \& CDM halo is a little bit smaller than the cases of standard RN as well as for the cases of Schwarzschild BH with halo and standard Schwarzschild BHs.\\\\
In the scenario of the URC DM halo, where \(0 \leq Q \leq 0.5\), the angular image position \(\theta_{\infty}\) falls within the range of \(15.369 \, \mu \text{as}\) to \(19.965 \, \mu \text{as}\) for \(M87^{*}\), and \(20.3 \, \mu \text{as}\) to \(26.3828 \, \mu \text{as}\) for \(SgrA^{*}\). Simultaneously, the angular image separation \(S\) spans from \(0.024 \, \mu \text{as}\) to \(0.1077 \, \mu \text{as}\) for \(M87^{*}\), and \(0.033 \, \mu \text{as}\) to \(0.1423 \, \mu \text{as}\) for \(SgrA^{*}\). The magnification \(r_{\text{mag}}\) varies between \(4.824\) and \(6.8218\) magnitudes for \(M87^{*}\) and \(4.823\) to \(6.822\) magnitudes for \(SgrA^{*}\). In the case of the CDM halo with \(0 \leq Q \leq 0.5\), the angular image position \(\theta_{\infty}\) ranges from \(15.368 \, \mu \text{as}\) to \(19.9644 \, \mu \text{as}\) for \(M87^{*}\) and \(20.3 \, \mu \text{as}\) to \(26.3825 \, \mu \text{as}\) for \(SgrA^{*}\). The angular image separation \(S\) covers \(0.024 \, \mu \text{as}\) to \(0.1077 \, \mu \text{as}\) for \(M87^{*}\) and \(0.033 \, \mu \text{as}\) to \(0.1423 \, \mu \text{as}\) for \(SgrA^{*}\). The magnification \(r_{\text{mag}}\) varies within \(4.8237\) to \(6.822\) magnitudes for \(M87^{*}\) and \(4.8237\) to \(6.822\) magnitudes for \(SgrA^{*}\). It is also observed that the Einstein ring radius  $\theta^E_{n}$ in the background of charged BH ($Q=0.2$) with both the context of URC DM halo and CDM halo is smaller than the cases of standard RN BH ($A=B=0,Q=0.2$), Schwarzschild BH with halo ($Q=0$) and standard Schwarzschild BH($A=0,B=0,Q=0$). The findings in our observation concluded that the charged BHs, with the presence of DM halo (URC \& CDM), might be more easily detectable in strong gravitational lensing observations using existing technology. Moreover, if the outermost image can be resolved, it will differentiate the BH with a DM halo from standard RN BH, Schwarzschild BH with a halo and standard Schwarzschild BH. This will allow for the characterization of the BHs with a DM halo using existing technology.\\\\
Another important astrophysical consequence is the time delay $\Delta T_{2,1}$ between two different relativistic images, which are decreased with the charge parameters $Q$ in the context of $M87^{*}$ and $Sgr A^{*}$ BHs with  URC \& CDM halo. It is further observed that the time delay $\Delta T_{2,1}$ is smaller than the cases of standard RN BH($Q=0.3$), Schwarzschild BH with URC \& CDM halo as well as standard Schwarzschild BH. Deviation of the time delay $\Delta T_{2,1}$ for Charged BH (Q=0.3)with URC halo in having the same mass and distance respectively from the standard RN BH ($A=B=0,Q=0.3$), Schwarzschild BH with halo ($Q=0$)  and standard Schwarzschild BH ($A=B=Q=0$) are $\sim 2.2 $ minutes,$\sim 1,129 $ minutes and$\sim 1130 $ minutes in the context of  $M87^*$ supermassive BH. Deviation of the time delay $\Delta T_{2,1}$ for Charged BH ($Q=0.3$)with CDM halo in having the same mass and distance respectively from the standard RN BH ($A=B=0,Q=0.3$), Schwarzschild BH with halo ($Q=0$) and standard Schwarzschild BH ($A=B=Q=0$) are $\sim 0.1 $ minutes,$\sim 1,128.6 $ minutes and$\sim 1128.7 $ minutes in the context of  $M87^*$ supermassive BH.\\\\
Therefore, the findings in our analysis suggest how the charged BH
with DM halo ( URC \& CDM), is distinguishable from the other
astrophysical BHs such as standard RN BH,
Schwarzschild BH with URC \& CDM halo as well as standard Schwarzschild BH via strong gravitational lensing observation. As a continuation of the present work, we will  study strong gravitational lensing using different DM profiles. We also intend to continue working on the charged rotating BHs. Lastly, we may investigate how different BH solutions of modified theories of gravity are affected by DM
halos.
\section{Acknowledgements}
N.U.M would like to thank  CSIR, Govt. of
India for providing Senior Research Fellowship (No. 08/003(0141))/2020-EMR-I). This research is partly supported by Research Grant F-FA-2021-510 of the Uzbekistan Ministry for Innovative Development. S.C. acknowledges the support of {\it Istituto Nazionale di Fisica Nucleare} (sez. di Napoli), {\it Iniziative Specifiche} QGSKY and MOONLIGHT2.

\bibliographystyle{unsrt}  
\bibliography{mybib}


\end{document}